\documentclass[final,onefignum,onetabnum]{siamart171218}

\usepackage{lipsum}
\usepackage{multicol}
\usepackage{amsfonts}
\usepackage{amssymb}
\usepackage{graphicx}
\usepackage{epstopdf}
\usepackage{algorithmic}
\usepackage{xspace}
\usepackage{stmaryrd}  
\usepackage{bm}
\usepackage{makecell}
\usepackage{graphicx,epstopdf}
\usepackage[caption=false]{subfig}
\usepackage{sidecap}
\usepackage{bbm}
\usepackage{cancel}
\usepackage[10pt]{moresize}
\ifpdf
  \DeclareGraphicsExtensions{.eps,.pdf,.png,.jpg}
\else
  \DeclareGraphicsExtensions{.eps}
\fi

\newsiamremark{remark}{Remark}
\newsiamremark{hypothesis}{Hypothesis}
\crefname{hypothesis}{Hypothesis}{Hypotheses}
\newsiamthm{claim}{Claim}
\newsiamthm{observation}{Observation}
\newsiamremark{notation}{Notation}

\headers{On mass conservation for ice sheets}{E. M. Cummings}

\title{On mass conservation for ice sheets\thanks{Submitted to the editors DATE.
}}

\author{Evan M.~Cummings\thanks{Alfred Wegener Institute, Bremerhaven, Germany (\email{evan.cummings@awi.de}).}}

\usepackage{amsopn}

\let\Gamma\varGamma
\let\Omega\varOmega
\let\Sigma\varSigma

\newcommand{\T}{^\intercal}
\renewcommand{\d}[1]{\ \mathrm{d}#1}
\newcommand{\diff}[1]{\mathrm{d}#1}
\newcommand{\unit}[1]{\hat{\underline{#1}}}

\newcommand{\rankone}[1]{\underline{#1}}
\newcommand{\ranktwo}[1]{\underline{\underline{#1}}}

\newcommand{\totder}[3][]{\frac{\mathrm{d}^{#1} #2}{\mathrm{d} #3^{#1} }}
\newcommand{\parder}[3][]{\frac{\partial^{#1} #2}{\partial #3^{#1} }}
\newcommand{\R}{\mathbb{R}}
\newcommand{\bigo}{\mathcal{O}}
\newcommand{\Lu}{\mathcal{L}}
\newcommand{\normal}{\rankone{\hat{n}}}
\newcommand{\jump}[1]{\left\llbracket {#1} \right\rrbracket}
\newcommand{\fs}{\mathring{F}} 
\newcommand{\smb}{\mathring{S}} 
\newcommand{\bmb}{\mathring{B}} 

\newcommand{\hairsp}{\hspace{1pt}}
\newcommand{\ie}{\emph{i.\hairsp{}e.}\xspace}
\newcommand{\eg}{\emph{e.\hairsp{}g.}\xspace}
\newcommand{\etc}{\emph{etc}}

\newcommand{\cf}{\textit{cf.}\xspace}

\newcommand{\ice}{\mathrm{i}} 
\newcommand{\water}{\mathrm{w}}
\newcommand{\air}{\mathrm{a}}
\newcommand{\liquid}{\mathrm{\ell}}
\newcommand{\solid}{\mathrm{s}}
\newcommand{\vapor}{\mathrm{v}}

\newcommand{\seawater}{\mathrm{sw}}

\newcommand{\e}{\mathrm{e}}

\newcommand{\bed}{\mathrm{B}}
\newcommand{\srf}{\mathrm{S}}

\newcommand{\ana}{\mathrm{a}}

\usepackage[final]{listings}

\usepackage{color}
\usepackage{xcolor}

\definecolor{bg}{rgb}{0.96,0.96,0.85}
\definecolor{deepblue}{rgb}{0,0,0.5}
\definecolor{deepred}{rgb}{0.6,0,0}
\definecolor{deepgreen}{rgb}{0,0.5,0}
\definecolor{gray}{gray}{0.5}

\colorlet{commentcolour}{cyan!40!black}
\colorlet{stringcolour}{magenta!40!blue}
\colorlet{keywordcolour}{red!90!black}
\colorlet{exceptioncolour}{yellow!50!red}
\colorlet{commandcolour}{blue!60!black}
\colorlet{numpycolour}{green!70!blue!90!black}
\colorlet{digitcolour}{cyan!70!blue!90!black}
\colorlet{literatecolour}{magenta!90!black}
\colorlet{promptcolour}{green!80!black}
\colorlet{specmethodcolour}{violet}
\colorlet{indendifiercolour}{green!70!white}

\newcommand{\literatecolour}{\textcolor{literatecolour}}

\newcommand\pythonstyle{\lstset{
language=python,
showtabs=true,
tab=,
tabsize=2,
basicstyle=\ttfamily\ssmall,
stringstyle=\color{stringcolour},
showstringspaces=false,
alsoletter={1234567890},
otherkeywords={\ , \}, \{, \%, \&, \|},
keywordstyle=\color{keywordcolour}\bfseries,
emph={and,break,class,continue,def,yield,del,elif ,else,%
except,exec,finally,for,from,global,if,import,in,%
lambda,not,or,pass,print,raise,return,try,while,assert},
emphstyle=\color{blue}\bfseries,
emph={[2]True, False, None},
emphstyle=[2]\color{keywordcolour},
emph={[3]object,type,isinstance,copy,deepcopy,zip,enumerate,reversed,list,len,dict,tuple,xrange,append,execfile,real,imag,reduce,str,repr},
emphstyle=[3]\color{commandcolour},
emph={Exception,NameError,IndexError,SyntaxError,TypeError,ValueError,OverflowError,ZeroDivisionError},
emphstyle=\color{exceptioncolour}\bfseries,
morestring=[s]{"""}{"""},
morestring=[s]{'''}{'''},
commentstyle=\color{commentcolour}\slshape,
emph={[4]ode, fsolve, sqrt, exp, sin, cos, arccos, pi,  array, norm, solve, dot, arange, , isscalar, max, sum, flatten, shape, reshape, find, any, all, abs, linspace, legend, quad, polyval,polyfit, hstack, concatenate,vstack,column_stack,empty,zeros,ones,rand,vander,grid,pcolor,eig,eigs,eigvals,svd,qr,tan,det,logspace,roll,min,mean,cumsum,cumprod,diff,vectorize,lstsq,cla,eye,xlabel,ylabel,squeeze,plot,median,std,hist},
emphstyle=[4]\color{numpycolour},
emph={[5]__init__,__add__,__mul__,__div__,__sub__,__call__,__getitem__,__setitem__,__eq__,__ne__,__nonzero__,__rmul__,__radd__,__repr__,__str__,__get__,__truediv__,__pow__,__name__,__future__,__all__},
emphstyle=[5]\color{specmethodcolour},
emph={[6]assert,range,yield},
emphstyle=[6]\color{keywordcolour}\bfseries,
literate=*%
{:}{{\literatecolour:}}{1}%
{=}{{\literatecolour=}}{1}%
{-}{{\literatecolour-}}{1}%
{+}{{\literatecolour+}}{1}%
{*}{{\literatecolour*}}{1}%
{/}{{\literatecolour/}}{1}%
{!}{{\literatecolour!}}{1}%
{[}{{\literatecolour[}}{1}%
{]}{{\literatecolour]}}{1}%
{<}{{\literatecolour<}}{1}%
{>}{{\literatecolour>}}{1}%
{>>>}{{\textcolor{promptcolour}{>>>}}}{1}%
,%
breaklines=true,
breakatwhitespace= true,
aboveskip=1ex,
frame=trbl,
rulecolor=\color{black!40},
backgroundcolor=\color{black!5}
}}

\lstnewenvironment{python}[1][]
{
  \pythonstyle
  \lstset{#1}
}
{}

\newcommand{\pythonexternal}[2][]
{{
  \pythonstyle
  \lstinputlisting[#1]{#2}
}}

\newcommand\pythoninline[1]
{{
  \pythonstyle
  \lstinline!#1!
}}

\crefname{lstlisting}{Listing}{listings}
\Crefname{lstlisting}{Listing}{Listings}

\ifpdf
\hypersetup{
  pdftitle={On mass conservation for ice sheets},
  pdfauthor={E. M. Cummings}
}
\fi

\begin{document}

\maketitle

\begin{center}
\small \textit{In memory of Professor Toma `Thomas' Tonev, 1945--2015.\footnote{For inspiring me to continue my studies in mathematics at the University of Montana.}}
\end{center}

\begin{abstract}
  A new continuum-mechanical formulation is proposed which encompasses all material processes within and surrounding an ice sheet.
  Using this formulation, the balance of mass and free-surface relations for ice sheets are derived and elaborated upon.
  The resulting three-dimensional mass-balance relation is then integrated vertically to produce an ice-sheet quasi-mass-balance relation in the horizontal plane, and is demonstrated to reduce velocity errors in regions with high magnitude surface gradients.
  An analytic velocity satisfying the corrected mass-balance relation is formulated as a means to verify future numerical ice-sheet models.
\end{abstract}



\begin{DOI}
  00.0000/000000000
\end{DOI}

\begin{multicols}{2}
\noindent
\textsc{Notation.}

\noindent
\begin{tabular}{lll}
$u$                             & rank-zero tensor (scalar) \\
$\rankone{u}$                   & rank-one tensor (vector) \\
$\ranktwo{u}$                   & rank-two tensor (matrix) \\
$\rankone{u}\T, \ranktwo{u}\T$  & transpose of tensor $\rankone{u}$ or $\ranktwo{u}$ \\
$\rankone{u}_{\perp}$, $\rankone{u}_{\Vert}$ & perp./tangent comp.~of $\rankone{u}$ \\
$\rankone{u} \cdot \rankone{v}$ & scalar product \\
$\bar{u}, \bar{\rankone{u}}$    & vertical average of $u$ or $\rankone{u}$ \\
$\unit{u}$                      & normal vector $\unit{u} = \rankone{u}/\Vert \rankone{u} \Vert$ \\
$\partial_i u$                  & partial deriv.~of $u$ w.r.t.~$i$ \\
$\diff_i u$                     & derivative of $u$ w.r.t.~$i$ \\
$\nabla u$                      & gradient of $u$ \\
$\nabla \cdot \rankone{u}$      & divergence of $\rankone{u}$ \\
$\dot{u}$                       & time derivative of $u$ \\
$\mathring{u}$                  & non-diff.~rate of change of $u$ \\
\end{tabular}

\vspace{2mm}
\noindent
\textsc{Empiricals.}

\noindent
\begin{tabular}{l l l l}
$\rho_{\air}$      & $\sim 1.2$    & kg m$^{-3}$  & air density \\
$\rho_{\ice}$      & $\sim 910$    & kg m$^{-3}$  & ice density \\
$\rho_{\water}$    & $\sim 1000$   & kg m$^{-3}$  & water density \\
$\rho_{\seawater}$ & $\sim 1028$   & kg m$^{-3}$  & seawater den. \\
\end{tabular}

\vspace{2mm}
\noindent
\textsc{Superscripts.}

\noindent
\begin{tabular}{c l}
$+$    & exterior region \\
$-$    & interior region \\
$\e$   & sub-element region \\
\end{tabular}

\columnbreak

\vspace{2mm}
\noindent
\textsc{Subscripts.}

\noindent
\begin{tabular}{l l l}
$x,y,z$     & Cartesian coordinate \\
$\solid$    & solid \\
$\liquid$   & liquid \\
$\vapor$    & vapor \\
$\ice$      & ice \\
$\water$    & water \\
$\seawater$ & seawater \\
$\air$      & air \\
$\bed$      & lower surface \\
$\srf$      & upper surface \\
\end{tabular}

\vspace{2mm}
\noindent
\textsc{Variables.}

\noindent
\begin{tabular}{lll}
$\omega$                 & --           & mass fraction \\
$\alpha$                 & --           & volume fraction \\
$m$                      & kg           & mass \\
$\rho$                   & kg m$^{-3}$  & mass density \\
$F$                      & m            & implicit surface \\
$x$                      & m            & position \\
$u$                      & m s$^{-1}$   & fluid velocity \\
$w$                      & m s$^{-1}$   & surface velocity \\
$q$                      & kg m$^{-2}$  & surface-mass density \\
$\sigma$                 & Pa           & stress \\
$\Omega$                 & m$^3$        & interior domain \\
$\Gamma$                 & m$^2$        & boundary of $\Omega$ \\
$\Sigma$                 & m$^2$        & discontinuity surface \\
\end{tabular}

\end{multicols}

\newpage

\section*{Introduction}

Continuum-mechanical formulations which describe the dynamics of ice sheets and glaciers are complicated due to varied environmental interactions, including the atmosphere, lithosphere, sub- and supra-surface lakes, the oceans, \etc; clearly-defined discontinuities within the ice-sheet interior; water transport within and surrounding the ice sheet; and conservation laws associated with compressible multi-phase flows.
In order to describe these processes, the basis of a new continuum-mechanical formulation for ice sheets---\ie mass conservation---is herein proposed.

It has been taken axiomatic within the glaciological community that the evolution of ice-sheet mass is governed by the equation
$\partial_t H + \nabla \cdot \left( H \rankone{\bar{u}} \right) = \smb + \bmb$,
with ice thickness $H$, vertically-averaged ice velocity $\rankone{\bar{u}}$, and accumulation/ablation functions over upper and lower surfaces $\smb$ and $\bmb$.
This equation is commonly stated without derivation, justification, or citations; \eg, equations (28) of \cite{larour_2012}, (1) of \cite{schoof_2007}, (1) of \cite{bueler_2007}, (6) of \cite{payne_1997}, (2) of \cite{rutt_2009}, (2.13) of \cite{leng_2014}, and (1) of \cite{morlighem_2011}.
The original use of this relation is unknown; however, one may conjecture that this relation has been formed by assuming that the ice-sheet thickness flux in the $xy$ plane can be described with a forcing term $\mathring{H} = \smb + \bmb$ that is independent of variations in the upper surface $S$ and lower surface $B$.
The results of \cite{hutter_1982} are herein revisited and lead to the conclusion that this assumption is inappropriate over regions associated with steep surface gradients.
The correct vertically-integrated mass-balance equation for incompressible materials such as ice is thereafter derived and elaborated upon.

The basis of any numerical model consists of an analytic solution; that is, the verification of computer-code integrity is necessary for numerical model development.
Therefore, in order for new numerical models of ice-sheet dynamics to make use of the concepts developed here, an analytic velocity is proposed which satisfies any combination of upper- and lower-surface mass-balance boundary conditions.

The paper is organized as follows.
A description of the cryosphere environment and associated interactions is first presented in \cref{sec_the_cryosphere}, followed by a description of the surface-mass balance in \cref{sec_surface_mass_balance}, an explanation of global- and local-mass conservation in \cref{sec_global_local_mass_conservation},
derivation of the kinematic free-surface relations in \cref{sec_free_surface}, derivation of the vertically-integrated mass-balance equation and error quantification associated with commonly-applied assumptions in \cref{sec_vertically_integrated_mass_balance}, and formulation of a generalized three-dimensional analytic solution for incompressible mass conservation incorporating both the upper and lower surface-mass-balance terms in \cref{sec_r3_analytic_solution}.
The paper is concluded with a short statement of intended future work in \cref{sec_future_work}.

\section{The multi-phase cryosphere} \label{sec_the_cryosphere}

Ice sheets and glaciers may be represented as a system of differential multi-phase equations of disperse flow (\cf \cite{brennen_2005}).
These equations were first stated for binary ice and water mixtures by \cite{hutter_1982}, which is a modification to concepts developed in \cite{fowler_1982}.
The formulation developed here incorporates statements of vapor conservation, which is required to describe surface processes such as evaporation, condensation, sublimation, and deposition (\cf \cite{box_2001}).
Within the ice interior, vapor conservation is needed to describe mechanical effects such as firn densification or closure of cavities between ice grains.

\begin{SCfigure}
  \centering
    \def\svgwidth{0.4\linewidth}
    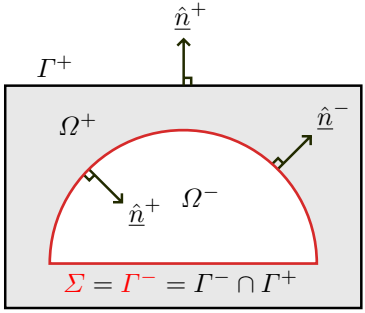
  \caption{Two-material domain with interior volume $\Omega^-$ and exterior volume $\Omega^+$ which share a common boundary $\Sigma$.  The exterior surface $\partial\Omega = \Gamma$ is in this case given by $\Gamma = \Gamma^+ / \left(\Gamma^+ \cap \Gamma^-\right)$; this analysis does not consider effects over $\Gamma$.}
  \label{fig_domain}
\end{SCfigure}

Consider the cryosphere domain $\Omega = \Omega(\rankone{x},t)$ composed of the ice-sheet interior $(-)$ and exterior $(+)$ domains $\Omega^{\pm} = \Omega^{\pm}(\rankone{x},t)$ such that $\Omega = \Omega^- \cup \Omega^+$ and $\Omega^- \cap \Omega^+ = \emptyset$.\footnote{Singular surfaces exist within both $(-)$ and $(+)$ domains and as such may be further decomposed.}
Further define the boundaries $\Gamma^{\pm} = \partial \Omega^{\pm}(\rankone{x},t)$, ice-sheet surface $\Sigma = \Gamma^- \cap \Gamma^+$, and outward-pointing-unit normals $\normal^{\pm}$ (\cref{fig_domain}).
Imagine within $\Omega$ an ice-mixture mass density $\rho = \rho(\rankone{x},t)$ which is composed of some amount of solid ($\solid$), liquid ($\liquid$), and vapor ($\vapor$) components, each moving relative to \emph{barycentric velocity} $\rankone{u} = \rankone{u}(\rankone{x},t)$ defined by \cite{moebius_1827} satisfying
\begin{align}
  \label{barycentric_velocity}
  \rho^{\pm} \rankone{u}^{\pm} = \rho_{\solid}^{\pm} \rankone{u}_{\solid}^{\pm} + \rho_{\liquid}^{\pm} \rankone{u}_{\liquid}^{\pm} + \rho_{\vapor}^{\pm} \rankone{u}_{\vapor}^{\pm},
  \hspace{15mm}
  \rho^{\pm} = \rho_{\solid}^{\pm} + \rho_{\liquid}^{\pm} + \rho_{\vapor}^{\pm},
\end{align}
where the \emph{extrinsic}-mass densities $\rho_k = \rho_k(\rankone{x},t)$, $k = \solid, \liquid, \vapor$ are defined as the component masses per unitary volume, and are connected bijectively to the constant empirically-derived ice ($\ice$), water ($\water$), and air ($\air$) \emph{intrinsic}-mass densities $\rho_q$ for $q = \ice, \water, \air$ via
\begin{align}
  \label{partial_densities}
  \rho_{\solid}^{\pm} = \alpha_{\solid}^{\pm} \rho_{\ice},
  \hspace{10mm}
  \rho_{\liquid}^{\pm} = \alpha_{\liquid}^{\pm} \rho_{\water},
  \hspace{10mm}
  \rho_{\vapor}^{\pm} = \alpha_{\vapor}^{\pm} \rho_{\air},
\end{align}
where $\alpha_k^{\pm} = \alpha_k^{\pm}(\rankone{x},t)$ is the \emph{volume fraction} of phase $k$ with property $\sum_k \alpha_k^{\pm} = 1$.

\begin{remark}
Stating the extrinsic-mass density of each phase in the manner of \cref{barycentric_velocity,partial_densities} allows the mass to vary continuously both within and surrounding the ice sheet domain $\Omega^-$.
For example, the atmosphere boundary will have $\rho_{\liquid}^+ = \rho_{\water}$ in regions in contact with collected surface water, the basal surface may have a range of values $\rho_{\liquid}^+ \in [0, \rho_{\water}]$ depending on the saturation of the sub-glacial aquifer, and $\rho_{\liquid}^+ = \rho_{\seawater}$ over regions in contact with the ocean.
\end{remark}

Division of \cref{barycentric_velocity}$_2$ by $\rho^{\pm}$ produces the \emph{equation of state}
\begin{align}
  \label{equation_of_state}
  1 = \omega_{\solid}^{\pm} + \omega_{\liquid}^{\pm} + \omega_{\vapor}^{\pm},
  \hspace{15mm}
  \omega_k^{\pm} = \rho_k^{\pm} / \rho^{\pm},
  \hspace{5mm}
  k = \solid, \liquid, \vapor;
\end{align}
where $\omega_k^{\pm} = \omega_k^{\pm}(\rankone{x},t)$ is the \emph{mass fraction} of phase $k$.
In addition, taking the derivative of mixture density \cref{barycentric_velocity}$_2$ with respect to time yields 
\begin{align}
  \label{mixture_density_source}
  \mathring{\rho} = \mathring{\rho}_{\solid}^{\pm} + \mathring{\rho}_{\liquid}^{\pm} + \mathring{\rho}_{\vapor}^{\pm} && \text{in } \Omega
\end{align}
with individual components $\mathring{\rho}_k$ in units of mass per unitary volume per time, and are clearly positive for mass gain and negative for mass loss.\footnote{\label{footnote_mathring}The notation $(\ \mathring{ }\ )$ denotes a time rate of change in the sense of \cref{def_generalized_conservation_equation}.}

In addition to mass fluctuations within $\Omega$ given by \cref{mixture_density_source}, there will be also be mass-phase changes on the surface $\Sigma$ due to interactions between the $(\pm)$ domains given by
\begin{align}
  \label{surface_mixture_density_source}
  \mathring{q} = \mathring{q}_{\solid} + \mathring{q}_{\liquid} + \mathring{q}_{\vapor} && \text{on } \Sigma
\end{align}
where $\mathring{q}_k$ are in units of mass per unitary surface area per time, and are positive for mass gain and negative for mass loss over the surface $\Sigma$.
The following proposition illustrates that source terms \cref{mixture_density_source,surface_mixture_density_source} are in fact homogeneous:

\begin{proposition}
\label{thm_homogeneous_total_mass_reaction}
The total rate of change of the mass density $\rho = \rho(\rankone{x},t)$ and surface-mass density $q = q(\rankone{x},t)$ given by \cref{mixture_density_source} and \cref{surface_mixture_density_source} are respectively $\mathring{\rho} = 0$ and $\mathring{q} = 0$ for all $\rankone{x}$ and $t$.
\end{proposition}


\begin{proof}
Considering all mass sources and sinks of partial densities \cref{partial_densities}, the total-reaction rates encompassing \cref{mixture_density_source,surface_mixture_density_source} for each phase are
\begin{align}
  \label{solid_mass_reaction}
  \mathring{\rho}_{\solid}^{\pm} &= \mathring{\rho}_{\liquid \solid}^{\pm} + \mathring{\rho}_{\vapor \solid}^{\pm},
  &\mathring{q}_{\solid} &= \mathring{q}_{\liquid \solid} + \mathring{q}_{\vapor \solid}  && \leftarrow \text{ solid} \\
  \label{liquid_mass_reaction}
  \mathring{\rho}_{\liquid}^{\pm} &= \mathring{\rho}_{\solid \liquid}^{\pm} + \mathring{\rho}_{\vapor \liquid}^{\pm},
  &\mathring{q}_{\liquid} &= \mathring{q}_{\solid \liquid} + \mathring{q}_{\vapor \liquid}  && \leftarrow \text{ liquid} \\
  \label{vapor_mass_reaction}
  \mathring{\rho}_{\vapor}^{\pm} &= \mathring{\rho}_{\solid \vapor}^{\pm} + \mathring{\rho}_{\liquid \vapor}^{\pm},
  &\mathring{q}_{\vapor} &= \mathring{q}_{\solid \vapor} + \mathring{q}_{\liquid \vapor}  && \leftarrow \text{ vapor},
\end{align}
Furthermore, let the first subscript $i$ and second subscript $j$ of $\mathring{\rho}_{ij}$ or $\mathring{q}_{ij}$ be the source and destination phases, respectively, such that the action-reaction pairs associated with \cref{solid_mass_reaction,liquid_mass_reaction,vapor_mass_reaction} are given by\footnote{In addition to the surface action-reaction pairs \cref{solid_liquid_reaction_pair,solid_vapor_reaction_pair,liquid_vapor_reaction_pair}, a significant quantity of rock debris may accumulate via interaction with the lithosphere; these processes are neglected in this work.}
\begin{align}
  \label{solid_liquid_reaction_pair}
  \mathring{\rho}_{\solid \liquid}^{\pm} = &- \mathring{\rho}_{\liquid \solid}^{\pm},
  &\mathring{q}_{\solid \liquid} = &- \mathring{q}_{\liquid \solid}  && \leftarrow \text{ melting/freezing} \\
  \label{solid_vapor_reaction_pair}
  \mathring{\rho}_{\solid \vapor}^{\pm}  = &- \mathring{\rho}_{\vapor \solid}^{\pm},
  &\mathring{q}_{\solid \vapor}  = &- \mathring{q}_{\vapor \solid}   && \leftarrow \text{ sublimation/deposition}  \\
  \label{liquid_vapor_reaction_pair}
  \mathring{\rho}_{\liquid \vapor}^{\pm}  = &- \mathring{\rho}_{\vapor \liquid}^{\pm},
  &\mathring{q}_{\liquid \vapor}  = &- \mathring{q}_{\vapor \liquid} && \leftarrow \text{ evaporation/condensation},
\end{align}
which are positive for increasing component $j$ and negative for decreasing $j$.
Finally, insertion of \cref{solid_liquid_reaction_pair,solid_vapor_reaction_pair,liquid_vapor_reaction_pair} into \cref{solid_mass_reaction,liquid_mass_reaction,vapor_mass_reaction} and taking the sum results in \cref{thm_homogeneous_total_mass_reaction}.
\end{proof}

\begin{remark}
\label{rmk_reaction}
The mass changes along surface $\Sigma$ encompassed by relations \cref{solid_mass_reaction,liquid_mass_reaction,vapor_mass_reaction} capture any and all imaginable sources of mass fluctuations due to environmental forcings.
For example, the ice sheet sliding against bedrock which melts due to friction or geothermal heat corresponds with $\mathring{q}_{\liquid \solid} < 0$, ocean water freezing to the surface corresponds with $\mathring{q}_{\liquid \solid} > 0$, snow deposited on the upper surface corresponds with $\mathring{q}_{\solid \vapor} < 0$, and water condensing on the ice-sheet surface during summer months corresponds with $\mathring{q}_{\liquid \vapor} < 0$.
These terms only represent changes in phase and do not in any way incorporate mechanisms of mass transport.
\end{remark}

\begin{remark}
The sign of basal melting/freezing rate $\mathring{q}_{\solid \liquid}$ defined by \cref{solid_liquid_reaction_pair} is commonly inverted---or equivalently, for reasons made apparent in the next section, by altering the definition of the \emph{outward}-pointing-unit normal $\normal$ to the \emph{inward}-pointing-unit normal along the basal surface $\Sigma_{\bed}$---in order to let positive values of basal melting/freezing rate $\mathring{q}_{\solid \liquid}$ be associated with mass loss.
This deviation from convention unnecessarily obfuscates the origin of this parameter.
\end{remark}

\section{Surface-mass balance} \label{sec_surface_mass_balance}

The mechanisms which control the dynamics of ice sheets involve environmental interactions; hence an accurate description of the ice-sheet surface is essential.
To this end, the balance of mass over the environmental interface $\Sigma$ is separately described here for each component mass, then combined to give a single multi-phase surface-mass-balance relation.

\subsection{Component surface-mass balance}
\label{sec_component_surface_mass_balance}

The transport of each solid ($\solid$), liquid ($\liquid$), and vapor ($\vapor$) component masses within $\Omega$ can only be governed by advection\footnote{The velocity $\rankone{u}$ represents the motion of the component masses.}, while along the surface of the ice sheet $\Sigma$, some fluctuation of each component mass will occur via interaction between the $(-)$ and $(+)$ environments (\cf \cref{rmk_reaction}).
Thus substitution of $\phi = \rho_k$, $\rankone{j} = \rankone{0}$, and $\mathring{f}_{\Sigma} = \mathring{q}_k$ in discontinuity equation \cref{thm_discontinuity_equation} produces\footnote{Relations \cref{component_mass_jump} evaluated at the lower surface are identical to Equations (9.121) and (9.124) of \cite{greve_2009} with $\rho_{\vapor} \equiv 0$.}
\begin{align}
  \label{component_mass_jump}
  \jump{\rho_k \left( \rankone{u}_k - \rankone{w} \right)} = &- \mathring{q}_k && \text{on } \Sigma, && k = \solid, \liquid, \vapor;
\end{align}
where density-flux terms $\mathring{q}_{\solid}$, $\mathring{q}_{\liquid}$, and $\mathring{q}_{\vapor}$ are given by phase reactions \cref{solid_mass_reaction,liquid_mass_reaction,vapor_mass_reaction}.
Applying jump discontinuity \cref{def_jump} to \cref{component_mass_jump} yields
\begin{align}
  \label{phase_surface_mass_balance_one}
  \rho_k^{\pm} \fs_{k}^{\pm} &= \rho_k^{\mp} \left( \rankone{u}_{k}^{\mp} - \rankone{w}^{\mp} \right) \cdot \normal^{\mp} + \mathring{q}_{k},
\end{align}
with \emph{phase surface-mass-balance} terms
\begin{align}
  \label{phase_surface_mass_balance}
  \fs_{k}^{\pm} &= \left( \rankone{w}^{\pm} - \rankone{u}_{k}^{\pm} \right) \cdot \normal^{\pm} && \text{on } \Sigma, && k = \solid, \liquid, \vapor
\end{align}
in units of length per time and are positive when mass flows into the $(\pm)$ domain and negative when mass flows out of the $(\pm)$ domain.

\begin{remark}
\label{rmk_impenetrable_surface_remark}
Ice-sheet basal surfaces in contact with the lithosphere may under sufficiently short time scales be considered static, meaning $\rankone{w}^{\pm} \cdot \normal^{\pm} = 0$.
In addition, in the event that the solid component of the lithosphere is not transported into the ice sheet, $\rankone{u}_{\solid}^+ \cdot \normal^+ = 0$.
Hence relations \cref{phase_surface_mass_balance_one,phase_surface_mass_balance} evaluated from the ice-sheet $(-)$ perspective with $k = \solid$ yields the Dirichlet component of the \emph{Navier boundary conditions} \cite{navier_1823} given by
\begin{align}
  \label{impenetrable_surface_mass_balance}
  \rankone{u}_{\solid}^- \cdot \normal^- = -\fs_{\solid}^- = - \mathring{q}_{\solid} / \rho_{\solid}^-;
\end{align}
therefore, in this case the solid mass of the ice sheet $\rho_s^-$ will have a component of velocity directed into the supporting bedrock.
In the event that mass is removed along the lower ice-sheet surface by melting, the height of the upper surface will experience a proportional drop in height with some amount of discrepancy caused by viscous forces within the ice that resist the gravitational force downward; that is, a change in the ice sheet velocity at the basal surface need first affect the distribution of stress $\ranktwo{\sigma}$ within the ice-sheet volume before its effects can reach the upper surface.\footnote{A complete explanation of this phenomena requires a description of momentum conservation that is beyond the scope of this work.}
However, this relationship is commonly simplified by some authors of thermo-mechanical ice-sheet models to the \emph{impenetrability} expression $\rankone{u}_{\solid}^- \cdot \normal^- \equiv 0$ (\eg, \cite{leng_2014}, \cite{larour_2012}).
In order to conserve mass, this simplification requires that the change in height corresponding to a mass change along the basal surface be subtracted from the upper surface height, thereby breaking the continuum.
Therefore, the assumption that $\rankone{u}_{\solid}^- \cdot \normal^- \equiv 0$ will induce significant errors to the ice-sheet momentum over regions corresponding with high surface-mass-balance magnitude.
Finally, note that the cost of imposing basal-mass-balance relation \cref{impenetrable_surface_mass_balance} in place of $\rankone{u}_{\solid}^- \cdot \normal^- = 0$ is insignificant.
\end{remark}

\subsection{Mixture surface-mass balance}
\label{sec_mixture_surface_mass_balance}

Using barycentric velocity $\rankone{u}^{\pm}$ from relation \cref{barycentric_velocity}, the total mass jump over the surface $\Sigma$ is given by the sum of each component jump of \cref{component_mass_jump}:
\begin{align}
  \label{mixture_mass_jump}
  \jump{ \rho \left( \rankone{u} - \rankone{w} \right) } = 0 && \text{on } \Sigma,
\end{align}
where \cref{thm_homogeneous_total_mass_reaction} has been used to homogenize the right-hand side.\footnote{Mixture jump \cref{mixture_mass_jump} is identical to Equations (2.15)$_1$ of \cite{hutter_1982} and (3.61) of \cite{greve_2009} with $\rho_{\vapor} \equiv 0$.}
Expanding this relationship using jump \cref{def_jump} produces
\begin{align}
  \label{surface_flux_one}
  \rho^{\pm} \fs^{\pm} &= \rho^{\mp} \left( \rankone{u}^{\mp} - \rankone{w}^{\mp} \right) \cdot \normal^{\mp},
\end{align}
with the \emph{mixture surface-mass balance} or \emph{accumulation/ablation function}
\begin{align}
  \label{surface_flux}
  \fs^{\pm} &= \left( \rankone{w}^{\pm} - \rankone{u}^{\pm} \right) \cdot \normal^{\pm} && \text{ on } \Sigma,
\end{align}
in units of length per time.
Identically to the source terms of component-balance relations \cref{phase_surface_mass_balance}, $\fs^{\pm}$ given by \cref{surface_flux} is positive for mass gain and negative for mass loss within the ($\pm$) domain.\footnote{$\fs^{\pm}$ is usually prescribed on the atmospheric surface from measurements and parameterized at the lateral and basal surfaces where data is more difficult to collect.}
An important observation regarding \cref{phase_surface_mass_balance} and \cref{surface_flux} follows.

\begin{proposition}
\label{thm_normal_fs}
The phase surface-mass balance $\fs_k^{\pm}$ and mixture surface-mass balance $\fs^{\pm}$ terms defined by \cref{phase_surface_mass_balance} and \cref{surface_flux} are the components of phase and mixture mass-balance vectors in the $\normal^{\pm}$ direction.
\end{proposition}

\begin{proof}
Let either phase or mixture surface-mass balance vector from either ($\pm$) perspective be $\rankone{r} = \rankone{w} - \rankone{u}$ (\cf \cref{rmk_surface_velocity_difference}).
The scalar product of $\rankone{r}$ with $\normal$ combined with relation \cref{phase_surface_mass_balance} or \cref{surface_flux} implies that $\fs = \rankone{r} \cdot \normal$.
\end{proof}

\begin{remark}
\label{smb_remark}
Continuing the line of reasoning began with \cref{rmk_reaction}, mixture surface-mass balance \cref{surface_flux} incorporates all mass-transport processes due to interaction of the ice sheet with its environment within the single parameter $\fs^{\pm}$.
These processes include phase transitions, densification of snow on the upper surface, basal-water flux due to internal-viscous heating, interaction with the basal-hydraulic system, accumulation of ice under a floating ice shelf due to super-cooled water that has frozen due to rising convection currents, \etc.
\end{remark}

\begin{remark}
If the extrinsic density of the liquid and vapor components on either side are negligible, $\rho_k^{\pm} \rankone{u}_k^{\pm} = \rankone{0}$ and $\mathring{q}_k = 0$ for $k = \liquid, \vapor$.
If in addition the lithosphere is immobile, $\rankone{w}^{\pm} \cdot \normal^{\pm} = \rankone{u}_{\solid}^+ \cdot \normal^+ = 0$; hence relation \cref{surface_flux_one} yields
$\fs^- = 0$
while relation \cref{surface_flux} yields
$\rankone{u}_{\solid}^- \cdot \normal^- = 0$.
These conditions are only appropriate for static surfaces which are below the temperature-melting point of ice, such as the interior of Antarctica and Greenland (\cf \cref{rmk_impenetrable_surface_remark}).
\end{remark}

\subsection{Ice-sheet surface-mass balance}
\label{sec_ice_surface_mass_balance}

Setting $(\pm) = (-) \iff (\mp) = (+)$, removing all $(-)$ superscripts from the ice variables, and collapsing the exterior $(+)$ domain to the ice-sheet surface $\Gamma \rightarrow \Sigma$, relation \cref{surface_flux} is identical to the expression\footnote{Surface-mass balance relation \cref{surface_mass_balance} is identical to Equations (5.19) and (5.29) of \cite{greve_2009} with $\rho_v \equiv 0$.}
\begin{align}
  \label{surface_mass_balance}
  \fs &= \rankone{w} \cdot \normal - \rankone{u} \cdot \normal &&\text{ on } \Gamma.
\end{align}
The exterior ($+$) domain exists only at ice-sheet surface $\Gamma$ for the remainder of this work; all exterior effects are henceforth enveloped within surface-mass balance $\fs$ of \cref{surface_mass_balance}.

\section{Global/local mass conservation} \label{sec_global_local_mass_conservation}

Consider the ice-sheet domain $\Omega(\rankone{x},t) \in \R^3$ with boundary $\Gamma = \partial \Omega(\rankone{x},t)$ and outward-pointing-unit normal $\normal$.
Substitution of mass density $\phi = \rho$ in Reynolds transport theorem (\cf \cref{thm_reynolds_transport}) yields
\begin{align}
  \label{mass_balance_one}
  \totder{}{t} \int_{\Omega} \rho \d{\Omega} = \int_{\Omega} \parder{\rho}{t} \d{\Omega} + \int_{\Gamma} \rho \rankone{w} \cdot \normal \d{\Gamma},
\end{align}
where $\rankone{w}$ is the velocity of the surface $\Gamma$.
In the event that the total ice-sheet mass in invariant with time, $\dot{m} = \diff_t \int_{\Omega} \rho \d{\Omega} = 0$ and mass-conservation relation \cref{mass_balance_one} combined with surface-mass balance \cref{surface_mass_balance} yields
\begin{align}
  \label{ice_sheet_mass_equilibrium}
  \totder{m}{t} = \int_{\Omega} \parder{\rho}{t} \d{\Omega} + \int_{\Gamma} \rho \left( \rankone{u} \cdot \normal + \fs \right) \d{\Gamma} = 0;
\end{align}
these are the requirements for ice-sheet-mass equilibrium.

\begin{remark}
\label{global_mass_balance_remark}
In the derivation of mass rate of change \cref{ice_sheet_mass_equilibrium}, the integrals were taken over the entire ice sheet and thus this relation represents the \emph{global} balance of mass and volume.
In addition, if the density is assumed constant, mass-conservation reduces to $\dot{V} = \int_{\Gamma} \rankone{w} \cdot \normal \d{\Gamma} =  \int_{\Gamma} ( \rankone{u} \cdot \normal + \fs ) \d{\Gamma}$; therefore, in this case conservation of mass reduces to a statement of conservation of volume.
\end{remark}

\subsection{Local incompressibility} \label{sec_local_incompressibility}

Consider within $\Omega$ an arbitrarily fixed ice-sheet material element $\Omega^{\e}(\rankone{x}) \subset \Omega$ with boundary $\partial \Omega^{\e} = \Gamma^{\e}$.
The constant volume of $\Omega^{\e}$ demands that density fluctuations caused by melting or freezing---induced by changes in pressure and heat flux---present a \emph{mass source} within $\Omega^{\e}$.
Therefore, for each component mass $k$, substitution of extrinsic mass density $\phi = \rho_k$; non-advective mass-density flux $\rankone{j} = \rankone{0}$; and internal mass-density source $\mathring{f} = \mathring{\rho}_k$ into the conservative form of differential-continuity equation \cref{def_conservative_continuity_equation} yields the \emph{component mass-balance} relations\footnote{Component balance \cref{component_mass_balance} is identical to Equation (2.8)$_1$ of \cite{hutter_1982} with $\rho_v \equiv 0$.}
\begin{align}
  \label{component_mass_balance}
  \parder{\rho_k}{t} + \nabla \cdot \left( \rho_k \rankone{u}_k \right) &= \mathring{\rho}_k && \text{in } \Omega, && k = \solid, \liquid, \vapor;
\end{align}
where phase-reaction rates $\mathring{\rho}_k$ are given by \cref{solid_mass_reaction,liquid_mass_reaction,vapor_mass_reaction}.

In analogy with mixture surface-mass balance \cref{mixture_mass_jump}, the sum of each component balance $k$ in \cref{component_mass_balance} produces the \emph{mixture mass-balance} relation\footnote{Mixture balance \cref{mixture_mass_balance} is identical to Equations (2.10)$_1$ of \cite{hutter_1982} and (3.58) of \cite{greve_2009} with $\rho_v \equiv 0$.}
\begin{align}
  \label{mixture_mass_balance}
  \parder{\rho}{t} + \nabla \cdot \left( \rho \rankone{u} \right) = 0 && \text{in } \Omega,
\end{align}
where $\rho \rankone{u}$ and $\rho$ are defined by \cref{barycentric_velocity} and the left-hand side has been homogenized using \cref{thm_homogeneous_total_mass_reaction}.

The chain rule (\cf \cref{thm_one_variable_chain_rule,total_derivative}) applied to \cref{mixture_mass_balance} produces
$\dot{\rho} + \rho \nabla \cdot \rankone{u} = 0$,
which, using mass density $\rho = m/V$ with mass $m(\rankone{x},t) = \int \rho \d{\Omega^{\e}}$ and constant volume $V(\rankone{x}) = \int 1 \d{\Omega^{\e}}$ yields
$V^{-1} \left( \dot{m} + m \nabla \cdot \rankone{u} \right) = 0$.
A final multiplication by $V$ and division by $m$ to this relation produces
$\frac{\dot{m}}{m} + \nabla \cdot \rankone{u} = 0$.
Therefore, provided that mass transport is conserved such that $\dot{m} = 0$, relation \cref{mixture_mass_balance} is reduced to\footnote{Relation \cref{incompressible_conservation_of_mass} is identical to Equations (2.11)$_1$ of \cite{hutter_1982} and (3.60) of \cite{greve_2009} with $\rho_v \equiv 0$.}
\begin{align}
  \label{incompressible_conservation_of_mass}
  \nabla \cdot \rankone{u} &= 0 &&\text{ in } \Omega;
\end{align}
the well-known \emph{incompressibility constraint}.

The firn layer consists of snow densifying under pressure to eventually become solid ice; in this region it is entirely possible that $\nabla \cdot u \neq 0$ and thus $\dot{m} \neq 0$.
Therefore, the fully \emph{compressible} mass-balance relation \cref{mixture_mass_balance} appropriately describes the firn layer.
Additionally, although the ice below the firn-ice transition surface may undergo phase transitions due to strain heat and thus in general $\partial_t \rho \not \equiv 0 \iff \partial_t m \not \equiv 0$, observations indicate (\cf \cite{greve_2009} and references therein) that an appropriate upper bound for the mass fraction of water is approximately $\omega_{\liquid}^{\max} = 0.05$.
Evaluating state equation \cref{equation_of_state} with $\omega_{\vapor} = 0$ implies $\omega_{\solid} = 1 - \omega_{\liquid}$; hence the maximum ice-sheet-mass density is $\rho^{\max} = \rho_{\ice} \left( 1 - \omega_{\liquid}^{\max} \right) + \rho_{\water} \omega_{\liquid}^{\max} \approx 914.5$ kg m$^{-3}$ and varies from pure ice by only $\approx 0.5\%$. 
Therefore, it is safe to assume that the ice-sheet mixture is effectively incompressible and that the assumptions leading to \cref{incompressible_conservation_of_mass} are indeed valid for regions below the firn layer.

\section{Free-surface equation} \label{sec_free_surface}

The results of this section are identical to that of \cite{hutter_1982}.
For reasons stated in the introduction, and for clarity of the analysis to follow, the fundamental free-surface relations are revisited.
To this end, assume that the surface of the ice-sheet body can be stated in implicit form in the sense of \cref{thm_implicit_surface}.
Then a function $F = F(\rankone{x},t)$ can be defined that satisfies $F = 0$ for all $\rankone{x} \in \Gamma$ and $t$.
In this case, application of the chain rule (\cf \cref{thm_one_variable_chain_rule} and \cref{total_derivative}) produces the purely \emph{kinematic} relation
\begin{align}
  \label{kinematic_surface}
  \totder{F}{t} &= \parder{F}{t} + \rankone{w} \cdot \nabla F = 0 && \text{ on } \Gamma.
\end{align}
Combining outward-pointing-unit normal \cref{def_normal_vector} and surface-mass balance \cref{surface_mass_balance} yields
$\rankone{w} \cdot \left( \nabla F / \Vert \nabla F \Vert \right) = \fs + \rankone{u} \cdot \left( \nabla F / \Vert \nabla F \Vert \right)$,
from whence the advective term of kinematic equation \cref{kinematic_surface} given by
$\rankone{w} \cdot \nabla F = \Vert \nabla F \Vert \fs + \rankone{u} \cdot \nabla F$
substituted back into \cref{kinematic_surface} results in\footnote{Relation \cref{intermediate_kinematic_surface} is identical to Equation (2.18) of \cite{hutter_1982}.}
\begin{align}
  \label{intermediate_kinematic_surface}
  \parder{F}{t} + \rankone{u} \cdot \nabla F &= - \Vert \nabla F \Vert \fs && \text{ on } \Gamma.
\end{align}

\begin{remark}
\label{rmk_breaking_waves}
By specifying the surface $F$ implicitly, the normal vector to $F$ could be stated analytically with constant $z$ component (\cf \cref{thm_implicit_surface}).
Thus non-linear hyperbolic equation \cref{intermediate_kinematic_surface} represents mass conservation of the ice-sheet exterior surface $\Gamma \subset \R^3$ projected onto the $xy$ plane in $\R^2$.
As a consequence of this projection, a surface $F$ derived by solving \cref{intermediate_kinematic_surface} lacks the capability to generate breaking waves, which are possible with relation \cref{kinematic_surface}; however, ice sheets do not naturally display these phenomena due to their highly-viscous and relatively slow-moving nature.
\end{remark}

\begin{remark}
\label{rmk_normal_kinematic_forcing}
Kinematic relation \cref{intermediate_kinematic_surface} is a description of the ice-boundary movement governed solely by the horizontal components\footnote{This is due to the fact that $\partial_z F = 0$.} of ice-sheet velocity $\rankone{u}$ and surface-mass balance $\fs$.
It is common practice to first solve for $\rankone{u}$ at time $t$ from momentum conservation, then solve \cref{intermediate_kinematic_surface} for an updated surface $F(\rankone{x}, t + \Delta t)$ some interval of time $\Delta t$ from $t$.
\end{remark}

Decomposition of the surface $\Gamma$ into $\Gamma = \Gamma_{\srf} \cup \Gamma_{\bed}$ with $\Gamma_{\srf} \cap \Gamma_{\bed} = \emptyset$, making use of outward-pointing-unit normal \cref{def_normal_vector}, and defining
$\smb = \{ \fs(\rankone{x})\ |\ \rankone{x} \in \Gamma_{\srf} \}$;
$\bmb = \{ \fs(\rankone{x})\ |\ \rankone{x} \in \Gamma_{\bed} \}$
partitions \cref{intermediate_kinematic_surface} into\footnote{Relations \cref{upper_free_surface,lower_free_surface} are identical to (2.20) and (2.25) of \cite{hutter_1982} and (5.21) and (5.31) of \cite{greve_2009}.}
\begin{align}
  \label{upper_free_surface}
  \parder{S}{t} - u_z + \rankone{u} \cdot \nabla S &= + \Vert \unit{k} - \nabla S \Vert \smb && \text{ on } \Gamma_{\srf} \\
  \label{lower_free_surface}
  \parder{B}{t} - u_z + \rankone{u} \cdot \nabla B &= - \Vert \nabla B - \unit{k} \Vert \bmb && \text{ on } \Gamma_{\bed},
\end{align}
where
$\unit{k} \cdot \rankone{u} = u_z$,
$\partial_t F_{\srf} = - \partial_t S$,
and
$\partial_t F_{\bed} = \partial_t B$
have been used.

\begin{SCfigure}
  \centering
    \def\svgwidth{0.6\linewidth}
    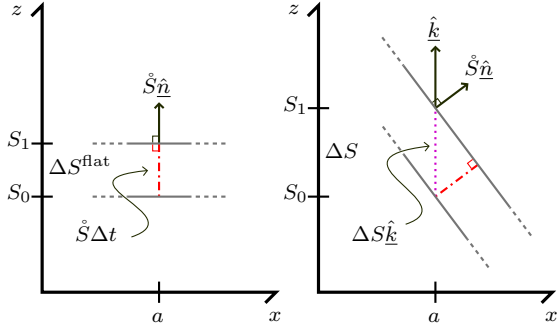
  \caption[surface-mass flux 1]{From the perspective of the surface height change at a point $x=a$, the accumulation/ablation-vector $\smb \normal$ of an inclined surface will have the effect of \emph{increasing} the surface height rate of change at $x=a$ beyond that of a surface with zero-gradient magnitude.  Note that the dashed-dotted lines are identical in length between the two plots with magnitude equal to $\smb \Delta t$ with deposition interval $\Delta t$.}
  \label{fig_surface_flux_1}
\end{SCfigure}

The relationship between the surface gradient and accumulation/ablation can be further illuminated as follows: consider relation \cref{upper_free_surface} with a constant rate of deposition on the upper surface, meaning $\smb > 0$, over an interval of time $\Delta t > 0$ with zero fluid velocity $\rankone{u} = \rankone{0}$.
In this case, the change in height of a completely flat surface is $\Delta S^{\mathrm{flat}} = \smb \Delta t$, while an inclined surface with gradient $\nabla S$ will experience a change of height $\Delta S$ satisfying the relation 
$\Delta S \unit{k} \cdot \normal = \smb \Delta t$\footnote{This is the projection of $\Delta S \unit{k}$ onto $\normal$.}
with outward-pointing-unit normal $\normal$ and $z$-coordinate basis vector $\unit{k}$ (\cf \cref{fig_surface_flux_1}).
Therefore, for an inclined surface,
$\Delta S = \smb \Delta t ( \unit{k} \cdot \normal )^{-1} = \smb \Delta t \Vert \unit{k} - \nabla S \Vert$,
which after division by $\Delta t$ and taking the limit as $\Delta t \rightarrow 0$ produces the instantaneous rate of change
$\partial_t S = \smb \Vert \unit{k} - \nabla S \Vert$;
a first-order non-linear partial-differential equation for $S(\rankone{x},t)$.
A similar line of reasoning will produce an identical relation for lower surface-mass balance $\bmb$ along $z=B(x,t)$.

\section{Mass balance in $\R^2$}
\label{sec_vertically_integrated_mass_balance}

Similar to \cref{sec_free_surface}, the results of this section are identical to that of \cite{greve_2009}.
For reasons stated in the introduction, and for clarity of the analysis to follow, the fundamental ice-thickness relation is revisited.

The ice-sheet domain of analysis is reduced from $\Omega \in \R^3$ to $\Omega \in \R^2$ by vertical integration.
In this case, it is helpful for the following assessment to define the vertically-averaged velocity, or \emph{balance velocity}
\begin{align}
  \label{balance_velocity}
  \bar{\rankone{u}} = \frac{1}{H} \int_z \rankone{u} \d{z} 
\end{align}
with components $\bar{\rankone{u}} = [\bar{u}_x\ \bar{u}_y\ \bar{u}_z]\T$.
The value of this quantity will be shown to be a useful description for mass balance in the $xy$-coordinate plane, as follows.

Integrating volume-conservation relation \cref{incompressible_conservation_of_mass} vertically produces
\begin{align}
  \label{first_vertical_integration}
  \int_B^S \nabla \cdot \rankone{u} \d{z} 
  = & \int_B^S \left( \parder{u_x}{x} + \parder{u_y}{y} + \parder{u_z}{z} \right) \d{z} = 0.
\end{align}
The first two integrals of \cref{first_vertical_integration} are derived using Leibniz's rule \cref{thm_leibniz_rule_corollary} and the last integral is evaluated using the first fundamental theorem of calculus (\cf \cref{thm_first_fundamental_theorem_of_calculus}):
\begin{align*}
  \int_B^S \parder{u_x}{x} \d{z} 
  = &\parder{}{x} \int_B^S u_x \d{z} + u_x |_B \parder{B}{x} - u_x |_S \parder{S}{x} \\
  \int_B^S \parder{u_y}{y} \d{z} 
  = &\parder{}{y} \int_B^S u_y \d{z} + u_y |_B \parder{B}{y} - u_x |_S \parder{S}{y} \\
  \int_B^S \parder{u_z}{z} \d{z}
  = & u_z (x,y,S) - u_z(x,y,B);
\end{align*}
thus \cref{first_vertical_integration} can be stated compactly as
\begin{align*}
  \nabla \cdot \left( \int_B^S \rankone{u} \d{z} \right) + \rankone{u} |_B \cdot \nabla B - \rankone{u} |_S \cdot \nabla S + u_z |_S - u_z |_B = 0,
\end{align*}
which on elimination of $\rankone{u} |_B \cdot \nabla B$ and $\rankone{u} |_S \cdot \nabla S$ via free-surface relations \cref{upper_free_surface} and \cref{lower_free_surface} and applying balance-velocity definition \cref{balance_velocity} yields
\begin{align*}
  \nabla \cdot \big( H \bar{\rankone{u}} \big) 
  &+ \left( - \Vert \nabla B - \unit{k} \Vert \bmb - \parder{B}{t} \right) 
   - \left( + \Vert \unit{k} - \nabla S \Vert \smb - \parder{S}{t}\right)
   = 0;
\end{align*}
combining this with thickness $H(x,y) = S(x,y) - B(x,y)$ results in the vertically-integrated mass-balance relation\footnote{Relation \cref{balance_velocity_equation_one} is identical to Equations (5.48) and (5.55) of \cite{greve_2009}.}
\begin{align}
  \label{balance_velocity_equation_one}
  \parder{H}{t} + \nabla \cdot \big( H \bar{\rankone{u}} \big) =& \mathring{H},
  \hspace{15mm}
  \mathring{H} = \Vert \unit{k} - \nabla S \Vert \smb + \Vert \nabla B - \unit{k} \Vert \bmb.
\end{align}

\begin{remark}
\label{rmk_smb_coefficients}
The components of source \cref{balance_velocity_equation_one}$_2$ include surface-mass-balance terms $\smb$ and $\bmb$ and clearly increase the rate of change of thickness with increasing accumulation or decreasing ablation.
In addition, the surface-normal-vector magnitudes (\cf \cref{def_normal_vector})
\begin{align*}
  \Vert \nabla F_{\srf} \Vert = \Vert \unit{k} - \nabla S \Vert \geq 1
  \hspace{8mm}
  \text{and}
  \hspace{8mm}
  \Vert \nabla F_{\bed} \Vert = \Vert \nabla B - \unit{k} \Vert \geq 1
\end{align*}
multiplicatively attached to these terms have the effect of increasing the time rate of change of the surface height in proportion with the magnitude of their respective surface gradients (\cf \cref{fig_surface_flux_1}).
This is a by-product of the choice of Cartesian coordinate system; that is, a surface area in the $xy$ plane with high surface-gradient magnitude will contribute a larger proportion to the change of ice thickness $H$ than an identical surface area in the $xy$ plane with negligible surface-gradient magnitude (\cref{fig_surface_flux_1}).
This effect is further demonstrated by inverting quasi-mass balance \cref{balance_velocity_equation_one} for $\smb$; division of both sides of \cref{balance_velocity_equation_one} by $\Vert \unit{k} - \nabla S \Vert \geq 1$ will decrease the solution for $\smb$ as the surface-gradient magnitude increases.
\end{remark}

\begin{remark}
\label{rmk_misconception}
The fact that balance relation \cref{balance_velocity_equation_one} appears similar in form to continuity equation \cref{def_conservative_continuity_equation} suggests a possible source of confusion surrounding its ubiquitous use.
Replacing $\phi = H$, $\rankone{u} = \rankone{\bar{u}}$, $\rankone{j} = 0$, and $\mathring{H} = \smb + \bmb$ in \cref{def_conservative_continuity_equation} makes sense intuitively: the thickness flux is balanced by the local rate of change of the ice-sheet thickness and accumulation or ablation with transport governed solely by advection.
However, this direct formulation of mass balance disregards the fact that the statement of mass conservation is inherently \emph{three dimensional} with mathematical consequences resulting from coordinate-reducing operations (\cf \cref{rmk_smb_coefficients}).
Similar to the use of \emph{Cauchy's postulate}\footnote{Cauchy's postulate states that a stress exerted by the environment on a body will be a function of not only the position and time, but also the geometry; that is, $\ranktwo{\sigma} = \ranktwo{\sigma}(\rankone{x}, t; \normal)$.} to derive momentum conservation or the \emph{Fourier-Stokes heat-flux theorem}\footnote{The Fourier-Stokes heat-flux theorem states that the heat flowing from a body will be dependent on the geometry; that is, $\rankone{q} = \rankone{q}(\rankone{x},t;\normal)$.} to derive energy conservation, the geometry are also inextricably connected with the definition of surface-mass-balance terms $\smb$ and $\bmb$ as derived from \cref{surface_mass_balance} (\cf \cref{thm_normal_fs}).
\end{remark}

\subsection{Error analysis}

\begin{figure}
  \centering
    \hspace*{\fill}%
      \subfloat[$\Vert \nabla S \Vert$]{\label{fig_ronne_grad_s} \includegraphics[width=0.49\linewidth]{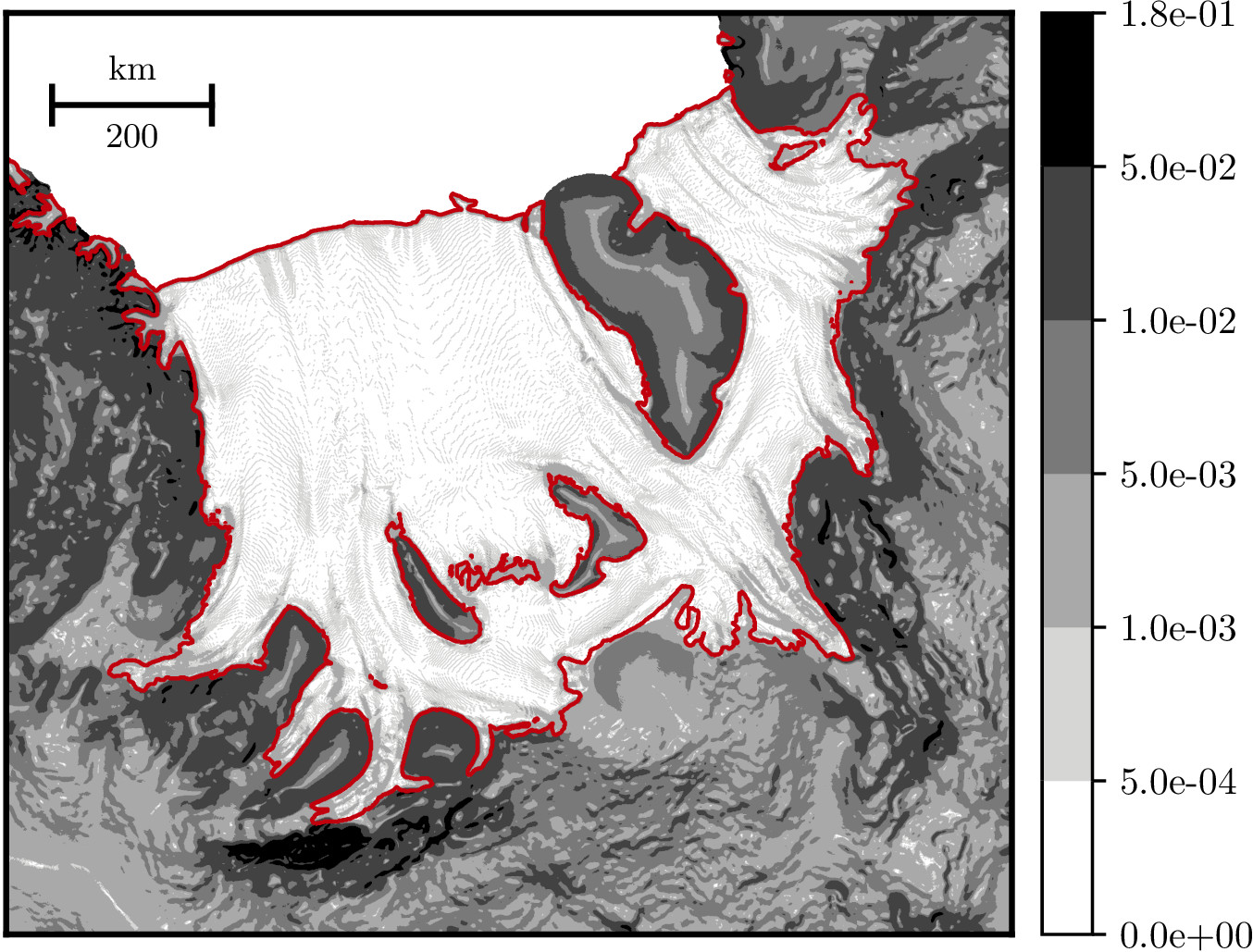}}
    \hfill
      \subfloat[$\Vert \nabla B \Vert$]{\label{fig_ronne_grad_b} \includegraphics[width=0.49\linewidth]{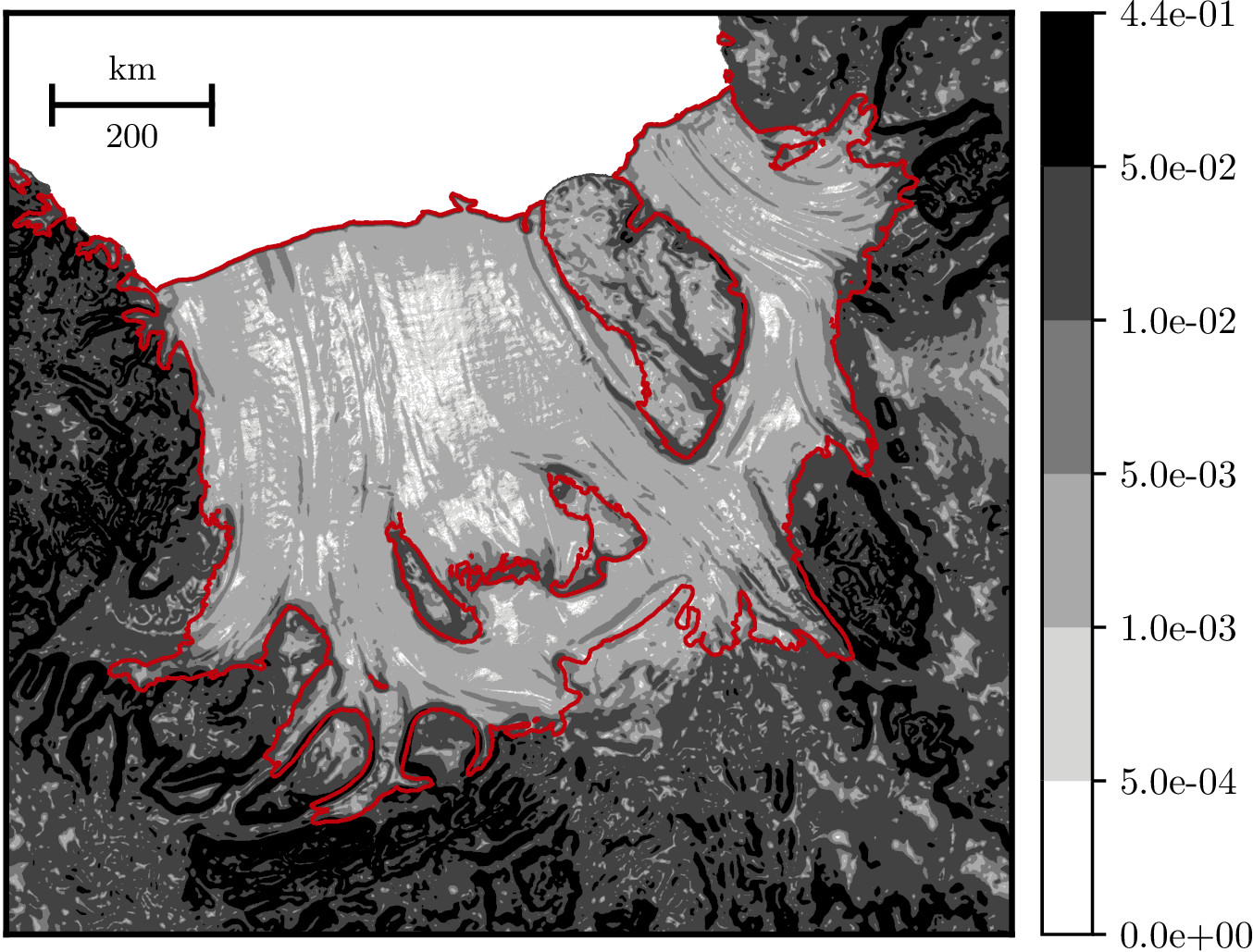}}
    \hspace*{\fill}%

    \hspace*{\fill}%
      \subfloat[$\Vert \nabla S \Vert$]{\label{fig_jakob_grad_s} \includegraphics[width=0.49\linewidth]{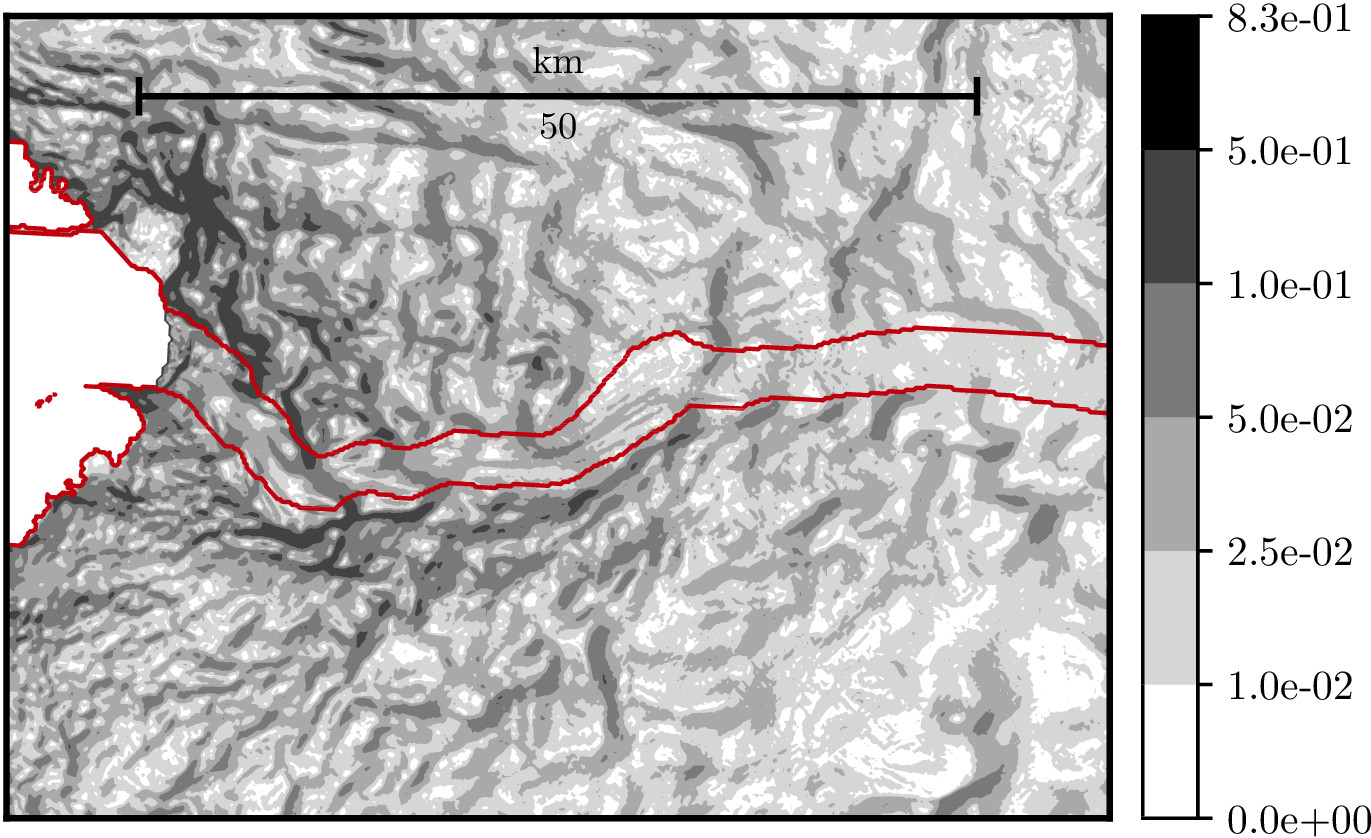}}
    \hfill
      \subfloat[$\Vert \nabla B \Vert$]{\label{fig_jakob_grad_b} \includegraphics[width=0.49\linewidth]{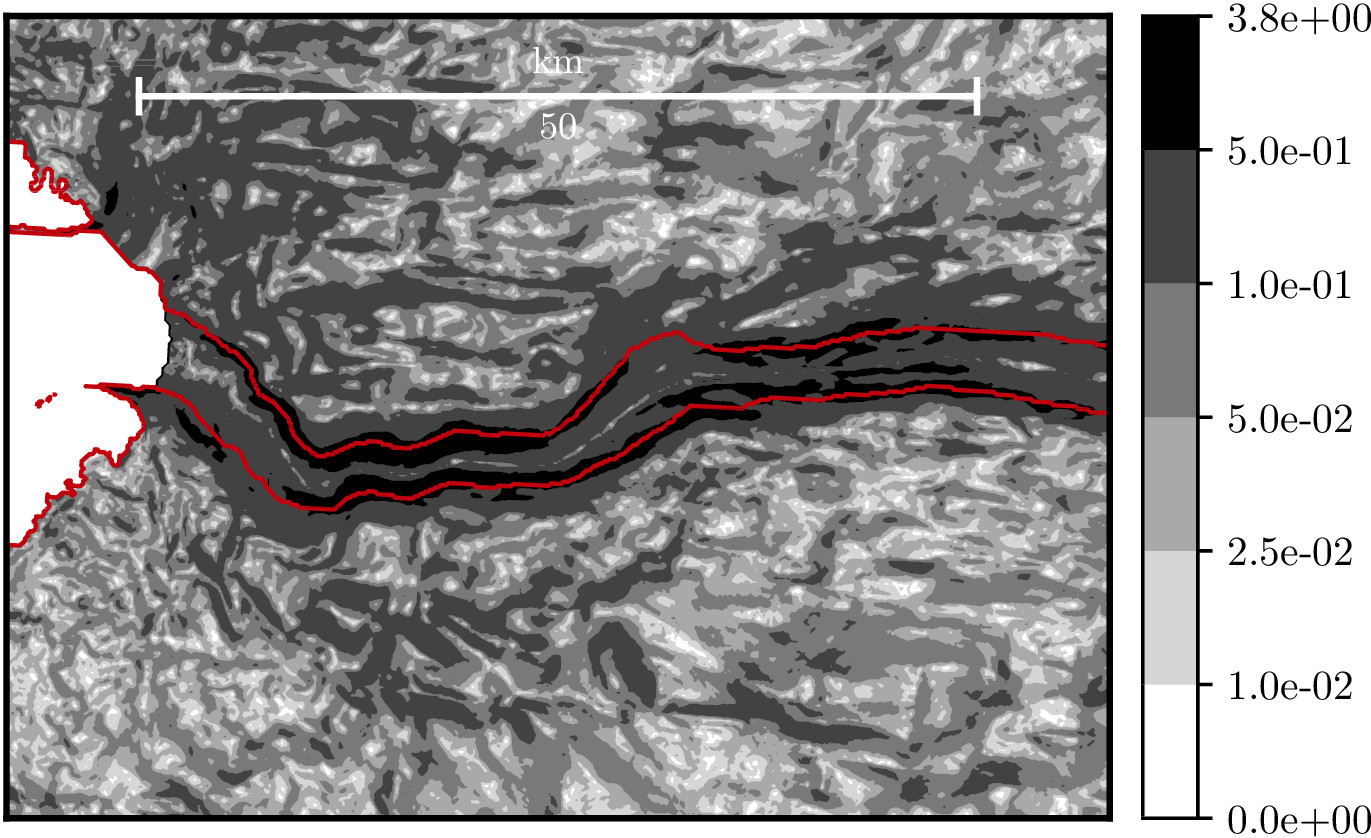}}
    \hspace*{\fill}%

  \caption{Unitless surface-gradient magnitude of the region surrounding the Filchner-Ronne-ice shelf with topography provided by \cite{fretwell_2013} (top) and region surrounding Jakobshavn glacier with topography provided by \cite{morlighem_2017} (bottom) calculated using methods provided by \cite{jones_2001}.  For purposes of visualization, a Gaussian filter with 2 km standard deviation has been applied to the Antarctica surfaces $S$ and $B$; the Jakobshavn topography have not been smoothed.  The periphery, grounding line, and -500 m depth contour of the main Jakobshavn ice channel are outlined red.}
  \label{fig_surface_gradients}
\end{figure}

If the upper and lower surfaces are relatively flat, it has been previously assumed that (\cf \cref{rmk_normal_kinematic_forcing}, \cref{rmk_smb_coefficients}, and \cref{rmk_misconception})
\begin{align}
  \label{common_assumptions}
  \Vert \unit{k} - \nabla S \Vert \equiv 1,
  \hspace{10mm}
  \Vert \nabla B - \unit{k} \Vert \equiv 1.
\end{align}
Assumption \cref{common_assumptions}$_1$ is a valid approximation near the ice-sheet divide---which is defined as regions which satisfy $\Vert \nabla S \Vert = 0$---or over some areas of floating ice, such as the Filchner-Ronne ice shelf (\cref{fig_ronne_grad_s}).
However, using \cref{thm_implicit_surface_area}, the surface areas of the upper surface, lower surface, and $xy$ plane are respectively
\begin{align}
  \label{ice_surface_areas}
  A_{\srf} \approx \sum_{i=1}^n \Vert \unit{k} - \nabla S_i \Vert \Delta x_i \Delta y_i,
  \hspace{5mm}
  A_{\bed} \approx \sum_{i=1}^n \Vert \nabla B_i - \unit{k} \Vert \Delta x_i \Delta y_i,
  \hspace{5mm}
  A_{xy} \approx \sum_{i=1}^n \Delta x_i \Delta y_i;
\end{align}
evaluated over Greenland and Antarctica, the values of $A_{\srf}$ and $A_{\bed}$ are on the order of hundreds of square kilometers larger than $A_{xy}$ (\cref{tab_surface_area}).
Therefore, assumptions \cref{common_assumptions} are likely appropriate over the majority of both the upper and lower surfaces of Greenland and Antarctica (\eg, \cref{fig_surface_gradients}).
However, the flanks of Jakobshavn's trench possess a abnormally high surface-gradient magnitude (\cref{fig_jakob_grad_b}); coupled with the fact that this area is characterized by very high magnitudes of basal velocity and basal-mass balance $\bmb$, assumptions \cref{common_assumptions} will induce a significant error in the forcing term of \cref{balance_velocity_equation_one}$_2$.

Additionally, the error $\varepsilon$ introduced to quasi-mass-balance forcing term \cref{balance_velocity_equation_one}$_2$ by assuming \cref{common_assumptions}---meaning $\mathring{H} = \smb + \bmb + \varepsilon$---is given by
\begin{align}
  \label{r2_assumption_error}
  \varepsilon = \varepsilon_{\srf} + \varepsilon_{\bed},
  \hspace{15mm}
  \varepsilon_{\srf} = \left( \Vert \unit{k} - \nabla S \Vert - 1 \right) \smb,
  \hspace{5mm}
  \varepsilon_{\bed} = \left( \Vert \nabla B - \unit{k} \Vert - 1 \right) \bmb.
\end{align}
The lower surface-mass balance $\bmb$ is more difficult to quantify; as such, continent-scale estimations of $\bmb$ are limited.
However, a variety of estimates have been generated for the upper surface-mass balance $\smb$ over both Greenland and Antarctica, thereby providing a means to estimate quasi-forcing error $\varepsilon_{\srf}$ of \cref{r2_assumption_error}$_2$.
The approximate error of the rate of change of the ice sheet surface calculated from \cref{r2_assumption_error}$_2$ is on the order of hundreds or thousands of meters-ice-equivalent per annum and may therefore be significant (\cref{tab_r2_error}).

Clearly, where $\varepsilon$ is much less than the other terms of \cref{balance_velocity_equation_one}, forcing assumptions \cref{common_assumptions} are valid.
However, the cost of computing the surface gradient is low and therefore the benefits of imposing these assumptions are unclear.
Additionally, the magnitude of the surface-normal vectors are on average largest near regions of fast flow, \eg, near the periphery of the ice sheet (\cref{fig_surface_gradients}).
These areas are associated with high magnitudes of both the surface gradient and surface-mass balance at both the upper and lower surfaces; thus both error terms \cref{r2_assumption_error}$_2$ and \cref{r2_assumption_error}$_3$ may be significant over these areas.

\begin{table*}[t]
\centering
\caption{Total surface area of Greenland and Antarctica calculated from \cref{ice_surface_areas} using a variety of data sets.}
\label{tab_surface_area}
\begin{tabular}{l|c|l|l|l|l}
\textbf{Continent} & \makecell{$\bm{S}, \bm{B}$ \\ \textbf{data source}} & \makecell{$\bm{\Delta x, \Delta y}$ \\ \textbf{(m)}} & \makecell{$\bm{A_{xy}}$ \\ \textbf{(km$^{\bf{2}}$)}} & \makecell{$\bm{A_{\srf} - A_{xy}}$ \\ \textbf{(km$^{\bf{2}}$)}} &  \makecell{$\bm{A_{\bed} - A_{xy}}$ \\ \textbf{(km$^{\bf{2}}$)}} \\ \hline
Antarctica & \cite{fretwell_2013}      & 1000 & $1.35 \times 10^{7}$ & $1814$               & $13774$               \\
Greenland  & \cite{bamber_2013}        & 1000 & $1.68 \times 10^{6}$ & $840$                & $3021$                \\
Greenland  & \cite{noeel_2016}         & 1000 & $1.66 \times 10^{6}$ & $522$                & --                    \\
Greenland  & \cite{morlighem_2017}     & 150  & $1.73 \times 10^{6}$ & $105237$             & $348223$              \\
\end{tabular}
\end{table*}

\begin{table*}[t]
\centering
\caption{Sources of $\smb$ and associated error $\varepsilon_{\srf}$ given by \cref{r2_assumption_error}$_2$ induced by assuming \cref{common_assumptions}$_1$.  Units are in meters-ice equivalent per annum.}
\label{tab_r2_error}
\begin{tabular}{l|c|c|l}
\textbf{Continent} & \makecell{\textbf{$\bm{\smb}$ data} \\ \textbf{source}} & \makecell{$\bm{S}$ \textbf{data} \\ \textbf{source}} & \makecell{$\bm{\sum \varepsilon_{\srf}}$ \\ \textbf{(m a$^{\bf{-1}}$)}} \\ \hline
Antarctica & \cite{vandeberg_2005} & \cite{fretwell_2013} & $1227$             \\
Antarctica & \cite{arthern_2006}   & \cite{fretwell_2013} & $609$              \\
Greenland  & \cite{burgess_2010}   & \cite{bamber_2013}   & $338$              \\
Greenland  & \cite{noeel_2016}     & \cite{noeel_2016}    & $32973$            \\
\end{tabular}
\end{table*}

\begin{SCfigure}
  \centering
    \def\svgwidth{0.6\linewidth}
    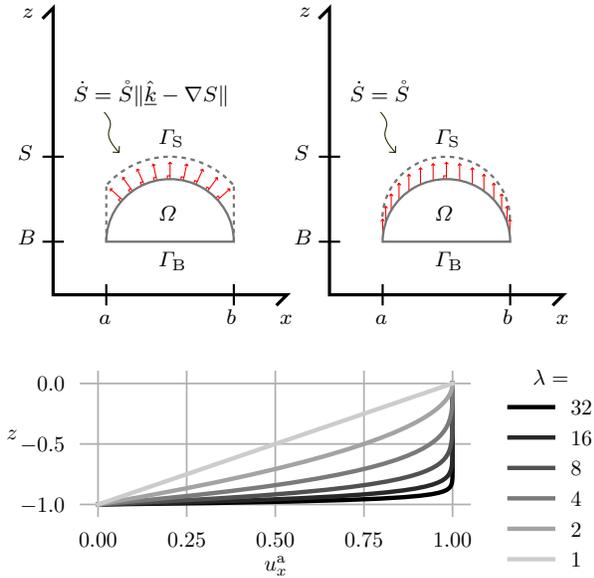
  \caption[surface-mass flux 2]{A two-dimensional-ice sheet with domain $\Omega = [a,b] \times [B,S]$ with uniform accumulation $\smb > 0$ along the upper surface $\Gamma_{\srf}$ and $\bmb = 0$ over the lower surface $\Gamma_{\bed}$.  After some interval of time $\Delta t$, a thickness layer of radially-uniform depth will be deposited (left).  An identical ice sheet which imposes assumptions \cref{common_assumptions} will have a uniform thickness layer in the $z$ direction (right).}
  \label{surface_flux_2_image}
\end{SCfigure}

Further consider a static half-sphere ice sheet with zero basal accumulation and uniform surface accumulation $\smb > 0$ (\cref{surface_flux_2_image}).
After a period of time, the imposition of assumptions \cref{common_assumptions} will generate a newly-deposited layer of ice with an unjustifiably lesser thickness at its margins than at its center.
In addition, the ice sheet flow will increase in magnitude as the surface gradient magnitude increases (\cf \cite{bueler_2007}, \cite{greve_2009}); therefore, the imposition of assumptions \cref{common_assumptions} will generate a non-physical augmentation of velocity that is of greatest magnitude near regions of high surface slope.

\begin{SCfigure}
   \centering
    \includegraphics[width=0.6\linewidth]{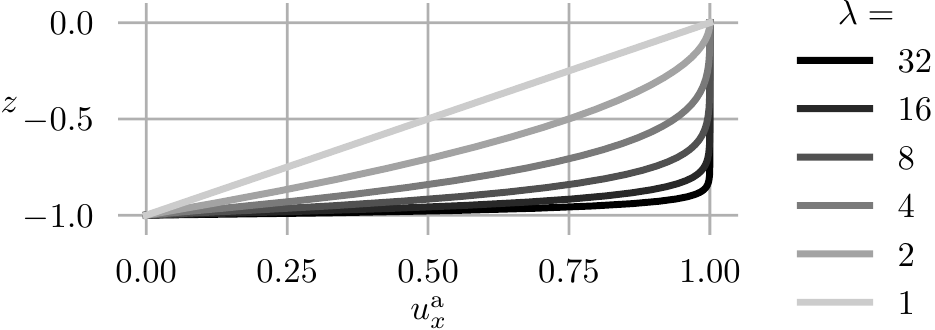}
  \caption{Analytic-$x$ component of velocity $u_x^{\ana}$ given by \cref{analytic_x_velocity} with $S=0$, $B=-1$, $u_{x \bed}^{\ana} = 0$, and $u_{x \srf}^{\ana} = 1$ over a selection of values of deformation parameter $\lambda$.}
  \label{fig_analytic_velocity}
\end{SCfigure}

\section{Analytic solution}
\label{sec_r3_analytic_solution}

An analytic solution satisfying incompressibility relation \cref{incompressible_conservation_of_mass} and free-surface equations \cref{upper_free_surface,lower_free_surface} provides the ability to verify and thereby guarantee the correct implementation of any numerical model associated with ice-sheet mass conservation.
The verification of ice-sheet momentum conservation models using with this velocity in turn makes possible the ability to quantify the effects associated with imposing $\rankone{u} \cdot \normal = 0$ over an impenetrable lower surface with non-zero surface accumulation or ablation (\cf \cref{rmk_impenetrable_surface_remark}); assumptions \cref{common_assumptions} when solving free-surface relations \cref{upper_free_surface,lower_free_surface} for an evolving surface $\Gamma$ (\cf \cref{rmk_normal_kinematic_forcing}); and assumptions \cref{common_assumptions} when inverting quasi-mass-balance relation \cref{balance_velocity_equation_one} (\cf \cref{rmk_smb_coefficients}).

Generalized analytic-velocity solutions proposed heretofore (\cf \cite{sargent_2010} and \cite{leng_2013}) did not incorporate the effects of basal-mass balance $\bmb$ and had been simplified using \cref{common_assumptions}.
Moreover, the boundary condition $\bmb = - \rankone{u} \cdot \normal$ has not been verified by ice-sheet momentum models up to this time; instead, the essential condition $\rankone{u} = \rankone{0}$ has been assumed.
It was first established by \cite{verfuerth_1985} that Galerkin implementations of the Navier boundary conditions converge at a sub-optimal rate.
In addition, it was discovered by \cite{urquiza_2014} that numerical results may fail to converge for domains with smooth boundaries for both Galerkin approximation techniques of \cite{nitsche_1970/71} and \cite{verfuerth_1991} which incorporate Navier boundary conditions.
The Navier boundary condition is a necessary component for any ice-sheet numerical model which specifies basal sliding.
Therefore, the analytic-velocity solution formulated here is the first \emph{completely} generalized and absolutely mass-conserving velocity for use as verification of momentum conservation and all boundary conditions associated with any two- or three-dimensional ice-sheet numerical model.\footnote{Two examples of two-dimensional models include the plane-strain and vertically-integrated $xy$-plane models; the former requires no modification other than the specification of inputs with zero gradient component in one of either the horizontal directions, the latter requires the analytic solutions provided here be integrated vertically.}

Following the procedure of \cite{sargent_2010} and \cite{leng_2013}, the vertical component of velocity $u_z$ is defined such that kinematic-boundary conditions \cref{upper_free_surface,lower_free_surface} are linearly interpolated with depth within the ice-sheet domain $\Omega$.
That is, the linearly-interpolated analytic vertical component of velocity is given by
\begin{align}
  \label{analytic_z_velocity}
  u_z^{\ana}(\rankone{x},t) = \xi_{\srf} u_{z \srf}^{\ana} + \xi_{\bed} u_{z \bed}^{\ana},
\end{align}
where
\begin{align}
  \label{relative_z}
  \xi_{\bed}(\rankone{x},t) = \frac{S(x,y,t) - z}{H(x,y,t)},
  \hspace{10mm}
  \xi_{\srf}(\rankone{x},t) = \frac{z - B(x,y,t)}{H(x,y,t)};
\end{align}
and where $u_{z \srf}^{\ana} = u_z^{\ana}(x,y,z=S,t)$ and $u_{z \bed}^{\ana} = u_z^{\ana}(x,y,z=B,t)$ are given by solving relations \cref{upper_free_surface,lower_free_surface} for $u_z$ at the upper and lower surfaces:
\begin{align}
  \label{analytic_z_velocity_S}
  u_{z \srf}^{\ana}(\rankone{x},t) = &- \Vert \unit{k} - \nabla S \Vert \smb + \parder{S}{t} + u_x^{\ana} \parder{S}{x} + u_y^{\ana} \parder{S}{y} \\
  \label{analytic_z_velocity_B}
  u_{z \bed}^{\ana}(\rankone{x},t) = &+ \Vert \nabla B - \unit{k} \Vert \bmb + \parder{B}{t} + u_x^{\ana} \parder{B}{x} + u_y^{\ana} \parder{B}{y}.
\end{align}
Analytic-velocity components $u_x = u_x^{\ana}(\rankone{x},t)$ and $u_y = u_y^{\ana}(\rankone{x},t)$ are to be determined in the following analysis.

\begin{remark}
Analytic-vertical velocity \cref{analytic_z_velocity} is a function of the geometry $S(\rankone{x},t)$ and $B(\rankone{x},t)$ and horizontal-velocity components $u_x^{\ana}(\rankone{x},t)$ and $u_y^{\ana}(\rankone{x},t)$ only.
Thus $u_z^{\ana}$ defined by \cref{analytic_z_velocity} is fully determined once $S$, $B$, $u_x^{\ana}$ and $u_y^{\ana}$ have been specified.
\end{remark}

For simplicity, the $x$ component of velocity is chosen to be
\begin{align}
  \label{analytic_x_velocity}
  u_x^{\ana}(\rankone{x},t) &= \left( u_{x \srf}^{\ana} - u_{x \bed}^{\ana} \right) \left( 1 - \xi_{\bed}^{\lambda} \right) + u_{x \bed}^{\ana},
\end{align}
where $\lambda \in \R$ controls the magnitude of the $x$-component of velocity with depth, $u_{x \bed}^{\ana}$ is the $x$ component of velocity at the lower surface $B$, and $u_{x \srf}^{\ana}$ is the $x$ component of velocity at the upper surface $S$ (\cref{fig_analytic_velocity}).

\begin{remark}
The value of $u_{x \bed}^{\ana}$ can be used to prescribe an amount of basal sliding, in turn accommodating the possibility to create a non-zero basal-mass balance $\bmb$ over static lower surfaces---characterized by $\rankone{w} \cdot \normal = 0$ on $\Gamma_{\bed}$---as indicated by \cref{surface_mass_balance}.
Hence this parameter may be used to verify the numerical implementation of the Dirichlet boundary condition $\bmb = - \rankone{u} \cdot \normal$ over basal surfaces (\cf \cref{rmk_impenetrable_surface_remark}).
\end{remark}

The final $y$ component of velocity $u_y^{\ana}$ must satisfy incompressibility relation \cref{incompressible_conservation_of_mass}, \ie,
\begin{align}
  \label{analytic_incompressible_conservation_of_mass}
  \parder{u_x^{\ana}}{x} + \parder{u_y^{\ana}}{y} + \parder{u_z^{\ana}}{z} &= 0.
\end{align}
Therefore, taking the derivative of analytic-$z$ velocity \cref{analytic_z_velocity} with respect to $z$ results in
\begin{align}
  \label{du_z_dz}
  \parder{u_z^{\ana}}{z}
  =&+ \frac{1}{H} \left( -\Vert \unit{k} - \nabla S \Vert \smb - \Vert \nabla B - \unit{k} \Vert \bmb + \parder{H}{t} + u_x^{\ana} \parder{H}{x} \right) \notag \\
  &+ \left( u_{x \srf}^{\ana} - u_{x \bed}^{\ana} \right) \lambda \xi_{\bed}^{\lambda} \frac{1}{H} \left( \frac{\xi_{\srf}}{\xi_{\bed}} \parder{S}{x} + \parder{B}{x} \right) \notag \\
  &+ \left( \xi_{\srf} \parder{S}{y} + \xi_{\bed} \parder{B}{y} \right) \parder{u_y^{\ana}}{z} + \frac{1}{H} \parder{H}{y} u_y^{\ana},
\end{align}
while differentiation of analytic-$x$ velocity \cref{analytic_x_velocity} with respect to $x$ results in
\begin{align}
  \label{du_x_dx}
  \parder{u_x^{\ana}}{x} 
  &= \left( 1 - \xi_{\bed}^{\lambda} \right) \parder{u_{x \srf}^{\ana}}{x} + \xi_{\bed}^{\lambda} \parder{u_{x \bed}^{\ana}}{x} - \left( u_{x \srf}^{\ana} - u_{x \bed}^{\ana} \right) \lambda \xi_{\bed}^{\lambda - 1} \parder{\xi_{\bed}}{x}.
\end{align}

Combining mass-conservation relation \cref{analytic_incompressible_conservation_of_mass} with velocity derivatives \cref{du_z_dz,du_x_dx} produces the linear first-order partial-differential equation for the unknown $y$ component of velocity $u_y^{\ana}$ (\cf supplementary material \cref{sec_sup_linear_pde})
\begin{align}
  \label{analytic_y_velocity_problem}
  A + G u_y^{\ana} + \parder{u_y^{\ana}}{y} + C \parder{u_y^{\ana}}{z} = 0,
\end{align}
where
\begin{align*}
  A(\rankone{x},t)
  =&+ \parder{u_{x \srf}^{\ana}}{x} + \frac{1}{H} \left( \parder{H}{t} + u_{x \srf}^{\ana} \parder{H}{x} -  \Vert \unit{k} - \nabla S \Vert \smb - \Vert \nabla B - \unit{k} \Vert \bmb \right) \\
   &+ \frac{(S - z)^{\lambda}}{H^{\lambda}} \left( \parder{u_{x \bed}^{\ana}}{x} - \parder{u_{x \srf}^{\ana}}{x} + \left( u_{x \srf}^{\ana} - u_{x \bed}^{\ana} \right) \left( \frac{1}{H} \parder{B}{x} \right) \right) \\
   &+ \frac{(S - z)^{\lambda - 1} \left( z - S \right) }{H^{\lambda + 1}} \left( u_{x \srf}^{\ana} - u_{x \bed}^{\ana} \right) \parder{S}{x} \\
  G(x,y,t)
  =& + \frac{1}{H} \parder{H}{y} \\
  C(\rankone{x},t)
  =&+ \xi_{\srf} \parder{S}{y} + \xi_{\bed} \parder{B}{y}.
\end{align*}

\begin{remark}
Due to the fact that each of $A$, $G$, and $C$ of equation \cref{analytic_y_velocity_problem} are known for all $\rankone{x}$ and $t$---and that only the $y$ and $z$ derivatives of $u_y^{\ana}$ are present---the unknown velocity $u_y^{\ana}$ is determined solely by its dependence on the $y$ and $z$ coordinates.
In addition, as the coefficients $A$, $G$, and $C$ do not depend on $u_y^{\ana}$, equation \cref{analytic_y_velocity_problem} is \emph{linear}; the authors of \cite{sargent_2010} and \cite{leng_2013} incorrectly describe their analogous relations as \emph{quasi linear}.
Regardless of the classification of the partial-differential equation, the appropriate method used to solve hyperbolic\footnote{Any first-order partial-differential equation is hyperbolic.} problem \cref{analytic_y_velocity_problem} is the \emph{method of characteristics} (\cf section 3.6.4 of \cite{chicone_2006}).
\end{remark}

Consider a normal vector to a manifold $M$ in the $yz$ plane defined by
\begin{align}
  \label{manifold_normal}
  \rankone{n}_M (y,z) = \left[ \hspace{3mm} 1\hspace{3mm} u_y^{\ana}(y,z)\hspace{3mm} \parder{u_y^{\ana}}{y}(y,z)\hspace{3mm} \parder{u_y^{\ana}}{z}(y,z) \hspace{3mm} \right]\T.
\end{align}
Problem \cref{analytic_y_velocity_problem} can thus be stated equivalently as
\begin{align}
  \label{analytic_y_velocity_problem_2}
  \rankone{n}_M \cdot \rankone{a} = 0,
  \hspace{10mm}
  \rankone{a} = \left[\hspace{2mm} A\hspace{2mm}  G\hspace{2mm}  1\hspace{2mm}  C \hspace{2mm} \right]\T. 
\end{align}
The manifold $M$ is called \emph{invariant} if and only if \cref{analytic_y_velocity_problem_2} holds, and in such a case the solution $u_y^{\ana}$ to \cref{analytic_y_velocity_problem_2} is a solution to partial-differential equation \cref{analytic_y_velocity_problem}.
Hence the following theorem is presented:

\begin{theorem}
\label{thm_invariant_manifold}
There exists an invariant manifold $M$ such that \cref{manifold_normal,analytic_y_velocity_problem_2} is a solution to partial-differential equation \cref{analytic_y_velocity_problem}.
\end{theorem}

\begin{proof}
Assume that a curve $s : \R \rightarrow \R^3$ in the $yz$ plane exists such that the solution to \cref{analytic_y_velocity_problem} may be parameterized by $u_y^{\ana}(s) = u_y^{\ana}(y(s), z(s))$.
In this case, application of the chain rule (\cf \cref{thm_one_variable_chain_rule}) produces
\begin{align*}
  \totder{u_y^{\ana}}{s} = \parder{u_y^{\ana}}{y} \totder{y}{s} + \parder{u_y^{\ana}}{z} \totder{z}{s}.
\end{align*}
Comparing this expression with problem \cref{analytic_y_velocity_problem} results in the observation that\footnote{The orbits of \cref{ode} are referred to as the \emph{characteristics} of partial-differential equation \cref{analytic_y_velocity_problem}.}
\begin{align}
  \label{ode}
  \totder{u_y^{\ana}}{s} = A + G u_y^{\ana}
  \hspace{2.5mm}
  \iff
  \hspace{2.5mm}
  \totder{y}{s} = -1,
  \hspace{5mm}
  \totder{z}{s} = -C,
\end{align}
and on elimination of $\d{s}$ produces the \emph{Lagrange-Charpit} equations
\begin{align}
  \label{lagrange_charpit}
  \diff{u_y^{\ana}} \left( A + G u_y^{\ana} \right)^{-1}
  \hspace{2mm}
  =
  \hspace{2mm}
  - \diff{y}
  \hspace{2mm}
  =
  \hspace{2mm}
  - \diff{z} C^{-1}.
\end{align}

The relationship between the two right-most differential terms of \cref{lagrange_charpit} can be stated as (\cf supplementary material \cref{sec_sup_z_coordinate})
$\partial_y \left( z / H \right) = \partial_y \left( B / H \right)$,
which integrated with respect to $y$ produces for some constant $z_0$,
\begin{align}
  \label{constant_0}
  \frac{z - B}{H} = \xi_{\srf} = z_0
  \hspace{5mm}
  \iff
  \hspace{5mm}
  z = z_0 H + B.
\end{align}

The relationship between the two left-most differential terms of \cref{lagrange_charpit} can be stated as (\cf supplementary material \cref{sec_sup_u_y_coordinate})
$\partial_y \left( H u_y^{\ana} \right) = -HA$,
which integrated with respect to $y$ produces for some constant $u_{y 0}$,
\begin{align}
  \label{constant_1}
  H u_y^{\ana} = - &\int_y HA \d{y} + u_{y 0}.
\end{align}
The integration constants embedded within \cref{constant_0} and \cref{constant_1} provide two invariant coordinates in the $yz$ plane defined by
\begin{align*}
  \phi_1(y,z) = z_0 = \xi_{\srf}
  \hspace{10mm}
  \text{and}
  \hspace{10mm}
  \phi_2(y,z) = u_{y0} = H u_y^{\ana} + \int_y H A \d{y};
\end{align*}
therefore, the manifold $M$ of \cref{analytic_y_velocity_problem_2} is invariant.
\end{proof}

The horizontal component of velocity $u_y^{\ana}$ may now be defined by a function $\vartheta = \vartheta(\phi_1, \phi_2)$ satisfying $\vartheta(\phi_1, \phi_2) = 0$.
This function may be arbitrarily specified in terms of the coordinates $\phi_1$ and $\phi_2$; for simplicity, let
$\vartheta (\phi_1, \phi_2) \equiv \{ \phi_1(y,z) = 0,\ \phi_2(y,z) = 0\} \iff z_0 = 0,\ u_{y0} = 0$.
In this case, solving \cref{constant_1} with $ u_{y0} = 0$ for $u_y^{\ana}$ produces 
\begin{align}
  \label{analytic_y_velocity}
  u_y^{\ana}(\rankone{x},t) 
  = - &\frac{1}{H} \int_y \left[ H \parder{u_{x \srf}^{\ana}}{x} + \parder{H}{t} + u_{x \srf}^{\ana} \parder{H}{x} - \Vert \unit{k} - \nabla S \Vert \smb - \Vert \nabla B - \unit{k} \Vert \bmb \right] \d{y} \notag \\
  - &\left( 1 - \frac{z-B}{H} \right)^{\lambda} \frac{1}{H} \int_y
     \left[ H \left( \parder{u_{x \bed}^{\ana}}{x} - \parder{u_{x \srf}^{\ana}}{x} \right) + \left( u_{x \srf}^{\ana} - u_{x \bed}^{\ana} \right) \parder{B}{x} \right] \d{y} \notag \\
  - &\left( 1 - \frac{z-B}{H} \right)^{\lambda - 1} \left( \frac{z-B}{H} - 1 \right) \frac{1}{H} \int_y
     \left[ \left( u_{x \srf}^{\ana} - u_{x \bed}^{\ana} \right) \parder{S}{x} \right] \d{y}.
\end{align}
This completes the derivation of an analytic velocity which satisfies mass conservation relations \cref{incompressible_conservation_of_mass,upper_free_surface,lower_free_surface}; namely, $\rankone{u}^{\ana} = [u_x^{\ana}\ u_y^{\ana}\ u_z^{\ana}]\T$ with components defined by \cref{analytic_z_velocity,analytic_x_velocity,analytic_y_velocity}.

\begin{remark}
\label{rmk_elliptic_integral}
The fourth and fifth integrals of \cref{analytic_y_velocity} given by
\begin{align*}
  \mathcal{I}_{\srf} = \frac{1}{H} \int_y \Vert \unit{k} - \nabla S \Vert \smb \d{y}
  \hspace{10mm}
  \text{and}
  \hspace{10mm}
  \mathcal{I}_{\bed} = \frac{1}{H} \int_y \Vert \nabla B - \unit{k} \Vert \bmb \d{y}
\end{align*}
are \emph{elliptic} for certain combinations of $S$, $\smb$, $B$, and $\bmb$.
Therefore, in order to derive an elementary representation for \cref{analytic_y_velocity}, these functions must be carefully chosen.
In particular, this complication will be avoided in the event that the set of functions $\{S, \smb, B, \bmb\}$ are specified to be independent of the $y$ coordinate.
\end{remark}

To complete the analysis, expression \cref{analytic_y_velocity} is evaluated at the upper and lower surfaces (\cf supplementary material \cref{sec_sup_u_y_surface}), producing 
\begin{align}
  \label{analytic_y_velocity_sb}
  u_{y \mathrm{k}}^{\ana}
  = &- \frac{1}{H} \int_y \left[ \parder{}{x} \left( H u_{x \mathrm{k}}^{\ana} \right) + \parder{H}{t} - \mathring{H} \right] \d{y},
  \hspace{5mm}
  \mathrm{k} = \srf,\bed,
\end{align}
where $\mathring{H}$ is quasi-mass-balance forcing term \cref{balance_velocity_equation_one}$_2$.
Finally, once the forms for the parameters have been chosen, the analytic balance velocity (\cf \cref{balance_velocity}) $\rankone{\bar{u}}^{\ana} = H^{-1} \int_B^S \rankone{u}^{\ana} \d{z}$ can be easily calculated along with the \emph{thickness flux}
\begin{align}
  \label{thickness_flux}
  \nabla \cdot \left(H \rankone{\bar{u}}^{\ana} \right)
  = \rankone{\bar{u}}^{\ana} \cdot \nabla H + H \nabla \cdot \rankone{\bar{u}}^{\ana}
  = \bar{u}_x^{\ana} \parder{H}{x} + \bar{u}_y^{\ana} \parder{H}{y} + H \parder{\bar{u}_x^{\ana}}{x} + H \parder{\bar{u}_y^{\ana}}{y}.
\end{align}

\subsection{Example calculation}

\begin{figure*}
  \centering
    \hspace*{\fill}%
      \subfloat[$u_{x \srf}^{\ana} = u_x^{\ana}(x,y,z = S)$]{\label{fig_r3_u_xs} \includegraphics[width=0.31\linewidth]{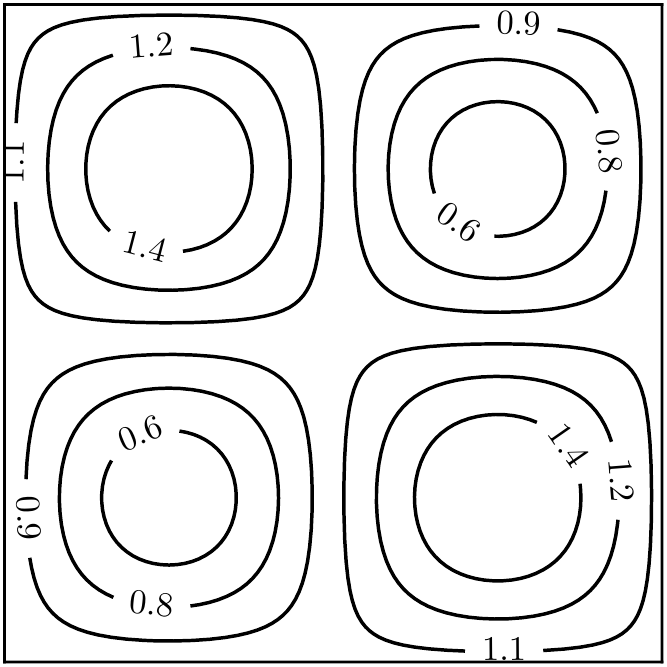}}
    \hfill
      \subfloat[$u_{y \srf}^{\ana} = u_y^{\ana}(x,y,z = S)$]{\label{fig_r3_u_ys} \includegraphics[width=0.31\linewidth]{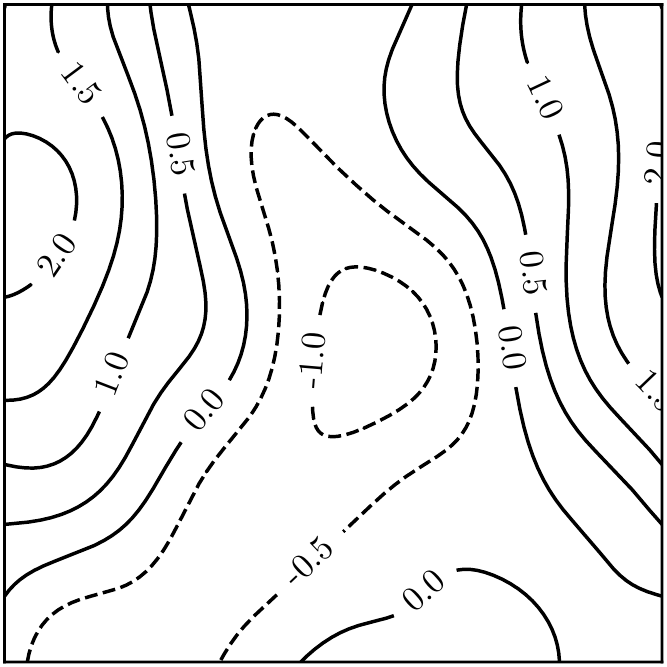}}
    \hfill
      \subfloat[$u_{z \srf}^{\ana} = u_z^{\ana}(x,y,z = S)$]{\label{fig_r3_u_zs} \includegraphics[width=0.31\linewidth]{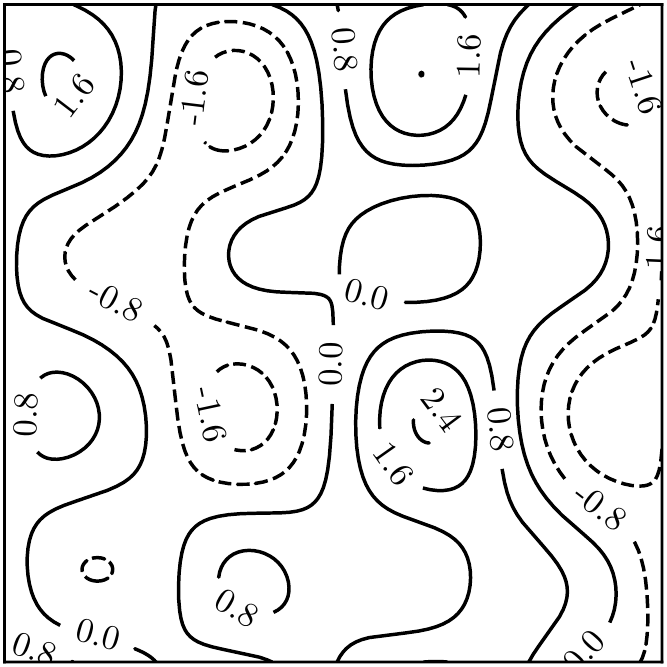}}
    \hspace*{\fill}%

    \vspace{1mm}

    \hspace*{\fill}%
      \subfloat[$u_{x \bed}^{\ana} = u_x^{\ana}(x,y,z = B)$]{\label{fig_r3_u_xb} \includegraphics[width=0.31\linewidth]{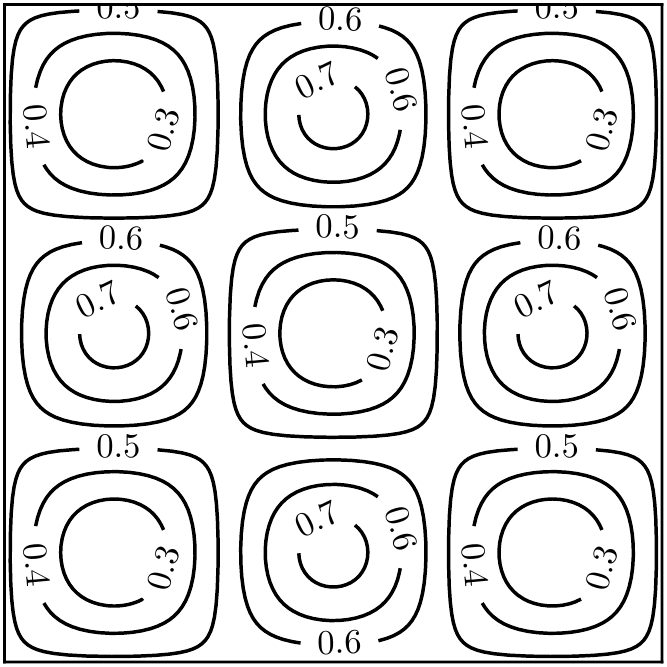}}
    \hfill
      \subfloat[$u_{y \bed}^{\ana} = u_y^{\ana}(x,y,z = B)$]{\label{fig_r3_u_yb} \includegraphics[width=0.31\linewidth]{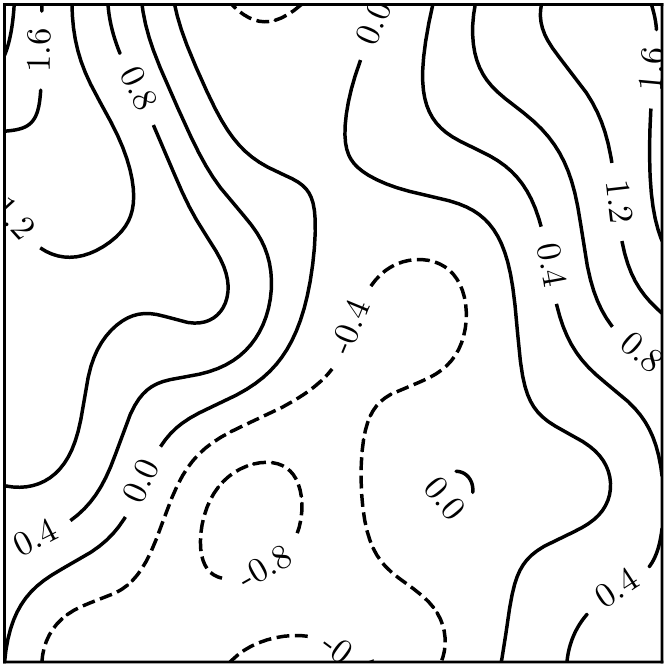}}
    \hfill
      \subfloat[$u_{z \bed}^{\ana} = u_z^{\ana}(x,y,z = B)$]{\label{fig_r3_u_zb} \includegraphics[width=0.31\linewidth]{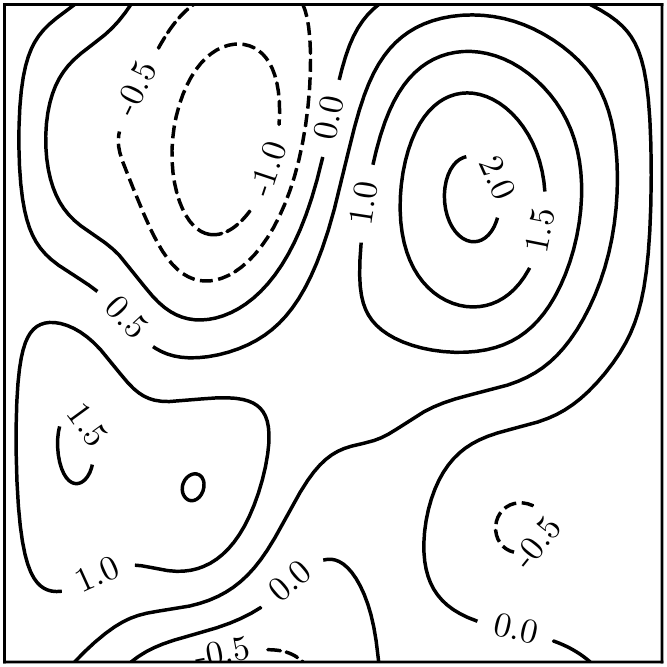}}
    \hspace*{\fill}%

    \vspace{1mm}

    \hspace*{\fill}%
      \subfloat[$\partial_t S$]{\label{fig_r3_dsdt} \includegraphics[width=0.31\linewidth]{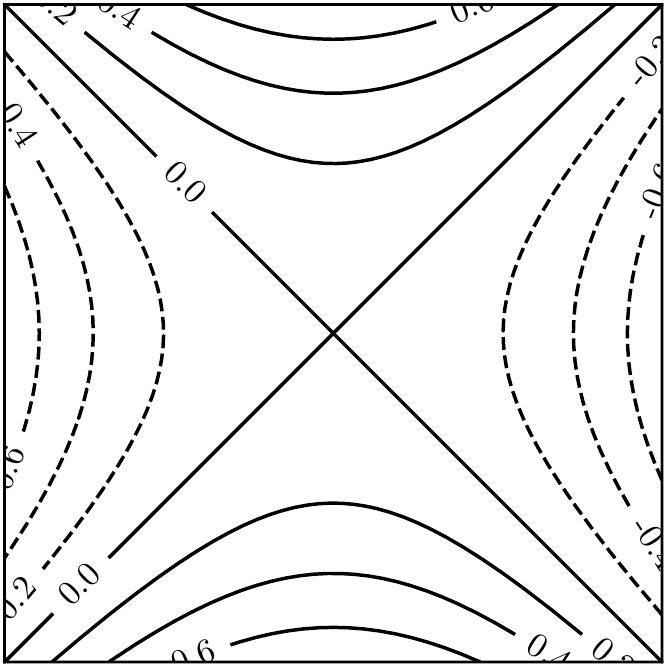}}
    \hfill
      \subfloat[$\partial_t B$]{\label{fig_r3_dbdt} \includegraphics[width=0.31\linewidth]{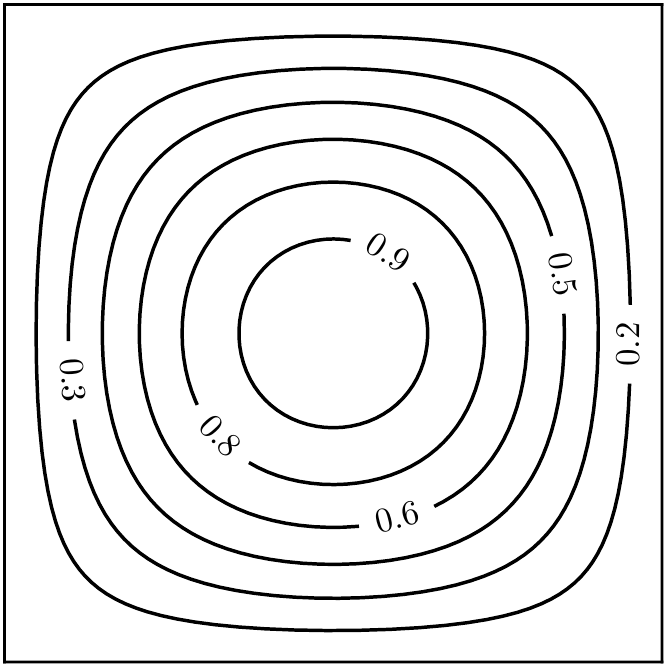}}
    \hfill
      \subfloat[$\partial_t H = \partial_t S - \partial_t B$]{\label{fig_r3_dhdt} \includegraphics[width=0.31\linewidth]{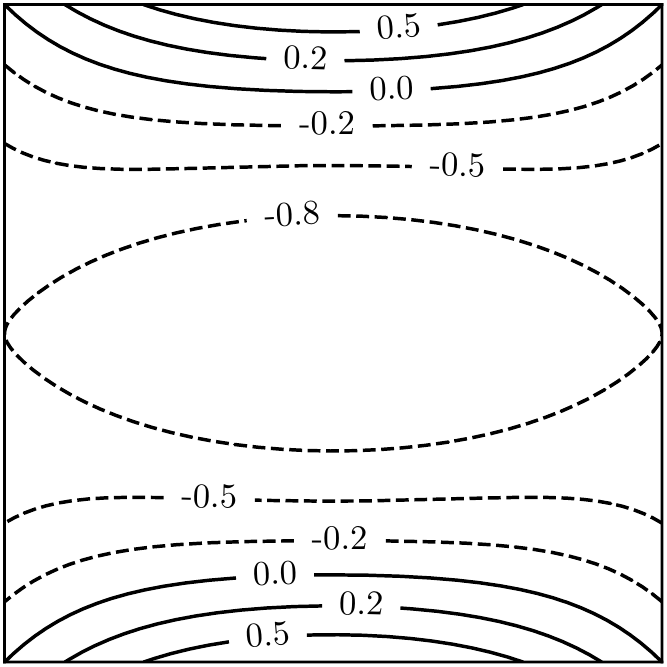}}
    \hspace*{\fill}%

    \vspace{1mm}
    
    \hspace*{\fill}%
      \subfloat[$\smb$]{\label{fig_r3_smb} \includegraphics[width=0.31\linewidth]{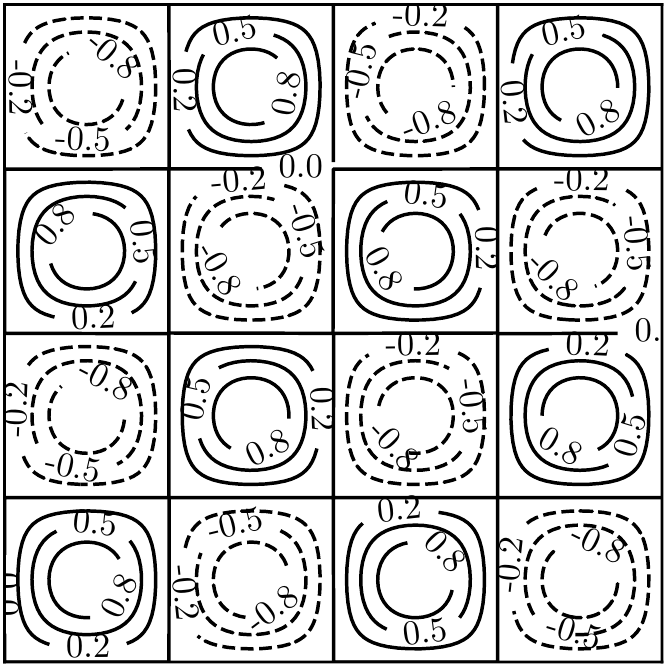}}
    \hfill
      \subfloat[$\bmb$]{\label{fig_r3_bmb} \includegraphics[width=0.31\linewidth]{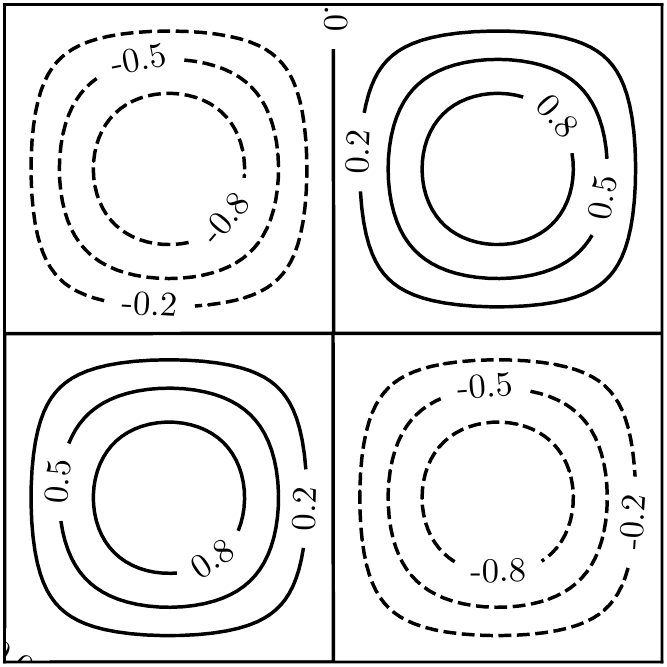}}
    \hfill
      \subfloat[$\nabla \cdot \left( H \rankone{\bar{u}}^{\ana} \right)$]{\label{fig_r3_div_hu} \includegraphics[width=0.31\linewidth]{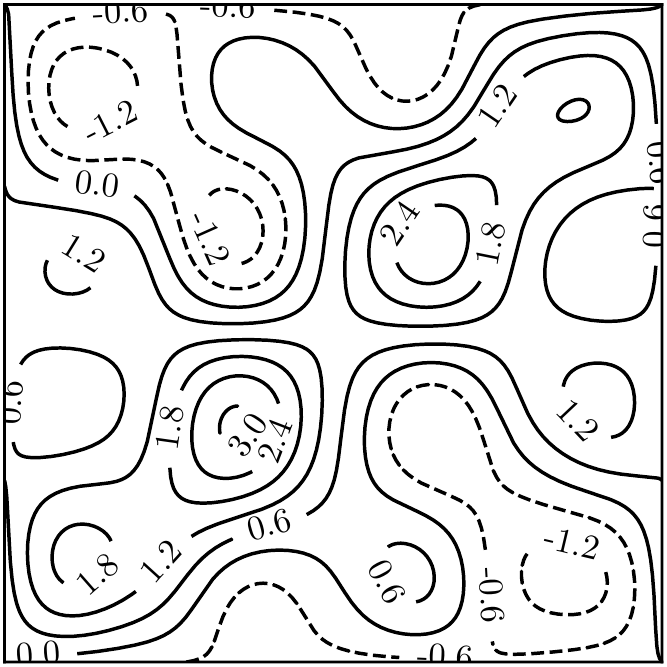}}
    \hspace*{\fill}%
  \caption{Inputs and analytic elements generated from relations \cref{analytic_z_velocity}, \cref{analytic_x_velocity}, and \cref{analytic_y_velocity} defined over the $xy$ plane with horizontal ($x$-axis) and vertical ($y$-axis) dimensions $\ell = 1$.}
  \label{fig_r3}
\end{figure*}

A specific realization of the solution derived in \cref{sec_r3_analytic_solution} is hereby generated (\cref{fig_r3}) over the ice-sheet domain $\Omega = [0,\ell] \times [0,\ell] \times [B,S] \subset \R^3$ with upper and lower surfaces\footnote{These surfaces were chosen independent from $y$ as suggested by \cref{rmk_elliptic_integral}.}
\begin{align*}
  S(x) = \frac{1}{10} \sin\left( \frac{3 \pi}{\ell} x \right) 
  \hspace{8mm}
  \text{and}
  \hspace{8mm}
  B(x) = S(x) - \frac{1}{2} + \frac{1}{10} \sin\left( \frac{2 \pi}{\ell} x \right).
\end{align*}
The sinusoidally-varying $x$ component of velocity at the upper and lower surfaces were chosen to be (\cref{fig_r3_u_xs,fig_r3_u_xb})
\begin{align*}
  u_{x \srf}^{\ana} = 1 - \frac{1}{2} \sin\left( \frac{2 \pi}{\ell} x \right) \sin\left( \frac{2 \pi}{\ell} y \right), 
  \hspace{10mm}
  u_{x \bed}^{\ana} = \frac{1}{2} - \frac{1}{4} \sin\left( \frac{3 \pi}{\ell} x \right) \sin\left( \frac{3 \pi}{\ell} y \right).
\end{align*}
The surface height rate of change at the upper and lower surfaces given respectively by (\cref{fig_r3_dsdt,fig_r3_dbdt})
\begin{align*}
  \parder{S}{t} =  3 \left( \left(y - \frac{\ell}{2} \right)^2 - \left(x - \frac{\ell}{2} \right)^2 \right)
  \hspace{8mm}
  \text{and}
  \hspace{8mm}
  \parder{B}{t} =  \sin\left( \frac{\pi}{\ell} x \right) \sin\left( \frac{\pi}{\ell} y \right)
\end{align*}
were chosen in order to generate an ice-sheet domain that is thinning over the interior and thickening in proximity to the $\pm y$ faces (\cref{fig_r3_dhdt}).
Finally, the upper and lower surface-mass balance terms were respectively chosen to be (\cref{fig_r3_smb,fig_r3_bmb})
\begin{align*}
  \smb = \sin\left( \frac{4 \pi}{\ell} x \right) \sin\left( \frac{4 \pi}{\ell} x \right)
  \hspace{8mm}
  \text{and}
  \hspace{8mm}
  \bmb = \sin\left( \frac{2 \pi}{\ell} x \right) \sin\left( \frac{2 \pi}{\ell} x \right)
\end{align*}
in order to optimally demonstrate the flexibility of the manufactured solution.

The analytic velocity components defined by \cref{analytic_z_velocity,analytic_x_velocity,analytic_y_velocity} were calculated using the open-source software SymPy \cite{meurer_2017} (\cf \cref{r3_script}).
Choosing $\lambda = 2$ and $\ell = 1$, the resulting analytic $y$ component of velocity derived from \cref{analytic_y_velocity} at the upper surface (\cref{fig_r3_u_ys}) and lower surface (\cref{fig_r3_u_yb}) contains both negative and positive values, with magnitude on the order of the $x$ component of velocity.
In addition, periodicity of $u_x^{\ana}$, $S$, $\partial_t S$, $\smb$, $B$, $\partial_t B$, and $\bmb$ are not translated to $u_y^{\ana}$ or $u_z^{\ana}$ due to the integration across $y$ used to form \cref{analytic_y_velocity} (\cref{fig_r3_u_zs,fig_r3_u_zb}).
Finally, the thickness flux (\cref{fig_r3_div_hu}) computed from \cref{thickness_flux} illustrates the complicated structures in mass flux resulting from incorporation of all aspects of mass conservation for ice sheets.

\begin{remark}
The fact that the analytic velocity described here is not periodic precludes the possibility of using this solution to verify periodic boundary conditions.
However, the input data can be chosen such that the resulting $u_y^{\ana}$ and therefore also $u_z^{\ana}$ are periodic.
In any case, the lateral-velocity boundary conditions for $u_y$ and $u_z$ can be specified to correspond with the analytic solutions $u_y^{\ana}$ and $u_z^{\ana}$, thereby making possible the verification of mass-conservation relations \cref{incompressible_conservation_of_mass}, \cref{upper_free_surface}, and \cref{lower_free_surface} within a numerical model; this is the main purpose of this section.
\end{remark}

\section{Future work}
\label{sec_future_work}

The concepts of mass balance derived here provide the basis of a new formulation for energy and momentum conservation for ice sheets.
Once formulated, these concepts will be used to describe conservation laws near the interface of ice and ocean.

\section*{Acknowledgments}

Special thanks to all involved with the development of the open-source software SymPy \cite{meurer_2017}, NumPy \cite{annoortvanvanderwalt_2011}, SciPy \cite{jones_2001}, Matplotlib \cite{hunter_2007}, IPython \cite{perez_2007}, and Inkscape \cite{inkscape}; without whose contributions this project would have been much more difficult to complete.

\DeclareRobustCommand{\VAN}[3]{#3}

\bibliographystyle{siamplain}
\bibliography{../../biblio.bib}

\newpage
\appendix

\section{Mathematical background}
\label{sec_sup_math_background}

\begin{SCfigure}
  \centering
    \includegraphics[width=0.5\linewidth]{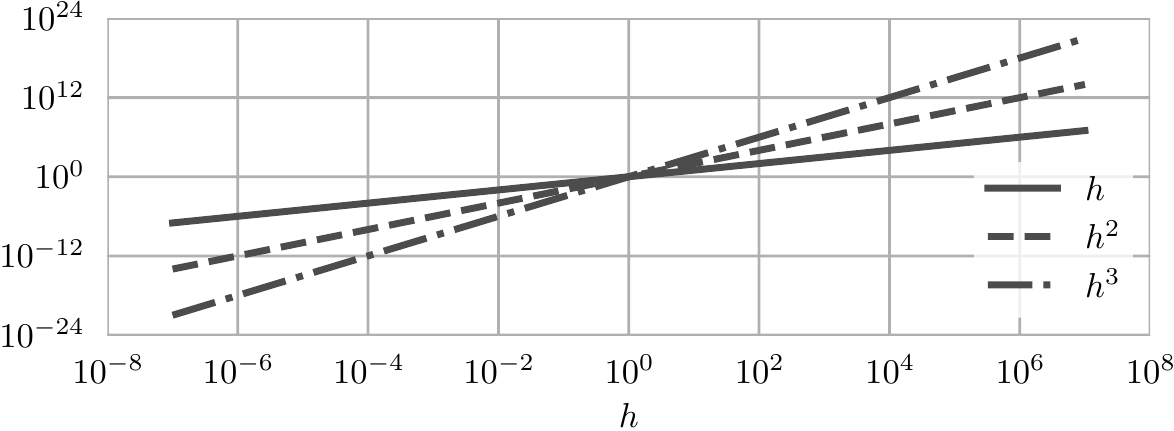}
  \caption{Behavior of the Taylor coefficients as $h$ decreases.  It is common to refer to the truncated Taylor series which does not include the $h^n$ terms as the \emph{$\mathrm{n}$th-order-Taylor-series approximation} which is clearly only accurate for $h \ll 1$.}
  \label{fig_taylor_perturbations}
\end{SCfigure}

This section introduces the relevant mathematical background and definitions utilized throughout the text.
For more information regarding higher-dimensional calculus, consult the work of \cite{larson_2010}.
For more information about general problems in applied mathematics and continuum mechanics, investigate the works of \cite{logan_2006} or \cite{salencon_2001}.
An alternative and closely-related continuum-mechanical formulation for ice-sheets has been provided by \cite{hutter_1982} and expanded upon by \cite{greve_2009}.

The mathematical theory for discontinuous materials are not as readily available as those for continuous materials; therefore, proofs of pertinent theorems have been deliberated.
As such, this section provides a new illustration of the origins of fundamental ice-sheet conservation laws and, moreover, continuum formulations for any discontinuous media.
The basis from which we begin is the following fundamental theorem presented by \cite{taylor_1715}:

\begin{theorem}[Taylor's theorem]
\label{thm_taylors_theorem}
Any real-valued function $f(x) \in \R$ with $x \in \R$ that is infinitely differentiable about a point $x+h$ with distance $h>0$ may be expressed as the infinite \emph{Taylor series}
\begin{align*}
  f(x+h) &= f(x) + f'(x) h + \frac{1}{2!} f''(x) h^2 + \frac{1}{3!} f'''(x) h^3 + \cdots,
\end{align*}
where $f'(x)$\footnote{The prime notation $f'(x)$ was coined by Joseph Louis Lagrange within the later half of the 18th century.} is the ratio of an infinitesimal change the function $f$ with respect to an infinitesimal change in its coordinate $x$.
\end{theorem}

\begin{proof}
Consult chapter 9 section 7 of \cite{larson_2010}.
\end{proof}

For scalar variables of multiple coordinates, \cref{thm_taylors_theorem} yields the following corollary:

\begin{corollary}
\label{thm_multi_dimensional_taylors_theorem}
Any multi-dimensional and real-valued function $f(\rankone{x}) \in \R$ with $\rankone{x} \in \R^n$ that is infinitely differentiable about a point $\rankone{x}+\rankone{h}$ where distance vector $\rankone{h}$ has magnitude $\Vert \rankone{h} \Vert > 0$ may be expressed as the infinite \emph{multi-dimensional Taylor series}
\begin{align*}
  f(\rankone{x}+\rankone{h}) &= f(\rankone{x}) + \nabla f(\rankone{x}) \cdot \rankone{h} + \frac{1}{2!} \rankone{h} \cdot \nabla^2 f(\rankone{x}) \cdot \rankone{h} + \cdots \hspace{2mm},
\end{align*}
where $\nabla f(\rankone{x})$ is the \emph{gradient} of the function $f$ with respect to its coordinates $\rankone{x}$.
\end{corollary}

Recalling that $0 < h \neq 0$, rearranging terms of \cref{thm_taylors_theorem} and division by $h$ produces
\begin{align*}
  \frac{ f(x+h) - f(x) }{h} = f'(x) + \bigo(h),
\end{align*}
which in the limit of small $h$ (\cref{fig_taylor_perturbations}) gives rise to the following definitions:

\begin{definition}[Derivative]
\label{def_derivative}
The \emph{derivative} of a real-valued function $f(x) \in \R$ with respect to $x \in \R$ is given by 
\begin{align*}
  f'(x) \equiv \totder{f}{x} = \lim_{h \rightarrow 0} \left\{ \frac{ f(x+h) - f(x) }{h} \right\}.
\end{align*}
\end{definition}

\begin{definition}[Partial derivative]
\label{def_partial_derivative}
The \emph{partial derivative} of a real-valued function $f(\rankone{x}) \in \R$ with respect to $x_i \in \rankone{x} \in \R^n$ is given by 
\begin{align*}
  \parder{f}{x_i} = \lim_{h \rightarrow 0} \left\{ \frac{ f(x_1, \ldots, x_i+h, \ldots, x_n) - f(\rankone{x}) }{h} \right\}.
\end{align*}
\end{definition}

Recalling that \cref{thm_multi_dimensional_taylors_theorem} required that the distance $\Vert \rankone{h} \Vert > 0$, rearranging terms of \cref{thm_multi_dimensional_taylors_theorem} and division by $\Vert \rankone{h} \Vert$ produces
\begin{align*}
  \frac{ f(\rankone{x}+\rankone{h}) - f(\rankone{x}) }{ \Vert \rankone{h} \Vert} = \nabla f(\rankone{x}) \cdot \hat{\rankone{h}} + \bigo\left( \Vert \rankone{h} \Vert \right),
\end{align*}
where $\hat{\rankone{h}} = \rankone{h} / \Vert \rankone{h} \Vert$ is a unit vector in the direction $\rankone{h}$.
Similar to the line of reasoning resulting in \cref{def_derivative}, taking the limit of small $\Vert \rankone{h} \Vert$ suggests the following definition:

\begin{definition}[Gradient]
\label{def_gradient}
The \emph{gradient} of a real-valued function $f(\rankone{x}) \in \R$ with respect to its vector of coordinates $\rankone{x} = [x_1\ x_2\ \cdots\ x_n]\T \in \R^n$ is given by 
\begin{align*}
  \parder{f}{\rankone{x}} &\equiv \nabla f = \left[ \parder{f}{x_1}\ \parder{f}{x_2}\ \cdots\ \parder{f}{x_n} \right]\T.
\end{align*}
\end{definition}

\begin{theorem}[Chain rule: one independent variable]
\label{thm_one_variable_chain_rule}
Let $f = f(\rankone{x})$ be a function composed of $n$ coordinates of a vector $\rankone{x} = \rankone{x}(t) \in \R^n$ each dependent on a single variable $t \in \R$.
Then the derivative of $f$ with respect to $t$ is
\begin{align*}
  \totder{f}{t} = \parder{f}{x_1}\totder{x_1}{t} + \parder{f}{x_2}\totder{x_2}{t} + \cdots + \parder{f}{x_n}\totder{x_n}{t}.
\end{align*}
\end{theorem}

\begin{proof}
Consult chapter 13 section 5 of \cite{larson_2010}.
\end{proof}

\begin{corollary}[Chain rule: two independent variables]
\label{thm_two_variable_chain_rule}
Let $f = f(\rankone{x})$ be a function composed of $n$ coordinates of a vector $\rankone{x} = \rankone{x}(t,s) \in \R^n$ each dependent on two variables $t,s \in \R$.
Then the derivative of $f$ with respect to $t$ is
\begin{align*}
  \parder{f}{t} = \parder{f}{x_1}\parder{x_1}{t} + \parder{f}{x_2}\parder{x_2}{t} + \cdots + \parder{f}{x_n}\parder{x_n}{t}
\end{align*}
and the derivative of $f$ with respect to $s$ is
\begin{align*}
  \parder{f}{s} = \parder{f}{x_1}\parder{x_1}{s} + \parder{f}{x_2}\parder{x_2}{s} + \cdots + \parder{f}{x_n}\parder{x_n}{s}.
\end{align*}
\end{corollary}

\begin{proof}
Consult chapter 13 section 5 of \cite{larson_2010}.
\end{proof}

\begin{remark}
The chain rule for one variable given by \cref{thm_one_variable_chain_rule} can be proved by linearizing the Taylor-series expansion of a function $f(\rankone{x}(t), t)$ about a point $t + \Delta t$; it is important to note that it is invariant with the choice of coordinate system and therefore provides a measure of total variance of a function with respect to an independent variable, commonly time.
Similarly, the chain rule for two variables given by \cref{thm_two_variable_chain_rule} can be proved by linearizing the Taylor-series expansion of a function $f(\rankone{x}(s,t), s,t)$ about the points $t + \Delta t$ and $s + \Delta s$.
\end{remark}

\begin{remark}
\label{total_derivative}
The chain rule for one independent variable is widely used in the field of continuum mechanics, which specifies that a variable $F = F(\rankone{h},t)$ in the \emph{Lagrangian frame} of reference defined from an initial-coordinate vector $\rankone{h}$ at time $t = 0$ has a counterpart $f = f(\rankone{x}, t)$ in the \emph{Eulerian frame} at the current-coordinate vector $\rankone{x} = \rankone{x}(\rankone{h}, t)$.
In addition, the function $\rankone{x}$ is defined such that a continuously-differentiable mapping back to initial coordinate $\rankone{h}$ exists for all $t$, thereby ensuring that $\rankone{x}(\rankone{h},t)$ may be inverted for the vector function $\rankone{h} = \rankone{h}(\rankone{x}, t)$.
Therefore,
$F(\rankone{h}, t) = f(\rankone{x}(\rankone{h}, t), t)$.
and
$f(\rankone{x}, t) = F(\rankone{h}(\rankone{x}, t), t)$.
Using the fact that the value of $\rankone{h} = \rankone{h}(\rankone{x}, t)$ will be identical at each instant $t$, the derivative of the function $F$ with respect to $t$ is given simply by\footnote{The overhead-dot notation $(\ \dot{ }\ )$ was first used by \cite{newton_1704} to specifically denote differentiation with respect to time.  Other rates of change are denoted herein by the notation $(\ \mathring{ }\ )$; these quantities are defined with relations for which chain-rule \cref{thm_one_variable_chain_rule} cannot be applied.}
\begin{align*}
  \totder{F}{t} = \dot{F}(t).
\end{align*}
However, the position of a point in the continuum will vary with time, and so $\rankone{x} = \rankone{x}(t)$.
Therefore, the derivative of the function $f$ with respect to $t$ is in this case given by
\begin{align*}
  \totder{f}{t} = \dot{f}(\rankone{x},t) = \parder{f}{t} \totder{t}{t} + \parder{f}{\rankone{x}} \totder{\rankone{x}}{t},
\end{align*}
which taken in combination with scalar-gradient \cref{def_gradient} and velocity-vector
$\dot{\rankone{x}} = \rankone{u} = \left[ \dot{x}\ \dot{y}\ \dot{z} \right]\T$
suggest the notation
\begin{align*}
  \dot{f} &= \parder{f}{t} + \rankone{u} \cdot \nabla f.
\end{align*}
This relation has been used often enough that it has been given many other names, including the \emph{substantial, convective, Lagrangian, material} or \emph{total} derivative, among others.
Additionally, the complete change in the state of $f$ described by the relation $\dot{f}$ is referred to as \emph{convection}; for example, Fourier's law of heat \emph{conduction} is specified with a second-order \emph{diffusion} operator $\Lu$ such that $\dot{f} = \Lu u$, and so convection encompasses both advective and diffusive processes.
Finally, the literature commonly refers to the second term $\rankone{u} \cdot \nabla f$ as the \emph{advective} component due to the fact that it is responsible for the transport of the quantity $f$ at speed $\rankone{u}$.\footnote{The process of advection can be better understood by consideration of fundamental-conservation equation \cref{def_fundamental_conservation_equation} and continuity equation \cref{def_conservative_continuity_equation} and \cref{def_nonconservative_continuity_equation}.}
\end{remark}

\begin{theorem}[First fundamental theorem of calculus]
\label{thm_first_fundamental_theorem_of_calculus}
If a given function $f(x)$ is continuous within a closed interval $[a,b]$ and $F(x)$ is an anti-derivative or indefinite integral of $f(x)$, then
\begin{align*}
  \int_a^b f(x) \d{x} = F(b) - F(a).
\end{align*}
\end{theorem}

\begin{proof}
Consult chapter 4 section 4 of \cite{larson_2010}.
\end{proof}

\begin{theorem}[Second fundamental theorem of calculus]
\label{thm_second_fundamental_theorem_of_calculus}
If $f(x)$ is a continuous function within an open interval $(a,b)$, an anti-derivative or indefinite integral $F(x)$ is defined as
\begin{align*}
  F(x) = \int_a^x f(t) \d{t} \hspace{5mm} \text{with inverse} \hspace{5mm} 
  \totder{F}{x} = f(x)
\end{align*}
at each point in $x \in (a,b)$.
\end{theorem}

\begin{proof}
Consult chapter 4 section 4 of \cite{larson_2010}.
\end{proof}

The ice-sheet surface can be completely described using the following theorem:

\begin{theorem}[Gradient of an implicit surface is normal to boundary]
\label{thm_implicit_surface}
Consider a differentiable surface $\Gamma = \Gamma(\rankone{x}) \in \R^3$ with curvature entirely represented by a differentiable function $z = S(x,y) \in \R^2$ with tangent surface $\nabla S$ in the $xy$ plane.
The surface $\Gamma$ can be represented implicitly as
\begin{align*}
  F(\rankone{x}) = 
  \left| z - S(x,y) \right| = 0 =
  \begin{cases}
    S(x,y) - z = 0 \\
    z - S(x,y) = 0
  \end{cases},
\end{align*}
and possesses the property that
$\nabla F \cdot \nabla S = 0$;
that is, the gradient of $F$ is normal to the surface $S$.
\end{theorem}

\begin{proof}
Let the vector-valued function $\rankone{s} = \rankone{s}(\rankone{x}(t))$ represent an arbitrary curve lying on the surface $z = S(x,y)$ for all $t$.
Then $F(\rankone{s}(t)) = 0$ for all $t$ and application of the chain rule (\cf \cref{thm_one_variable_chain_rule}) yields
\begin{align*}
  \totder{F}{t} = &\parder{F}{x} \totder{x}{t} + \parder{F}{y} \totder{y}{t} + \parder{F}{z} \totder{z}{t} = 0.
\end{align*}
Using the facts that $\diff_t x = \partial_x S$, $\diff_t y = \partial_y S$, and $\diff_t z = \partial_z S = 0$, gradient notation $\nabla f = [\partial_x f\ \partial_y f\ \partial_z f]\T$ (\cf \cref{def_gradient}) yields \cref{thm_implicit_surface}.
\end{proof}

When the boundary $\Gamma$ of a volume $\Omega$ can be decomposed into two surfaces, one coordinate can be eliminated from the normal vector using \cref{thm_implicit_surface}, as follows:

\begin{definition}[Outward-pointing unit-normal vector]
\label{def_normal_vector}
Consider a body $\Omega(\rankone{x},t) \in \R^3$ with differentiable surface $\Gamma = \partial \Omega(\rankone{x},t)$ composed of two surfaces $\Gamma_{\srf}$ and $\Gamma_{\bed}$ where $\Gamma_{\srf} \cup \Gamma_{\bed} = \Gamma$ and $\Gamma_{\srf} \cap \Gamma_{\bed} = \emptyset$.
Decompose $\Gamma$ using intrinsically-defined upper and lower surfaces (\cf \cref{thm_implicit_surface})
\begin{align*}
  F_{\srf}(\rankone{x},t) = z - S(x,y,t) = 0 
  \hspace{5mm}
  \text{and}
  \hspace{5mm}
  F_{\bed}(\rankone{x},t) = B(x,y,t) - z = 0
\end{align*}
for some pair of functions $S,B \in \R^2$ defined such that $B \leq S$ for all $x$, $y$, and $t$.
It follows that the gradient of the upper and lower surface at any instant $t$ is given by
$\nabla F_{\srf} = \unit{k} - \nabla S$
and
$\nabla F_{\bed} = \nabla B - \unit{k}$,
where $\unit{k}$ is the unit vector pointing in the direction of increasing $z$ coordinate.
The \emph{outward-pointing unit-normal vectors} over these surfaces are respectively 
\begin{align*}
  \normal_{\srf} = \frac{ \nabla F_{\srf} }{\Vert \nabla F_{\srf} \Vert}
  \hspace{10mm} \text{and} \hspace{10mm}
  \normal_{\bed} = \frac{ \nabla F_{\bed} }{\Vert \nabla F_{\bed} \Vert},
\end{align*}
with gradient magnitudes given by the $L^2(\Gamma)$ norm; \ie,
\begin{align*}
  \Vert \nabla F_{\srf}(\rankone{x}) \Vert &= \left( \unit{k} \cdot \unit{k} - 2 \unit{k} \cdot \nabla S + \nabla S \cdot \nabla S \right)^{\frac{1}{2}}
  = \left( 1 + \left( \parder{S}{x} \right)^2 + \left( \parder{S}{y} \right)^2 \right)^{\frac{1}{2}} \\ 
  \Vert \nabla F_{\bed}(\rankone{x}) \Vert &= \left( \nabla B \cdot \nabla B - 2 \unit{k} \cdot \nabla B + \unit{k} \cdot \unit{k} \right)^{\frac{1}{2}}
  = \left( 1 + \left( \parder{B}{x} \right)^2 + \left( \parder{B}{y} \right)^2 \right)^{\frac{1}{2}},
\end{align*}
where $\partial_z S = \partial_z B = 0$ and $\unit{k} \cdot \unit{k} = 1$ has been used.
\end{definition}

The following theorem illuminates the fact that the implicit surface-gradient norms defined in \cref{def_normal_vector} are integral to the concept of surface area:

\begin{theorem}[Area of an implicit surface]
\label{thm_implicit_surface_area}
For a continuous and differentiable function $S = S(x,y) \in \R^2$ defined within a region $R$ in the $xy$ plane, the area of the surface $S$ given by the implicit surface defined by \cref{thm_implicit_surface} is
\begin{align*}
  A = \int_R \d{S} = \int_R \Vert \unit{k} - \nabla S \Vert \d{R} \approx \sum_{i=1}^n \Vert \unit{k} - \nabla S_i \Vert \Delta x_i \Delta y_i,
\end{align*} 
for a discretization of $R$ into $n$ rectangles with dimensions $\Delta x_i$ and height $\Delta y_i$ for all $i = 1,2,\ldots,n$.
\end{theorem}

\begin{proof}
A region $A_i$ of an arbitrary parallelogram $i$ of the surface discretization $S$ defined in \cref{thm_implicit_surface_area} have sides given by the vectors
$\rankone{u}_i = \begin{bmatrix} \Delta x_i & 0 & \Delta x_i \partial_x S_i \end{bmatrix}\T$,
and
$\rankone{v}_i = \begin{bmatrix} 0 & \Delta y_i & \Delta y_i \partial_y S_i \end{bmatrix}\T$.
Hence the area of $A_i$ is given by
\begin{align*}
\Vert \rankone{u}_i \times \rankone{v}_i \Vert = \left(1 + \left( \parder{S_i}{x} \right)^2 + \left( \parder{S_i}{y} \right)^2 \right)^{\frac{1}{2}} \Delta x_i \Delta y_i.
\end{align*}
Using the normal-vector magnitude of \cref{def_normal_vector}, summing over all $i$ parallelograms, and taking the limit as $\Delta x_i, \Delta y_i \rightarrow 0$ produces \cref{thm_implicit_surface_area}.
\end{proof}
 
\begin{theorem}[Leibniz's rule]
\label{thm_leibniz_rule}
Leibniz formula, referred to as \emph{Leibniz's rule for differentiating an integral with respect to a parameter that appears in the integrand and in the limits of integration}, states that if $f(x,y)$ and $\partial_x f(x,y)$ are both continuous over the finite domain $y \in [a(x),b(x)]$,
\begin{align*}
  \totder{}{x} \int_{a(x)}^{b(x)} f(x,y) \d{y} &= \int_{a(x)}^{b(x)} \parder{f}{x}(x,y) \d{y}
  + f(x,b(x)) \totder{b}{x} - f(x,a(x))\totder{a}{x}.
\end{align*}
\end{theorem}
\begin{proof}
Let
$$\mathcal{I}(x,a(x),b(x)) = \int_{a(x)}^{b(x)} f(x,y) \d{y}.$$
Using one-variable chain rule \cref{thm_one_variable_chain_rule}, the derivative of $\mathcal{I}$ with respect to $x$ is
$$\totder{\mathcal{I}}{x} = \parder{\mathcal{I}}{x} \totder{x}{x} + \parder{\mathcal{I}}{a} \totder{a}{x} + \parder{\mathcal{I}}{b} \totder{b}{x}.$$
Next, due to the fact that integration is performed over the coordinate $y$, the linear operations of partial-$x$ differentiation and integration over $y$ may be safely exchanged.
That is,
\begin{align*}
  \parder{\mathcal{I}}{x} 
  &= \lim_{\Delta x \rightarrow 0} \frac{1}{\Delta x} \left( \int_{a(x)}^{b(x)} f(x + \Delta x,y) \d{y} - \int_{a(x)}^{b(x)} f(x,y) \d{y} \right) \\
  &= \lim_{\Delta x \rightarrow 0} \int_{a(x)}^{b(x)} \left( \frac{f(x + \Delta x,y) - f(x,y)}{\Delta x} \right) \d{y} 
   = \int_{a(x)}^{b(x)} \parder{f}{x}(x,y) \d{y}.
\end{align*}
Next, using first fundamental theorem of calculus \cref{thm_first_fundamental_theorem_of_calculus}, if $F(x,y)$ is the indefinite integral of $f(x,y)$ with respect to $y$ on $y \in [a,b]$,
\begin{align*}
  \parder{\mathcal{I}}{a} &= \parder{}{a} \Big[ F(x,b(x)) - F(x,a(x)) \Big] = - \parder{}{a} F(x,a(x)) \\
  \parder{\mathcal{I}}{b} &= \parder{}{b} \Big[ F(x,b(x)) - F(x,a(x)) \Big] = + \parder{}{b} F(x,b(x)),
\end{align*}
and using second fundamental theorem of calculus \cref{thm_second_fundamental_theorem_of_calculus},
\begin{align*}
  \left. \parder{}{a} F(x,y) \right|_{y=a} &= \left( \parder{F}{y} \parder{y}{a} \right)_{y=a} = f(x,a(x)) \\
  \left. \parder{}{b} F(x,y) \right|_{y=b} &= \left( \parder{F}{y} \parder{y}{b} \right)_{y=b} = f(x,b(x)).
\end{align*}
Combining the above relations results in \cref{thm_leibniz_rule}.
\end{proof}

\begin{corollary}[Leibniz's rule for two independent variables]
\label{thm_leibniz_rule_corollary}
If a function $f = f(x,y,z)$ and $\partial_x f$ are both continuous over a finite domain $z \in [a(x,y),b(x,y)]$,
\begin{align*}
  \parder{}{x} \int_{a(x,y)}^{b(x,y)} f(x,y,z) \d{z} = &+ \int_{a(x,y)}^{b(x,y)} \parder{f}{x}(x,y,z) \d{z} \notag \\
  &+ f(x,y,b(x,y)) \parder{b}{x}
   - f(x,y,a(x,y)) \parder{a}{x}.
\end{align*}
\end{corollary}

\begin{proof}
Let
$$\mathcal{I}(x,y,a(x,y),b(x,y)) = \int_{a(x,y)}^{b(x,y)} f(x,y,z) \d{z}.$$
Using the one-variable chain rule (\cf \cref{thm_one_variable_chain_rule}), the derivative of $\mathcal{I}$ with respect to $x$ is
\begin{align*}
  \totder{\mathcal{I}}{x} = &\parder{\mathcal{I}}{x} \totder{x}{x} + \parder{\mathcal{I}}{y} \totder{y}{x} + \parder{\mathcal{I}}{a} \totder{a}{x} + \parder{\mathcal{I}}{b} \totder{b}{x}.
\end{align*}
Due to the fact that integration is performed over the coordinate $z$, the linear operations of partial-$x$ differentiation and integration over $z$ may be safely exchanged (\cf \cref{thm_leibniz_rule}), such that
\begin{align*}
  \parder{\mathcal{I}}{x} &= \int_{a(x,y)}^{b(x,y)} \parder{f}{x}(x,y,z) \d{y}.
\end{align*}
Using the first fundamental theorem of calculus (\cf \cref{thm_first_fundamental_theorem_of_calculus}), if $F(x,y,z)$ is the indefinite integral of $f(x,y,z)$ with respect to $z$ over $z \in [a,b]$,
\begin{align*}
  \parder{\mathcal{I}}{a} &= \parder{}{a} \Big[ F(x,y,b)) - F(x,y,a) \Big] = - \parder{}{a} F(x,y,a) \\
  \parder{\mathcal{I}}{b} &= \parder{}{b} \Big[ F(x,y,b)) - F(x,y,a) \Big] = + \parder{}{b} F(x,y,b),
\end{align*}
and using the second fundamental theorem of calculus (\cf \cref{thm_second_fundamental_theorem_of_calculus}),
\begin{align*}
  \left. \parder{}{a} F(x,y,z) \right|_{z=a} &= \left( \parder{F}{z} \parder{z}{a} \right)_{z=a} = f(x,y,a(x,y)) \\
  \left. \parder{}{b} F(x,y,z) \right|_{z=b} &= \left( \parder{F}{z} \parder{z}{b} \right)_{z=b} = f(x,y,b(x,y)).
\end{align*}
Combining the above relations with the facts that $\diff_x y = 0$, $\diff_x b = \partial_x b$, $\diff_x a = \partial_x a$, and $\diff_x \mathcal{I} = \partial_x \mathcal{I}$ produces \cref{thm_leibniz_rule_corollary}.
\end{proof}

\begin{theorem}[Divergence theorem]
\label{thm_divergence_theorem}
The integral of the divergence of a differentiable-vector-field $\rankone{j}$ within an open and bounded volume $\Omega = \Omega(\rankone{x})$ is equal to the integral of the outward flux of the vector field across the surface of the volume $\partial \Omega = \Gamma(\rankone{x})$:
\begin{align*}
  \int_{\Omega} \nabla \cdot \rankone{j} \d{\Omega} = \int_{\Gamma} \rankone{j} \cdot \normal \d{\Gamma}.
\end{align*}
\end{theorem}

\begin{proof}
Consult chapter 15 section 7 of \cite{larson_2010}.
\end{proof}

\begin{theorem}[Reynolds transport theorem]
\label{thm_reynolds_transport}
Within an arbitrary time-evolving volume $\Omega = \Omega(\rankone{x},t) \in \R^3$ with boundary $\partial \Omega = \Gamma(\rankone{x},t)$, \emph{Leibniz's rule in three dimensions}---better known in continuum mechanics as \emph{Reynolds transport theorem} \cite{reynolds_1903}---states that for a given continuous quantity $\phi$,
\begin{align*}
  \totder{}{t} \int_{\Omega(t)} \phi \d{\Omega}(t) = \int_{\Omega(t)} \parder{\phi}{t} \d{\Omega(t)} + \int_{\Gamma(t)} \phi \rankone{w} \cdot \normal \d{\Gamma(t)},
\end{align*}
where $\rankone{w}$ is the velocity of surface $\Gamma$ and $\normal$ is the outward-facing unit-normal vector for $\Gamma$.
\end{theorem}

\begin{proof}
Consult chapter III section 4 of \cite{salencon_2001}.
\end{proof}

\begin{corollary}
\label{thm_constant_volume_reynolds}
If the volume defined in \cref{thm_reynolds_transport} remains constant---\ie, if $\Omega = \Omega(\rankone{x})$, $\partial \Omega = \Gamma(\rankone{x})$ and $\rankone{w} = \rankone{0}$---Reynolds-transport \cref{thm_reynolds_transport} reduces to
\begin{align*}
  \totder{}{t} \int_{\Omega} \phi \d{\Omega} = \int_{\Omega} \parder{\phi}{t} \d{\Omega}.
\end{align*}
\end{corollary}

\begin{proof}
Time-invariant volumes are characterized by $\rankone{w} = \rankone{0}$ on $\Gamma$ and therefore $\int_{\Gamma} \phi \rankone{w} \cdot \normal \d{\Gamma} = 0$; this expression applied to \cref{thm_reynolds_transport} yields \cref{thm_constant_volume_reynolds}.
\end{proof}

\begin{definition}[Fundamental conservation equation]
\label{def_fundamental_conservation_equation}
Consider an arbitrary fixed volume $\Omega^{\e} = \Omega^{\e}(\rankone{x}) \subset \Omega(\rankone{x}, t)$ with boundary $\partial \Omega^{\e} = \Gamma^{\e}(\rankone{x})$ and quantity $\phi = \phi(\rankone{x},t)$.
Then
\begin{align*}
  \begin{matrix}
    \text{the total} \\
    \text{rate of change} \\
    \text{of quantity $\phi$} \\
    \text{in $\Omega^{\e}$}
  \end{matrix} \hspace{2.5mm} = \hspace{2.5mm} 
  \begin{matrix}
    \text{the inward flux} \\
    \text{of $\phi$ across the} \\
    \text{boundary $\Gamma^{\e}$} \\
  \end{matrix} \hspace{2.5mm} + \hspace{2.5mm} 
  \begin{matrix}
    \text{the generation} \\
    \text{of quantity $\phi$} \\
    \text{within $\Omega^{\e}$}
  \end{matrix}, 
\end{align*}
or mathematically, the \emph{fundamental-conservation equation}
\begin{align*}
  \totder{}{t} \int_{\Omega^{e}} \phi \d{\Omega^{\e}} = - \int_{\Gamma^{\e}} \left( \phi \rankone{u} + \rankone{j} \right) \cdot \normal \d{\Gamma^{\e}} + \int_{\Omega^{\e}} \mathring{f} \d{\Omega^{\e}},
\end{align*}
with volumetric-source term $\mathring{f}$, material velocity $\rankone{u}$, advective flux $\phi \rankone{u}$, non-advective flux $\rankone{j}$, and outward-pointing unit-normal vector $\normal$.
\end{definition}

\begin{definition}[Differential continuity equation in conservative form]
\label{def_conservative_continuity_equation}
Provided that the quantity $\phi$ is differentiable, applying Reynolds transport \cref{thm_constant_volume_reynolds} to the time-derivative term and divergence \cref{thm_divergence_theorem} to the surface integral of \cref{def_fundamental_conservation_equation} results in
\begin{align*}
  \int_{\Omega^{\e}} \parder{\phi}{t} \d{\Omega^{\e}} + \int_{\Omega^{\e}} \nabla \cdot \left( \phi \rankone{u} + \rankone{j} \right) \d{\Omega^{\e}} &= \int_{\Omega^{\e}} \mathring{f} \d{\Omega^{\e}}.
\end{align*}
Therefore, due to the fact that the domain $\Omega^{\e} \subset \Omega$ was taken arbitrarily and by linearity of the integral operator, the \emph{continuity equation in conservative form} is given by
\begin{align*}
  \parder{\phi}{t} + \nabla \cdot \left( \phi \rankone{u} + \rankone{j} \right) = \mathring{f} && \text{in } \Omega.
\end{align*}
\end{definition}

\begin{definition}[Differential continuity equation in non-conservative form]
  \label{def_nonconservative_continuity_equation}
The advective-flux divergence term in \cref{def_conservative_continuity_equation} can be expanded using the differential-product rule\footnote{$\nabla \cdot \left( \phi \rankone{u} \right) = \rankone{u} \cdot \nabla \phi + \phi \nabla \cdot \rankone{u}$} to
\begin{align*}
  \parder{\phi}{t} + \rankone{u} \cdot \nabla \phi + \phi \nabla \cdot \rankone{u} + \nabla \cdot \rankone{j} = \mathring{f} && \text{in } \Omega.
\end{align*}
This can be then be reduced by applying chain rule \cref{thm_one_variable_chain_rule} to get the \emph{continuity equation in non-conservative form}
\begin{align*}
  \totder{\phi}{t} + \phi \nabla \cdot \rankone{u} + \nabla \cdot \rankone{j} = \mathring{f} && \text{in } \Omega.
\end{align*}
\end{definition}

Continuity equation \cref{def_conservative_continuity_equation} is so named for the fact that in its derivation the quantity $\phi$ had been required to not contain discontinuities within the domain $\Omega$; this is needed in order to apply the divergence theorem (\cf \cref{thm_divergence_theorem}) which is applicable for continuous material properties only.
In addition, while the integral statement of conservation used in its derivation (\cf \cref{def_fundamental_conservation_equation}) does not utilize the divergence theorem and thus does not require the material be continuous within $\Omega$, additional theory is required in order to state an analogous integral relation for discontinuous media.
Materials which contain discontinuous properties are characterized by \emph{singular surfaces}; these are the interfaces across which a jump in material properties exists.
The following operator will simplify future notation:

\begin{definition}[Jump discontinuity]
\label{def_jump}
Consider a volume $\Omega^- = \Omega^-(\rankone{x}) \in \R^3$ with boundary $\Gamma^- = \partial \Omega^-(\rankone{x})$ and outward-pointing unit-normal vector $\normal^-$ in contact with another volume $\Omega^+ = \Omega^+(\rankone{x}) \in \R^3$ with boundary $\Gamma^+ = \partial \Omega^+(\rankone{x})$ and outward-pointing unit-normal vector $\normal^+$ (\cf \cref{fig_domain}). 
The jump of a scalar quantity $\phi = \phi(\rankone{x})$ and vector quantity $\rankone{\phi} = \rankone{\phi}(\rankone{x})$ across the shared boundary $\Sigma = \Gamma^+ \cap \Gamma^-$ are respectively
\begin{align*}
  \jump{\phi} = \phi^- \normal^- + \phi^+ \normal^+
  \hspace{10mm}
  \text{and}
  \hspace{10mm}
  \jump{\rankone{\phi}} = \rankone{\phi}^- \cdot \normal^- + \rankone{\phi}^+ \cdot \normal^+,
\end{align*}
where the normal vectors satisfy $\normal^+ = -\normal^-$ for all $\rankone{x} \in \Sigma$.
\end{definition}

Hence a scalar material property $\phi$ is discontinuous across a surface $\Sigma$ when $\jump{\phi} \neq 0$.
Jump discontinuity \cref{def_jump} can be used to neatly describe the divergence theorem for discontinuous media, as follows:

\begin{theorem}[Discontinuous divergence]
\label{thm_discontinuous_divergence_theorem}
The integral of the divergence of a vector field $\rankone{j} = \rankone{j}(\rankone{x}, t)$ possessing a discontinuity along a surface $\Sigma$ within an open and bounded volume $\Omega = \Omega(\rankone{x}, t)$ with exterior surface $\Gamma$ is given by
\begin{align*}
  \int_{\Gamma} \rankone{j} \cdot \normal \d{\Gamma} = \int_{\Omega} \nabla \cdot \rankone{j} \d{\Omega} - \int_{\Sigma} \jump{\rankone{j}} \d{\Sigma},
\end{align*}
where the jump operator $\jump{\cdot}$ is given by \cref{def_jump}.
\end{theorem}

\begin{proof}
Applying the continuous divergence theorem (\cf \cref{thm_divergence_theorem}) to both regions and taking the sum yields
\begin{align*}
  \int_{\Omega} \left( \nabla \cdot \rankone{j}^- + \nabla \cdot \rankone{j}^+ \right) \d{\Omega} = 
   \int_{\Gamma} \left( \rankone{j}^- \cdot \normal^- + \rankone{j}^+ \cdot \normal^+ \right) \d{\Gamma} 
  + \int_{\Sigma} \left( \rankone{j}^- \cdot \normal^- + \rankone{j}^+ \cdot \normal^+ \right) \d{\Sigma}.
\end{align*}
Making use of the facts that $\rankone{j}^+ = \rankone{0}$ in $\overline{\Omega^-}$, and $\rankone{j}^- = \rankone{0}$ in $\overline{\Omega^+}$,
\begin{align*}
  \int_{\Omega} \nabla \cdot \rankone{j} \d{\Omega} = 
  \int_{\Gamma} \rankone{j} \cdot \normal \d{\Gamma}
  + \int_{\Sigma} \left( \rankone{j}^- \cdot \normal^- + \rankone{j}^+ \cdot \normal^+ \right) \d{\Sigma}.
\end{align*}
Finally, applying jump discontinuity \cref{def_jump} and rearranging terms results in \cref{thm_discontinuous_divergence_theorem}.
\end{proof}

Discontinuous divergence \cref{thm_discontinuous_divergence_theorem} may be used to relate the total rate of change of an integrated quantity which contains an evolving discontinuity surface via the following theorem:

\begin{theorem}[Generalized Reynolds transport]
\label{thm_generalized_reynolds_transport_theorem}
Within an arbitrary time-evolving volume $\Omega = \Omega(\rankone{x},t)$ with exterior surface $\Gamma$, a quantity $\phi = \phi(\rankone{x},t)$ discontinuous across a surface $\Sigma \in \Omega$ will obey
\begin{align*}
  \totder{}{t} \int_{\Omega} \phi \d{\Omega} = \int_{\Omega} \parder{\phi}{t} \d{\Omega} + \int_{\Gamma} \phi \rankone{w} \cdot \normal \d{\Gamma} + \int_{\Sigma} \jump{\phi \rankone{w}} \d{\Sigma},
\end{align*}
where $\rankone{w} = \rankone{w}(\rankone{x},t)$ is the propagation velocity of exterior surface $\Gamma$ and discontinuity surface $\Sigma$; vector $\normal$ is the outward-facing-unit normal; and the jump operator $\jump{\cdot}$ is given by \cref{def_jump}.
\end{theorem}

\begin{proof}
Applying the continuous form of Reynolds transport theorem (\cf \cref{thm_reynolds_transport}) to both regions and taking the sum results in
\begin{align*}
  \totder{}{t} \int_{\Omega} \left( \phi^- + \phi^+ \right) \d{\Omega} = &+ \int_{\Omega} \left( \parder{\phi^-}{t} + \parder{\phi^+}{t} \right) \d{\Omega} \\
  &+ \int_{\Gamma} \left( \phi^- \rankone{w}^- \cdot \normal^-  + \phi^+ \rankone{w}^+ \cdot \normal^+ \right) \d{\Gamma} \\
  &+ \int_{\Sigma} \left( \phi^- \rankone{w}^- \cdot \normal^-  + \phi^+ \rankone{w}^+ \cdot \normal^+ \right) \d{\Sigma}.
\end{align*}
Using the facts that $\phi^+ = 0$ in $\overline{\Omega^-}$ and $\phi^- = 0$ in $\overline{\Omega^+}$,
\begin{align*}
  \totder{}{t} \int_{\Omega} \phi \d{\Omega} &= \int_{\Omega} \parder{\phi}{t} \d{\Omega} + \int_{\Gamma} \phi \rankone{w} \cdot \normal \d{\Gamma}
  + \int_{\Sigma} \left( \phi^- \rankone{w}^- \cdot \normal^-  + \phi^+ \rankone{w}^+ \cdot \normal^+ \right) \d{\Sigma}.
\end{align*}
Finally, applying jump discontinuity \cref{def_jump} with $\rankone{\phi} = \phi \rankone{w}$ results in \cref{thm_generalized_reynolds_transport_theorem}.
\end{proof}

\begin{corollary}
\label{thm_constant_volume_generalized_reynolds}
If the outer surface of the volume defined in \cref{thm_generalized_reynolds_transport_theorem} remains constant---\ie, if $\Omega = \Omega(\rankone{x})$ and $\Gamma = \partial \Omega(\rankone{x})$---generalized Reynolds transport \cref{thm_generalized_reynolds_transport_theorem} reduces to
\begin{align*}
  \totder{}{t} \int_{\Omega} \phi \d{\Omega} = \int_{\Omega} \parder{\phi}{t} \d{\Omega} + \int_{\Sigma} \jump{\phi \rankone{w}} \d{\Sigma}.
\end{align*}
\end{corollary}

\begin{proof}
A time-invariant volume $\Omega$ implies that $\rankone{w} = \rankone{0}$ on $\Gamma$ and therefore $\int_\Gamma \phi \rankone{w} \cdot \normal \d{\Gamma} = 0$, which substituted into \cref{thm_generalized_reynolds_transport_theorem} produces the result.
\end{proof}

A surface $\Gamma$ may have a surface velocity $\rankone{w}$ with non-zero magnitude and also satisfy the conditions of \cref{thm_constant_volume_generalized_reynolds}; this will happen, for example, if the surface velocity $\rankone{w}$ is everywhere tangential to $\Gamma$---meaning $\rankone{w}_{\perp} = \left(\rankone{w} \cdot \normal \right) \normal = \rankone{0}$---such as a rigid ball rolling down an inclined plane.
In this case, all points on the surface $\Gamma$ will move relative to a stationary observer with identical velocity, while the volume of $\Omega$ remains constant.
Ice sheets and glaciers may also have a non-zero-surface velocity with constant volume, but in this case volume equilibrium will depend on accumulation and ablation along the ice-sheet exterior.
In order to quantify this relation, a generalization of the integral conservation equation (\cf \cref{def_fundamental_conservation_equation}) for discontinuous media must be formulated, as follows:

\begin{definition}[Generalized conservation equation]
\label{def_generalized_conservation_equation}
Consider an arbitrarily-fixed volume $\Omega^{\e}(\rankone{x}) \subset \Omega(\rankone{x}, t)$ with boundary $\Gamma^{\e} = \partial \Omega^{\e}(\rankone{x})$ and discontinuous field $\phi = \phi(\rankone{x},t)$ across a surface $\Sigma \in \Omega$.
Then
\begin{align*}
  \begin{matrix}
    \text{the total} \\
    \text{rate of change} \\
    \text{of quantity $\phi$} \\
    \text{within $\Omega^{\e}$}
  \end{matrix} \hspace{2.5mm} = \hspace{2.5mm} 
  \begin{matrix}
    \text{the inward flux} \\
    \text{of $\phi$ across the} \\
    \text{boundary $\Gamma^{\e}$} \\
  \end{matrix} \hspace{2.5mm} + \hspace{2.5mm} 
  \begin{matrix}
    \text{the generation} \\
    \text{or depletion} \\
    \text{of quantity $\phi$} \\
    \text{within $\Omega^{\e}$}
  \end{matrix} \hspace{2.5mm} + \hspace{2.5mm} 
  \begin{matrix}
    \text{fluctuations} \\
    \text{of quantity $\phi$} \\
    \text{on $\Sigma^{\e}$.}
  \end{matrix}
\end{align*}
Mathematically, this is the \emph{generalized-conservation equation}
\begin{align*}
  \totder{}{t} \int_{\Omega^{e}} \phi \d{\Omega^{\e}} = - \int_{\Gamma^{\e}} \left( \phi \rankone{u} + \rankone{j} \right) \cdot \normal \d{\Gamma^{\e}} + \int_{\Omega^{\e}} \mathring{f}_{\Omega} \d{\Omega^{\e}} + \int_{\Sigma^{\e}} \mathring{f}_{\Sigma} \d{\Sigma^{\e}},
\end{align*}
with volumetric-source term $\mathring{f}_{\Omega}$, surface-source term $\mathring{f}_{\Sigma}$, material velocity $\rankone{u}$, advective flux $\phi \rankone{u}$, non-advective flux $\rankone{j}$, and outward-pointing unit-normal vector $\normal$.
\end{definition}

\begin{remark}
\label{rmk_surface_velocity_difference}
The velocity $\rankone{u}$ of \cref{def_generalized_conservation_equation} is the velocity of a material flowing within or through $\Omega^{\e}$, while the velocity $\rankone{w}$ of \cref{thm_generalized_reynolds_transport_theorem} is the velocity of the outer material surface $\Gamma$ or discontinuity surface $\Sigma$.
In the event that the quantity of interest $\phi$ is the material density $\rho$, regions where the surface $\Gamma^{\e}$ of \cref{def_generalized_conservation_equation} intersects the outer surface $\Gamma$ will have $\rankone{u} = \rankone{w}$ in the event that mass is not gained or lost along $\Gamma$.
However, if there is an additional fluctuation of mass unrelated to the fluid velocity $\rankone{u}$ along $\Gamma \cap \Gamma^{\e}$, there will clearly be a vector with non-zero magnitude defined by the difference $\rankone{r} = \rankone{w} - \rankone{u}$ with $\Vert \rankone{r} \Vert \neq 0$ on $\Gamma \cap \Gamma^{\e}$.
\end{remark}

The addition of the source term $\mathring{f}_{\Sigma}$ to the right-hand side of \cref{def_generalized_conservation_equation} accounts for the additional fluctuations in quantity $\phi$ due to interaction between the materials on either side of $\Sigma^{\e}$.
In the context of ice sheets, this term represents interactions between the ice sheet and its environment or discontinuities within the ice interior.
Interior discontinuity surfaces exist at glacier flow margins, where fractures have occurred due to extreme levels of stress, crevasses\footnote{Crevasses are defined as interior regions of the ice containing non-ice material; as such, these areas may be designated as exterior discontinuities.}, \etc.
However, the transition from compacted snow---referred to as firn---to solid ice is gradual such that both $\jump{\rho} = \jump{\rankone{u}} = 0$, and is therefore not discontinuous. 

The transition between cold ice---which is entirely frozen---and temperate ice containing water has been considered discontinuous, such that $\jump{\rho} \neq 0$ and $\jump{\rankone{u}} = 0$ (\cf \cite{greve_2009}).
This assumption implies that cold ice may present an impenetrable boundary to water flowing from the temperate ice mixture.
Unlike the clearly-defined ice and ice-free boundary, it is natural to assume that the transition between cold and temperate ice is gradual; for example, micro cavities or imperfections in the ice represent a medium from which water may flow.
Furthermore, water located between ice-grain boundaries will freeze from the grain surface into the center of the water cavity; on average, the density of this ice-water mixture will be characterized by a smooth density gradient, and therefore one may consider the possibility that $\jump{\rho} = 0$ at this surface.\footnote{Jump operator \cref{def_jump} represents the \emph{limiting} values of a quantity near a surface.}
Regardless, the following theorem is used to define the boundary conditions for any quantity from either interior or exterior domain at the material interface $\Sigma$:

\begin{theorem}[Discontinuity equation]
\label{thm_discontinuity_equation}
A field $\phi = \phi(\rankone{x},t)$ defined within an arbitrarily-fixed volume $\Omega^{\e}(\rankone{x}) \subset \Omega(\rankone{x}, t)$ with boundary $\Gamma^{\e} = \partial \Omega^{\e}(\rankone{x})$ which is differentiable everywhere except possibly across $\Sigma^{\e} \in \Omega^{\e}$ will satisfy the \emph{discontinuity equation}
\begin{align*}
  \jump{\phi \left( \rankone{u} - \rankone{w} \right) + \rankone{j}} = - \mathring{f}_{\Sigma} && \text{on } \Sigma^{\e},
\end{align*}
with material velocity $\rankone{u}$, $\Sigma^{\e}$-surface velocity $\rankone{w}$, non-advective flux $\rankone{j}$, source of $\phi$ on $\Sigma^{\e}$ denoted $\mathring{f}_{\Sigma}$ which is positive for increasing $\phi$ and negative for decreasing $\phi$, and jump operator $\jump{\cdot}$ given by \cref{def_jump}.
\end{theorem}

\begin{proof}
Decompose the volume $\Omega^{\e}$ along the discontinuity surface $\Sigma^{\e}$ such that $\Omega^{\e} = \Omega^{\e^-} \cup\ \Omega^{\e^+}$, $\Omega^{\e^-} \cap\ \Omega^{\e^+} = \emptyset$ and $\Sigma^{\e} = \partial\Omega^{\e^-} \cap\ \partial\Omega^{\e^+}$ (\cref{fig_domain}).
Applying generalized Reynolds transport \cref{thm_constant_volume_generalized_reynolds} to the left-hand side of generalized conservation equation \cref{def_generalized_conservation_equation} yields
\begin{align*}
  \int_{\Omega^{\e}} \parder{\phi}{t} \d{\Omega^{\e}} + \int_{\Sigma^{\e}} \jump{\phi \rankone{w}} \d{\Sigma^{\e}} = - \int_{\Gamma^{\e}} \left( \phi \rankone{u} + \rankone{j} \right) \cdot \normal \d{\Gamma^{\e}} 
  + \int_{\Omega^{\e}} \mathring{f}_{\Omega} \d{\Omega^{\e}} + \int_{\Sigma^{\e}} \mathring{f}_{\Sigma} \d{\Sigma^{\e}}.
\end{align*}
Applying discontinuous divergence \cref{thm_discontinuous_divergence_theorem} to the surface integral on the right-hand side and rearranging terms results in the \emph{integral discontinuity equation}
\begin{align*}
  \int_{\Omega^{\e}} \left[ \parder{\phi}{t} + \nabla \cdot \left( \phi \rankone{u} + \rankone{j} \right) - \mathring{f}_{\Omega} \right] \d{\Omega^{\e}} = \int_{\Sigma^{\e}} \left( \jump{\phi \left( \rankone{u} - \rankone{w} \right) + \rankone{j}} + \mathring{f}_{\Sigma} \right) \mathrm{d}{\Sigma^{\e}}.
\end{align*}
Finally, recall that continuity equation \cref{def_conservative_continuity_equation} demands that the left-hand side of this relation be zero over regions $\Omega^{\e^{\pm}}$; coupling this with the fact that the integration domain $\Sigma^{\e}$ was arbitrarily chosen, the integrand of the right-hand side is zero, thus implying \cref{thm_discontinuity_equation}.\footnote{This result can also be obtained by taking the limit of the integral-discontinuity equation as the volume $V = |\Omega^{\e}| \rightarrow 0$.}
\end{proof}

\begin{remark}
Clearly, \cref{thm_discontinuity_equation} also applies for material properties $\phi$ that are continuous across $\Sigma^{\e}$; such properties satisfy \cref{thm_discontinuity_equation} with $\mathring{f}_{\Sigma} = 0$.
Note that \cref{thm_discontinuity_equation} provide the boundary conditions for $\phi^{\pm}$ stated as functions of the material properties from the ($\mp$) perspective---that is, $\rankone{u}^{\mp}$, $\rankone{w}^{\mp}$, and $\rankone{j}^{\mp}$---and the source term $\mathring{f}_{\Sigma}$, which may itself depend on any property from either ($\pm$) perspective.
\end{remark}

All of the definitions and theorems presented in this section have been constructed from pure mathematical reasoning---referred to as first-principle derivations---without imposing assumptions, simplifications, or empirical relationships; hence these relations are \emph{exact} and represent the ideal starting point for an analysis of conservation properties of any given media, including ice sheets.

\section{Analytic solution in $\R^3$}
\label{sec_sup_r3_analytic_solution}

This section provides all relevant calculations associated with \cref{sec_r3_analytic_solution}.

\subsection{Linear hyperbolic equation \cref{analytic_y_velocity_problem}}
\label{sec_sup_linear_pde}

The final $y$ component of velocity $u_y^{\ana}$ must satisfy incompressibility-relation \cref{analytic_incompressible_conservation_of_mass}.
This section derives Equation \cref{du_z_dz,du_x_dx,analytic_y_velocity_problem} in detail.

To begin, the following derivatives will be required: 
\begin{align}
  \label{dHdx}
  \parder{H}{x}          &= \parder{S}{x} - \parder{B}{x} \\
  \label{dHdt}
  \parder{H}{t}          &= \parder{S}{t} - \parder{B}{t} \\
  \label{d_xi_b_dx}
  \parder{\xi_{\bed}}{x} &= \frac{1}{H^2} \left( \parder{S}{x} H - \parder{H}{x} \left(S - z \right) \right) \\
  \label{d_xi_b_dz}
  \parder{\xi_{\bed}}{z} &= - \frac{1}{H} \\
  \label{d_xi_s_dz}
  \parder{\xi_{\srf}}{z} &= + \frac{1}{H} \\
  \label{dwsdz}
  \parder{u_{z \srf}^{\ana}}{z} &= \parder{u_x^{\ana}}{z} \parder{S}{x} + \parder{u_y^{\ana}}{z} \parder{S}{y} \\
  \label{dwbdz}
  \parder{u_{z \bed}^{\ana}}{z} &= \parder{u_x^{\ana}}{z} \parder{B}{x} + \parder{u_y^{\ana}}{z} \parder{B}{y}.
\end{align}

The derivative of analytic-$z$ velocity \cref{analytic_z_velocity} with respect to $z$ is
\begin{align*}
  \parder{u_z^{\ana}}{z} = &\parder{\xi_{\srf}}{z} u_{z \srf}^{\ana} + \xi_{\srf} \parder{u_{z \srf}^{\ana}}{z} + \parder{\xi_{\bed}}{z} u_{z \bed}^{\ana} + \xi_{\bed} \parder{u_{z \bed}^{\ana}}{z},
\end{align*}
and applying \cref{d_xi_b_dz,d_xi_s_dz},
\begin{align*}
  \parder{u_z^{\ana}}{z} = &\frac{u_{z \srf}^{\ana}}{H} + \xi_{\srf} \parder{u_{z \srf}^{\ana}}{z} - \frac{u_{z \bed}^{\ana}}{H} + \xi_{\bed} \parder{u_{z \bed}^{\ana}}{z}.
\end{align*}
Expanding this relation using \cref{analytic_z_velocity_S,analytic_z_velocity_B,dwsdz,dwbdz} produces
\begin{align*}
  \parder{u_z^{\ana}}{z} =&+ \frac{1}{H} \left( -\Vert \hat{\rankone{k}} - \nabla S \Vert \smb + \parder{S}{t} + u_x^{\ana} \parder{S}{x} + u_y^{\ana} \parder{S}{y} \right) \\
  &- \frac{1}{H} \left( \Vert \nabla B - \hat{\rankone{k}} \Vert \bmb + \parder{B}{t} + u_x^{\ana} \parder{B}{x} + u_y^{\ana} \parder{B}{y} \right) \\
  &+ \xi_{\srf} \left( \parder{u_x^{\ana}}{z} \parder{S}{x} + \parder{u_y^{\ana}}{z} \parder{S}{y} \right) + \xi_{\bed} \left( \parder{u_x^{\ana}}{z} \parder{B}{x} + \parder{u_y^{\ana}}{z} \parder{B}{y} \right),
\end{align*}
and applying \cref{dHdx,dHdt},
\begin{align*}
  \parder{u_z^{\ana}}{z} =
  &+ \frac{1}{H} \left( -\Vert \hat{\rankone{k}} - \nabla S \Vert \smb - \Vert \nabla B - \hat{\rankone{k}} \Vert \bmb + \parder{H}{t} + u_x^{\ana} \parder{H}{x} + u_y^{\ana} \parder{H}{y} \right) \\
  &+ \xi_{\srf} \left( \parder{u_x^{\ana}}{z} \parder{S}{x} + \parder{u_y^{\ana}}{z} \parder{S}{y} \right) + \xi_{\bed} \left( \parder{u_x^{\ana}}{z} \parder{B}{x} + \parder{u_y^{\ana}}{z} \parder{B}{y} \right),
\end{align*}
The derivative of analytic-$x$ velocity $u_x^{\ana}$ defined by \cref{analytic_x_velocity} with respect to $z$ is calculated with the help of \cref{d_xi_b_dz}
\begin{align*}
  \parder{u_x^{\ana}}{z}
  =&+ \parder{}{z} \left( \left( u_{x \srf}^{\ana} - u_{x \bed}^{\ana} \right) \left( \cancel{1} - \xi_{\bed}^{\lambda} \right) \right) + \cancel{\parder{u_{x \bed}^{\ana}}{z}} \\
  =&- \left( u_{x \srf}^{\ana} - u_{x \bed}^{\ana} \right) \lambda \xi_{\bed}^{\lambda - 1} \parder{\xi_{\bed}}{z} \\
  =&+ \left( u_{x \srf}^{\ana} - u_{x \bed}^{\ana} \right) \lambda \xi_{\bed}^{\lambda - 1} \frac{1}{H}
\end{align*}
which substituting into the above yields
\begin{align*}
  \parder{u_z^{\ana}}{z} =
  &+ \frac{1}{H} \left( -\Vert \hat{\rankone{k}} - \nabla S \Vert \smb - \Vert \nabla B - \hat{\rankone{k}} \Vert \bmb + \parder{H}{t} + u_x^{\ana} \parder{H}{x} + u_y^{\ana} \parder{H}{y} \right) \\
  &+ \left( u_{x \srf}^{\ana} - u_{x \bed}^{\ana} \right) \lambda \xi_{\srf} \xi_{\bed}^{\lambda - 1} \frac{1}{H} \parder{S}{x} + \xi_{\srf} \parder{u_y^{\ana}}{z} \parder{S}{y} \\
  &+ \left( u_{x \srf}^{\ana} - u_{x \bed}^{\ana} \right) \lambda \xi_{\bed}^{\lambda} \frac{1}{H} \parder{B}{x} + \xi_{\bed} \parder{u_y^{\ana}}{z} \parder{B}{y},
\end{align*}
which on rearrangement yields
\begin{align*}
  \parder{u_z^{\ana}}{z} =
  &+ \frac{1}{H} \left( -\Vert \hat{\rankone{k}} - \nabla S \Vert \smb - \Vert \nabla B - \hat{\rankone{k}} \Vert \bmb + \parder{H}{t} + u_x^{\ana} \parder{H}{x} \right) \\
  &+ \left( u_{x \srf}^{\ana} - u_{x \bed}^{\ana} \right) \lambda \xi_{\bed}^{\lambda} \frac{1}{H} \left( \frac{\xi_{\srf}}{\xi_{\bed}} \parder{S}{x} + \parder{B}{x} \right) \\
  &+ \left( \xi_{\srf} \parder{S}{y} + \xi_{\bed} \parder{B}{y} \right) \parder{u_y^{\ana}}{z} + \frac{1}{H} \parder{H}{y} u_y^{\ana};
\end{align*}
Equation \cref{du_z_dz} has thus been derived.

Differentiation of analytic-$x$ velocity \cref{analytic_x_velocity} with respect to $x$,
\begin{align*}
  \label{d_ux_dx_one}
  \parder{u_x^{\ana}}{x} 
  =& + \parder{}{x} \left( \left( u_{x \srf}^{\ana} - u_{x \bed}^{\ana} \right) \left( 1 - \xi_{\bed}^{\lambda} \right) \right) + \parder{u_{x \bed}^{\ana}}{x} \\
  =&+ \left( \parder{u_{x \srf}^{\ana}}{x} - \parder{u_{x \bed}^{\ana}}{x} \right) \left( 1 - \xi_{\bed}^{\lambda} \right) - \left( u_{x \srf}^{\ana} - u_{x \bed}^{\ana} \right) \parder{\xi_{\bed}^{\lambda}}{x} + \parder{u_{x \bed}^{\ana}}{x} \\
  =&+ \parder{u_{x \srf}^{\ana}}{x} - \cancel{\parder{u_{x \bed}^{\ana}}{x}} - \left( \parder{u_{x \srf}^{\ana}}{x} - \parder{u_{x \bed}^{\ana}}{x} \right) \xi_{\bed}^{\lambda} - \left( u_{x \srf}^{\ana} - u_{x \bed}^{\ana} \right) \lambda \xi_{\bed}^{\lambda - 1} \parder{\xi_{\bed}}{x} + \cancel{\parder{u_{x \bed}^{\ana}}{x}} \\
  =&+ \parder{u_{x \srf}^{\ana}}{x} - \left( \parder{u_{x \srf}^{\ana}}{x} - \parder{u_{x \bed}^{\ana}}{x} \right) \xi_{\bed}^{\lambda} - \left( u_{x \srf}^{\ana} - u_{x \bed}^{\ana} \right) \lambda \xi_{\bed}^{\lambda - 1} \parder{\xi_{\bed}}{x} \\
  =&+ \left( 1 - \xi_{\bed}^{\lambda} \right) \parder{u_{x \srf}^{\ana}}{x} + \xi_{\bed}^{\lambda} \parder{u_{x \bed}^{\ana}}{x} - \left( u_{x \srf}^{\ana} - u_{x \bed}^{\ana} \right) \lambda \xi_{\bed}^{\lambda - 1} \parder{\xi_{\bed}}{x};
\end{align*}
Equation \cref{du_x_dx} has thus been derived.

Therefore, combining mass-conservation relation \cref{analytic_incompressible_conservation_of_mass} with velocity derivatives \cref{du_z_dz,du_x_dx} produces
\begin{align*}
  0 
  =&+ \parder{u_x^{\ana}}{x} + \parder{u_z^{\ana}}{z} + \parder{u_y^{\ana}}{y} \\
  =&+ \left( 1 - \xi_{\bed}^{\lambda} \right) \parder{u_{x \srf}^{\ana}}{x} + \xi_{\bed}^{\lambda} \parder{u_{x \bed}^{\ana}}{x} - \left( u_{x \srf}^{\ana} - u_{x \bed}^{\ana} \right) \lambda \xi_{\bed}^{\lambda - 1} \parder{\xi_{\bed}}{x} \\
  &+ \frac{1}{H} \left( -\Vert \hat{\rankone{k}} - \nabla S \Vert \smb - \Vert \nabla B - \hat{\rankone{k}} \Vert \bmb + \parder{H}{t} + u_x^{\ana} \parder{H}{x} \right) \\
  &+ \left( u_{x \srf}^{\ana} - u_{x \bed}^{\ana} \right) \lambda \xi_{\bed}^{\lambda} \frac{1}{H} \left( \frac{\xi_{\srf}}{\xi_{\bed}} \parder{S}{x} + \parder{B}{x} \right) \\
  &+ \left( \xi_{\srf} \parder{S}{y} + \xi_{\bed} \parder{B}{y} \right) \parder{u_y^{\ana}}{z} + \frac{1}{H} \parder{H}{y} u_y^{\ana} + \parder{u_y^{\ana}}{y},
\end{align*}
and using derivative \cref{d_xi_b_dx} in place of $\partial_x \xi_{\bed}$,
\begin{align*}
  0 
  =&+ \left( 1 - \xi_{\bed}^{\lambda} \right) \parder{u_{x \srf}^{\ana}}{x} + \xi_{\bed}^{\lambda} \parder{u_{x \bed}^{\ana}}{x} \\
  &+ \frac{1}{H} \left( -\Vert \hat{\rankone{k}} - \nabla S \Vert \smb - \Vert \nabla B - \hat{\rankone{k}} \Vert \bmb + \parder{H}{t} + u_x^{\ana} \parder{H}{x} \right) \\
  &+ \left( u_{x \srf}^{\ana} - u_{x \bed}^{\ana} \right) \lambda \xi_{\bed}^{\lambda} \frac{1}{H} \left( \frac{\xi_{\srf}}{\xi_{\bed}} \parder{S}{x} + \parder{B}{x} - \frac{H}{\xi_{\bed}} \parder{\xi_{\bed}}{x} \right) \\
  &+ \left( \xi_{\srf} \parder{S}{y} + \xi_{\bed} \parder{B}{y} \right) \parder{u_y^{\ana}}{z} + \frac{1}{H} \parder{H}{y} u_y^{\ana} + \parder{u_y^{\ana}}{y}.
\end{align*}
Isolating this relation with respect to $u_y^{\ana}$ yields
\begin{align*}
  A + G u_y^{\ana} + \parder{u_y^{\ana}}{y} + C \parder{u_y^{\ana}}{z} = 0,
\end{align*}
where
\begin{align}
  \label{a_one}
  A(\rankone{x},t)
  =&+ \left( 1 - \xi_{\bed}^{\lambda} \right) \parder{u_{x \srf}^{\ana}}{x} + \xi_{\bed}^{\lambda} \parder{u_{x \bed}^{\ana}}{x} \notag \\
  &+ \frac{1}{H} \left( -\Vert \hat{\rankone{k}} - \nabla S \Vert \smb - \Vert \nabla B - \hat{\rankone{k}} \Vert \bmb + \parder{H}{t} + u_x^{\ana} \parder{H}{x} \right) \notag \\
  &+ \left( u_{x \srf}^{\ana} - u_{x \bed}^{\ana} \right) \lambda \xi_{\bed}^{\lambda} \frac{1}{H} \left( \frac{\xi_{\srf}}{\xi_{\bed}} \parder{S}{x} + \parder{B}{x} - \frac{H}{\xi_{\bed}} \parder{\xi_{\bed}}{x} \right) \\
  \label{g_one}
  G(x,y,t)
  =& + \frac{1}{H} \parder{H}{y} \\
  \label{c_one}
  C(\rankone{x},t)
  =&+ \xi_{\srf} \parder{S}{y} + \xi_{\bed} \parder{B}{y}.
\end{align}
The coefficient $A$ may be decomposed such that $A(\rankone{x},t) = A_1(x,y,t) + A_2(\rankone{x},t)$ with
\begin{align}
  \label{a_1_one}
  A_1(x,y,t)
  =&+ \parder{u_{x \srf}^{\ana}}{x} + \frac{1}{H} \left( \parder{H}{t} - \Vert \hat{\rankone{k}} - \nabla S \Vert \smb - \Vert \nabla B - \hat{\rankone{k}} \Vert \bmb \right)
\end{align}
and
\begin{align}
  \label{a_2_one}
  A_2(\rankone{x},t)
  =&+ \xi_{\bed}^{\lambda} \left( \parder{u_{x \bed}^{\ana}}{x} - \parder{u_{x \srf}^{\ana}}{x} \right) + u_x^{\ana} \frac{1}{H} \parder{H}{x} \notag \\
  &+ \left( u_{x \srf}^{\ana} - u_{x \bed}^{\ana} \right) \lambda \xi_{\bed}^{\lambda} \frac{1}{H} \left( \frac{\xi_{\srf}}{\xi_{\bed}} \parder{S}{x} + \parder{B}{x} - \frac{H}{\xi_{\bed}} \parder{\xi_{\bed}}{x} \right).
\end{align}
Applying \cref{analytic_x_velocity,relative_z,d_xi_b_dx} produces
\begin{align*}
  A_2(\rankone{x},t)
  =&+ \left( \frac{S - z}{H} \right)^{\lambda} \left( \parder{u_{x \bed}^{\ana}}{x} - \parder{u_{x \srf}^{\ana}}{x} \right) + \left( \left( u_{x \srf}^{\ana} - u_{x \bed}^{\ana} \right) \left( 1 - \left( \frac{S - z}{H} \right)^{\lambda} \right) + u_{x \bed}^{\ana} \right) \frac{1}{H} \parder{H}{x} \\
  &+ \left( u_{x \srf}^{\ana} - u_{x \bed}^{\ana} \right) \lambda \left( \frac{S - z}{H} \right)^{\lambda} \frac{1}{H} \left( \left( \frac{z - B}{S - z} \right) \parder{S}{x} + \parder{B}{x} - \left( \frac{H^2}{S - z} \right) \frac{1}{H^2} \left( \parder{S}{x} H - \parder{H}{x} \left(S - z \right) \right) \right).
\end{align*}
Simplifying once,
\begin{align*}
  A_2(\rankone{x},t)
  =&+ \left( \frac{S - z}{H} \right)^{\lambda} \left( \parder{u_{x \bed}^{\ana}}{x} - \parder{u_{x \srf}^{\ana}}{x} \right) \\
   &+ \left( \left( u_{x \srf}^{\ana} - u_{x \bed}^{\ana} \right) \left( 1 - \left( \frac{S - z}{H} \right)^{\lambda} \right) + u_{x \bed}^{\ana} \right) \frac{1}{H} \parder{H}{x} \\
  &+ \left( u_{x \srf}^{\ana} - u_{x \bed}^{\ana} \right) \lambda \left( \frac{S - z}{H} \right)^{\lambda} \frac{1}{H} \left( \frac{z - B}{S - z} \right) \parder{S}{x} \\
  &+ \left( u_{x \srf}^{\ana} - u_{x \bed}^{\ana} \right) \lambda \left( \frac{S - z}{H} \right)^{\lambda} \frac{1}{H} \parder{B}{x} \\
  &- \left( u_{x \srf}^{\ana} - u_{x \bed}^{\ana} \right) \lambda \left( \frac{S - z}{H} \right)^{\lambda} \frac{1}{H} \left( \frac{1}{S - z} \right) \left( \parder{S}{x} H - \parder{H}{x} \left(S - z \right) \right),
\end{align*}
and again
\begin{align*}
  A_2(\rankone{x},t)
  =&+ \left( \frac{S - z}{H} \right)^{\lambda} \left( \parder{u_{x \bed}^{\ana}}{x} - \parder{u_{x \srf}^{\ana}}{x} \right) \\
   &+ \left( u_{x \srf}^{\ana} - u_{x \bed}^{\ana} \right) \left( 1 - \left( \frac{S - z}{H} \right)^{\lambda} \right) \frac{1}{H} \parder{H}{x} \\
   &+ u_{x \bed}^{\ana} \frac{1}{H} \parder{H}{x} \\
  &+ \left( u_{x \srf}^{\ana} - u_{x \bed}^{\ana} \right) \lambda \left( \frac{S - z}{H} \right)^{\lambda} \frac{1}{H} \left( \frac{z - B}{S - z} \right) \parder{S}{x} \\
  &+ \left( u_{x \srf}^{\ana} - u_{x \bed}^{\ana} \right) \lambda \left( \frac{S - z}{H} \right)^{\lambda} \frac{1}{H} \parder{B}{x} \\
  &- \left( u_{x \srf}^{\ana} - u_{x \bed}^{\ana} \right) \lambda \left( \frac{S - z}{H} \right)^{\lambda} \left( \frac{1}{S - z} \right) \parder{S}{x} \\
  &+ \left( u_{x \srf}^{\ana} - u_{x \bed}^{\ana} \right) \lambda \left( \frac{S - z}{H} \right)^{\lambda} \frac{1}{H} \parder{H}{x},
\end{align*}
and again
\begin{align*}
  A_2(\rankone{x},t)
  =&+ \left( \frac{S - z}{H} \right)^{\lambda} \left( \parder{u_{x \bed}^{\ana}}{x} - \parder{u_{x \srf}^{\ana}}{x} \right) \\
   &+ \left( u_{x \srf}^{\ana} - \cancel{u_{x \bed}^{\ana}} \right) \frac{1}{H} \parder{H}{x} \\
   &- \left( u_{x \srf}^{\ana} - u_{x \bed}^{\ana} \right) \left( \frac{S - z}{H} \right)^{\lambda} \frac{1}{H} \parder{H}{x} \\
   &+ \cancel{u_{x \bed}^{\ana} \frac{1}{H} \parder{H}{x}} \\
  &+ \left( u_{x \srf}^{\ana} - u_{x \bed}^{\ana} \right) \lambda \left( \frac{S - z}{H} \right)^{\lambda} \frac{1}{H} \left( \frac{z - B}{S - z} \right) \parder{S}{x} \\
  &+ \left( u_{x \srf}^{\ana} - u_{x \bed}^{\ana} \right) \lambda \left( \frac{S - z}{H} \right)^{\lambda} \frac{1}{H} \parder{B}{x} \\
  &- \left( u_{x \srf}^{\ana} - u_{x \bed}^{\ana} \right) \lambda \left( \frac{S - z}{H} \right)^{\lambda} \left( \frac{1}{S - z} \right) \parder{S}{x} \\
  &+ \left( u_{x \srf}^{\ana} - u_{x \bed}^{\ana} \right) \lambda \left( \frac{S - z}{H} \right)^{\lambda} \frac{1}{H} \parder{H}{x},
\end{align*}
and again
\begin{align*}
  A_2(\rankone{x},t)
  =&+ \left( \frac{S - z}{H} \right)^{\lambda} \left( \parder{u_{x \bed}^{\ana}}{x} - \parder{u_{x \srf}^{\ana}}{x} \right) \\
   &+ u_{x \srf}^{\ana} \frac{1}{H} \parder{H}{x} \\
   &- \left( u_{x \srf}^{\ana} - u_{x \bed}^{\ana} \right) \lambda \left( \frac{S - z}{H} \right)^{\lambda} \frac{1}{\lambda} \frac{1}{H} \parder{H}{x}  \\
  &+ \left( u_{x \srf}^{\ana} - u_{x \bed}^{\ana} \right) \lambda \left( \frac{S - z}{H} \right)^{\lambda} \frac{1}{H} \left( \frac{z - B}{S - z} \right) \parder{S}{x} \\
  &+ \left( u_{x \srf}^{\ana} - u_{x \bed}^{\ana} \right) \lambda \left( \frac{S - z}{H} \right)^{\lambda} \frac{1}{H} \parder{B}{x} \\
  &- \left( u_{x \srf}^{\ana} - u_{x \bed}^{\ana} \right) \lambda \left( \frac{S - z}{H} \right)^{\lambda} \left( \frac{1}{S - z} \right) \parder{S}{x} \\
  &+ \left( u_{x \srf}^{\ana} - u_{x \bed}^{\ana} \right) \lambda \left( \frac{S - z}{H} \right)^{\lambda} \frac{1}{H} \parder{H}{x}.
\end{align*}
Isolating the terms not dependent on $z$ within $A_1$ and substituting $\zeta = \left( u_{x \srf}^{\ana} - u_{x \bed}^{\ana} \right) \lambda \left( \frac{S - z}{H} \right)^{\lambda}$, the elements of coefficient $A$ given by \cref{a_1_one,a_2_one} are now
\begin{align}
  \label{a_1_two}
  A_1(x,y,t)
  =&+ \parder{u_{x \srf}^{\ana}}{x} + \frac{1}{H} \left( \parder{H}{t} - \Vert \hat{\rankone{k}} - \nabla S \Vert \smb - \Vert \nabla B - \hat{\rankone{k}} \Vert \bmb \right) + u_{x \srf}^{\ana} \frac{1}{H} \parder{H}{x} \\
  \label{a_2_two}
  A_2(\rankone{x},t)
  =&+ \left( \frac{S - z}{H} \right)^{\lambda} \left( \parder{u_{x \bed}^{\ana}}{x} - \parder{u_{x \srf}^{\ana}}{x} \right) \notag \\
   &+ \zeta \left( 
    - \frac{1}{\lambda} \frac{1}{H} \parder{H}{x}
    + \frac{1}{H} \frac{z - B}{S - z} \parder{S}{x}
    + \frac{1}{H} \parder{B}{x}
    - \frac{1}{S - z} \parder{S}{x}
    + \frac{1}{H} \parder{H}{x}
      \right).
\end{align}
Further simplification of the $A_2$ coefficient,
\begin{align*}
  A_2(\rankone{x},t)
  =&+ \left( \frac{S - z}{H} \right)^{\lambda} \left( \parder{u_{x \bed}^{\ana}}{x} - \parder{u_{x \srf}^{\ana}}{x} \right) \\
   &+ \zeta \left(
    - \frac{1}{\lambda} \frac{1}{H} \parder{H}{x}
    + \frac{1}{H} \frac{z - B}{S - z} \parder{S}{x}
    + \cancel{\frac{1}{H} \parder{B}{x}}
    - \frac{1}{S - z} \parder{S}{x}
    + \frac{1}{H} \parder{S}{x}
    - \cancel{\frac{1}{H} \parder{B}{x}},
      \right)
\end{align*}
and again
\begin{align*}
  A_2(\rankone{x},t)
  =&+ \left( \frac{S - z}{H} \right)^{\lambda} \left( \parder{u_{x \bed}^{\ana}}{x} - \parder{u_{x \srf}^{\ana}}{x} \right) \\
   &+ \zeta \left(
    - \frac{1}{\lambda} \frac{1}{H} \parder{S}{x}
    + \frac{1}{\lambda} \frac{1}{H} \parder{B}{x}
    + \frac{1}{H} \frac{z - B}{S - z} \parder{S}{x}
    - \frac{1}{S - z} \parder{S}{x}
    + \frac{1}{H} \parder{S}{x}
      \right),
\end{align*}
and again
\begin{align*}
  A_2(\rankone{x},t)
  =&+ \left( \frac{S - z}{H} \right)^{\lambda} \left( \parder{u_{x \bed}^{\ana}}{x} - \parder{u_{x \srf}^{\ana}}{x} \right) \\
   &+ \zeta \left(
      \left(
    - \frac{1}{\lambda} \frac{1}{H}
    + \frac{1}{H} \frac{z - B}{S - z}
    - \frac{1}{S - z}
    + \frac{1}{H}
      \right) \parder{S}{x}
    + \frac{1}{\lambda} \frac{1}{H} \parder{B}{x}
      \right),
\end{align*}
and again
\begin{align*}
  A_2(\rankone{x},t)
  =&+ \left( \frac{S - z}{H} \right)^{\lambda} \left( \parder{u_{x \bed}^{\ana}}{x} - \parder{u_{x \srf}^{\ana}}{x} \right) \\
   &+ \zeta \left(
      \left(
    - \frac{1}{\lambda} \frac{1}{H} \frac{S - z}{S - z} 
    + \frac{\lambda}{\lambda} \frac{1}{H} \frac{z - B}{S - z}
    - \frac{\lambda}{\lambda} \frac{H}{H} \frac{1}{S - z}
    + \frac{\lambda}{\lambda} \frac{1}{H} \frac{S - z}{S - z} 
      \right) \parder{S}{x}
    + \frac{1}{\lambda} \frac{1}{H} \parder{B}{x}
      \right),
\end{align*}
and again
\begin{align*}
  A_2(\rankone{x},t)
  =&+ \left( \frac{S - z}{H} \right)^{\lambda} \left( \parder{u_{x \bed}^{\ana}}{x} - \parder{u_{x \srf}^{\ana}}{x} \right) \\
   &+ \zeta \left(
      \left( 
      \frac{
    - (S - z)
    + \lambda (z - B)
    - \lambda H 
    + \lambda (S - z)
      }{\lambda H (S - z)}
      \right) \parder{S}{x}
    + \frac{1}{\lambda} \frac{1}{H} \parder{B}{x}
      \right),
\end{align*}
and again
\begin{align*}
  A_2(\rankone{x},t)
  =&+ \left( \frac{S - z}{H} \right)^{\lambda} \left( \parder{u_{x \bed}^{\ana}}{x} - \parder{u_{x \srf}^{\ana}}{x} \right) \\
   &+ \zeta \left(
      \left( 
      \frac{
    - S + z
    + \cancel{\lambda z} 
    - \cancel{\lambda B}
    - \cancel{\lambda S}
    + \cancel{\lambda B}
    + \cancel{\lambda S}
    - \cancel{\lambda z}
      }{\lambda H (S - z)}
      \right) \parder{S}{x}
    + \frac{1}{\lambda} \frac{1}{H} \parder{B}{x}
      \right),
\end{align*}
and again
\begin{align*}
  A_2(\rankone{x},t)
  =&+ \left( \frac{S - z}{H} \right)^{\lambda} \left( \parder{u_{x \bed}^{\ana}}{x} - \parder{u_{x \srf}^{\ana}}{x} \right)
    + \zeta \left(
      \left( 
      \frac{z- S}{\lambda H (S - z)}
      \right) \parder{S}{x}
    + \frac{1}{\lambda} \frac{1}{H} \parder{B}{x}
      \right).
\end{align*}
Re-substitution of $\zeta = \left( u_{x \srf}^{\ana} - u_{x \bed}^{\ana} \right) \lambda \left( \frac{S - z}{H} \right)^{\lambda}$ results in
\begin{align*}
  A_2(\rankone{x},t)
  =&+ \left( \frac{S - z}{H} \right)^{\lambda} \left( \parder{u_{x \bed}^{\ana}}{x} - \parder{u_{x \srf}^{\ana}}{x} \right) \\
   &+ \left( u_{x \srf}^{\ana} - u_{x \bed}^{\ana} \right) \lambda \left( \frac{S - z}{H} \right)^{\lambda} \left( \frac{z- S}{\lambda H (S - z)} \right) \parder{S}{x} \\
   &+ \left( u_{x \srf}^{\ana} - u_{x \bed}^{\ana} \right) \lambda \left( \frac{S - z}{H} \right)^{\lambda} \left( \frac{1}{\lambda} \frac{1}{H} \parder{B}{x} \right),
\end{align*}
so that simplifying again
\begin{align*}
  A_2(\rankone{x},t)
  =&+ \left( \frac{S - z}{H} \right)^{\lambda} \left( \parder{u_{x \bed}^{\ana}}{x} - \parder{u_{x \srf}^{\ana}}{x} \right) \\
   &+ \left( \frac{S - z}{H} \right)^{\lambda} \left( u_{x \srf}^{\ana} - u_{x \bed}^{\ana} \right) \cancel{\lambda} \left( \frac{z- S}{\cancel{\lambda} H (S - z)} \right) \parder{S}{x} \\
   &+ \left( \frac{S - z}{H} \right)^{\lambda} \left( u_{x \srf}^{\ana} - u_{x \bed}^{\ana} \right) \cancel{\lambda} \left( \frac{1}{\cancel{\lambda}} \frac{1}{H} \parder{B}{x} \right),
\end{align*}
and again
\begin{align*}
  A_2(\rankone{x},t)
  =&+ \left( \frac{S - z}{H} \right)^{\lambda} \left( \parder{u_{x \bed}^{\ana}}{x} - \parder{u_{x \srf}^{\ana}}{x} + \left( u_{x \srf}^{\ana} - u_{x \bed}^{\ana} \right) \left( \frac{1}{H} \parder{B}{x} \right) \right) \\
   &+ \left( \frac{S - z}{H} \right)^{\lambda} \left( \frac{z- S}{H (S - z)} \right) \left( u_{x \srf}^{\ana} - u_{x \bed}^{\ana} \right) \parder{S}{x},
\end{align*}
and again
\begin{align*}
  A_2(\rankone{x},t)
  =&+ \frac{(S - z)^{\lambda}}{H^{\lambda}} \left( \parder{u_{x \bed}^{\ana}}{x} - \parder{u_{x \srf}^{\ana}}{x} + \left( u_{x \srf}^{\ana} - u_{x \bed}^{\ana} \right) \left( \frac{1}{H} \parder{B}{x} \right) \right) \\
   &+ \frac{(S - z)^{\lambda - 1} \left( z - S \right) }{H^{\lambda + 1}} \left( u_{x \srf}^{\ana} - u_{x \bed}^{\ana} \right) \parder{S}{x}.
\end{align*}
Therefore, combining \cref{g_one,c_one,a_1_two,a_2_two} with $A = A_1 + A_2$, relation \cref{analytic_y_velocity_problem} has been derived.

\subsection{Invariant coordinate \cref{constant_0}}
\label{sec_sup_z_coordinate}

The relationship between the two right-most differential terms from \cref{lagrange_charpit} can be stated as
\begin{align*}
  \totder{z}{y} &= \frac{1}{H} \left( z \parder{H}{y} + S \parder{B}{y} - B \parder{S}{y} \right) \\
  \totder{z}{y} - \frac{z}{H} \parder{H}{y} &= \frac{1}{H} \left( S \parder{B}{y} - B \parder{S}{y} \right) \\
  \frac{1}{H} \totder{z}{y} - \frac{z}{H^2} \parder{H}{y} &= \frac{1}{H^2} \left( S \parder{B}{y} - B \parder{S}{y} \right) \\
  \frac{1}{H} \totder{z}{y} - \frac{z}{H^2} \parder{H}{y} &= \frac{1}{H^2} \left( S \parder{B}{y} - B \parder{B}{y} - B \parder{S}{y} + B \parder{B}{y} \right) \\
  \frac{1}{H^2} \left( H \totder{z}{y} - z \parder{H}{y} \right) &= \frac{1}{H^2} \left( H \parder{B}{y} - B \parder{H}{y} \right),
\end{align*}
applying the differential quotient rule to both sides yields
\begin{align*}
  \parder{}{y}\left( \frac{z}{H} \right) &= \parder{}{y} \left( \frac{B}{H} \right).
\end{align*}
Integrating with respect to $y$ produces relation \cref{constant_0}.

\subsection{Invariant coordinate \cref{constant_1}}
\label{sec_sup_u_y_coordinate}

The relationship between the two left-most differential terms from \cref{lagrange_charpit} can be stated as
\begin{align*}
  \totder{u_y^{\ana}}{y} &= -A - G u_y^{\ana} \\
  \totder{u_y^{\ana}}{y} + \frac{u_y^{\ana}}{H} \parder{H}{y} &= -A \\
  H \totder{u_y^{\ana}}{y} + u_y^{\ana} \parder{H}{y} &= -HA \\
  \parder{}{y} \left( H u_y^{\ana} \right) &= -HA,
\end{align*}
which integrated with respect to $y$ produces for some constant $u_{y 0}$,
\begin{align*}
  H u_y^{\ana} = &- \int_y HA \d{y} + u_{y 0} \\
  = &- \int_y H A_1(x,y,t) \d{y} - \int_y H A_2(x,y,z,t) \d{y} + u_{y 0} \\
  = &- \int_y H \left[ \parder{u_{x \srf}^{\ana}}{x} + \cancel{\frac{1}{H}} \left( \parder{H}{t} - \Vert \hat{\rankone{k}} - \nabla S \Vert \smb - \Vert \nabla B - \hat{\rankone{k}} \Vert \bmb \right) + u_{x \srf}^{\ana} \cancel{\frac{1}{H}} \parder{H}{x} \right] \d{y} \\
  &- \int_y H 
     \left[ \frac{(S - z)^{\lambda}}{H^{\lambda}} \left( \parder{u_{x \bed}^{\ana}}{x} - \parder{u_{x \srf}^{\ana}}{x} + \left( u_{x \srf}^{\ana} - u_{x \bed}^{\ana} \right) \left( \frac{1}{H} \parder{B}{x} \right) \right) \right] \d{y} \\
  &- \int_y H 
     \left[ \frac{(S - z)^{\lambda - 1} \left( z - S \right) }{H^{\lambda + 1}} \left( u_{x \srf}^{\ana} - u_{x \bed}^{\ana} \right) \parder{S}{x} \right] \d{y} + u_{y 0}. 
\end{align*}
Simplifying this relation,
\begin{align*}
  H u_y^{\ana} 
  = &- \int_y \left[ H \parder{u_{x \srf}^{\ana}}{x} + \parder{H}{t} - \Vert \hat{\rankone{k}} - \nabla S \Vert \smb - \Vert \nabla B - \hat{\rankone{k}} \Vert \bmb + u_{x \srf}^{\ana} \parder{H}{x} \right] \d{y} \\
  &- \int_y
     \left[ \frac{(S - z)^{\lambda}}{H^{\lambda-1}} \left( \parder{u_{x \bed}^{\ana}}{x} - \parder{u_{x \srf}^{\ana}}{x} + \left( u_{x \srf}^{\ana} - u_{x \bed}^{\ana} \right) \left( \frac{1}{H} \parder{B}{x} \right) \right) \right] \d{y} \\
  &- \int_y
     \left[ \frac{(S - z)^{\lambda - 1} \left( z - S \right) }{H^{\lambda}} \left( u_{x \srf}^{\ana} - u_{x \bed}^{\ana} \right) \parder{S}{x} \right] \d{y} + u_{y 0} 
\end{align*}
and substituting $z = z_0 H + B$,
\begin{align*}
  H u_y^{\ana} 
  = &- \int_y \left[ H \parder{u_{x \srf}^{\ana}}{x} + \parder{H}{t} - \Vert \hat{\rankone{k}} - \nabla S \Vert \smb - \Vert \nabla B - \hat{\rankone{k}} \Vert \bmb + u_{x \srf}^{\ana} \parder{H}{x} \right] \d{y} \\
  &- \int_y
     \left[ \frac{(S - z_0 H - B)^{\lambda}}{H^{\lambda-1}} \left( \parder{u_{x \bed}^{\ana}}{x} - \parder{u_{x \srf}^{\ana}}{x} + \left( u_{x \srf}^{\ana} - u_{x \bed}^{\ana} \right) \left( \frac{1}{H} \parder{B}{x} \right) \right) \right] \d{y} \\
  &- \int_y
     \left[ \frac{(S - z_0 H - B)^{\lambda - 1} \left( z_0 H + B - S \right) }{H^{\lambda}} \left( u_{x \srf}^{\ana} - u_{x \bed}^{\ana} \right) \parder{S}{x} \right] \d{y} + u_{y 0}, 
\end{align*}
this can be further simplified to
\begin{align*}
  H u_y^{\ana} 
  = &- \int_y \left[ H \parder{u_{x \srf}^{\ana}}{x} + \parder{H}{t} - \Vert \hat{\rankone{k}} - \nabla S \Vert \smb - \Vert \nabla B - \hat{\rankone{k}} \Vert \bmb + u_{x \srf}^{\ana} \parder{H}{x} \right] \d{y} \\
  &- \int_y
     \left[ \frac{(H - z_0 H)^{\lambda}}{H^{\lambda-1}} \left( \parder{u_{x \bed}^{\ana}}{x} - \parder{u_{x \srf}^{\ana}}{x} + \left( u_{x \srf}^{\ana} - u_{x \bed}^{\ana} \right) \left( \frac{1}{H} \parder{B}{x} \right) \right) \right] \d{y} \\
  &- \int_y
     \left[ \frac{(H - z_0 H)^{\lambda - 1} \left( z_0 H - H \right) }{H^{\lambda}} \left( u_{x \srf}^{\ana} - u_{x \bed}^{\ana} \right) \parder{S}{x} \right] \d{y} + u_{y 0},
\end{align*}
and again
\begin{align*}
  H u_y^{\ana} 
  = &- \int_y \left[ H \parder{u_{x \srf}^{\ana}}{x} + \parder{H}{t} - \Vert \hat{\rankone{k}} - \nabla S \Vert \smb - \Vert \nabla B - \hat{\rankone{k}} \Vert \bmb + u_{x \srf}^{\ana} \parder{H}{x} \right] \d{y} \\
  &- \int_y
     \left[ \frac{H^{\lambda}(1 - z_0)^{\lambda}}{H^{\lambda-1}} \left( \parder{u_{x \bed}^{\ana}}{x} - \parder{u_{x \srf}^{\ana}}{x} + \left( u_{x \srf}^{\ana} - u_{x \bed}^{\ana} \right) \left( \frac{1}{H} \parder{B}{x} \right) \right) \right] \d{y} \\
  &- \int_y
     \left[ \frac{H^{\lambda - 1}(1 - z_0)^{\lambda - 1} H(z_0 - 1) }{H^{\lambda}} \left( u_{x \srf}^{\ana} - u_{x \bed}^{\ana} \right) \parder{S}{x} \right] \d{y} + u_{y 0},
\end{align*}
and again
\begin{align*}
  H u_y^{\ana} 
  = &- \int_y \left[ H \parder{u_{x \srf}^{\ana}}{x} + \parder{H}{t} - \Vert \hat{\rankone{k}} - \nabla S \Vert \smb - \Vert \nabla B - \hat{\rankone{k}} \Vert \bmb + u_{x \srf}^{\ana} \parder{H}{x} \right] \d{y} \\
  &- \int_y
     \left[ \frac{(1 - z_0)^{\lambda}}{H^{-1}} \left( \parder{u_{x \bed}^{\ana}}{x} - \parder{u_{x \srf}^{\ana}}{x} + \left( u_{x \srf}^{\ana} - u_{x \bed}^{\ana} \right) \left( \cancel{\frac{1}{H}} \parder{B}{x} \right) \right) \right] \d{y} \\
  &- \int_y
     \left[ \frac{\cancel{H^{\lambda}}(1 - z_0)^{\lambda - 1} (z_0 - 1) }{\cancel{H^{\lambda}}} \left( u_{x \srf}^{\ana} - u_{x \bed}^{\ana} \right) \parder{S}{x} \right] \d{y} + u_{y 0}, 
\end{align*}
and again
\begin{align*}
  H u_y^{\ana} 
  = &- \int_y \left[ H \parder{u_{x \srf}^{\ana}}{x} + \parder{H}{t} - \Vert \hat{\rankone{k}} - \nabla S \Vert \smb - \Vert \nabla B - \hat{\rankone{k}} \Vert \bmb + u_{x \srf}^{\ana} \parder{H}{x} \right] \d{y} \\
  &- (1 - z_0)^{\lambda} \int_y
     \left[ H \left( \parder{u_{x \bed}^{\ana}}{x} - \parder{u_{x \srf}^{\ana}}{x} \right) + \left( u_{x \srf}^{\ana} - u_{x \bed}^{\ana} \right) \parder{B}{x} \right] \d{y} \\
  &- (1 - z_0)^{\lambda - 1} (z_0 - 1) \int_y
     \left[ \left( u_{x \srf}^{\ana} - u_{x \bed}^{\ana} \right) \parder{S}{x} \right] \d{y} + u_{y 0} 
\end{align*}
and substituting $z_0 = (z - B) / H $ produces \cref{constant_1}.

\subsection{Surface expressions \cref{analytic_y_velocity_sb}}
\label{sec_sup_u_y_surface}

Evaluating expression \cref{analytic_y_velocity} at the upper surface,
\begin{align}
  \label{analytic_y_velocity_s}
  u_y^{\ana}(x,y,z=S,t) 
  = &- \frac{1}{H} \int_y \left[ H \parder{u_{x \srf}^{\ana}}{x} + \parder{H}{t} - \Vert \hat{\rankone{k}} - \nabla S \Vert \smb - \Vert \nabla B - \hat{\rankone{k}} \Vert \bmb + u_{x \srf}^{\ana} \parder{H}{x} \right] \d{y} - \frac{1}{H} \vartheta \left( \xi_{\srf} \right) \notag \\
  = &- \frac{1}{H} \int_y \left[ \parder{}{x} \left( H u_{x \srf}^{\ana} \right) + \parder{H}{t} - \Vert \hat{\rankone{k}} - \nabla S \Vert \smb - \Vert \nabla B - \hat{\rankone{k}} \Vert \bmb \right] \d{y} - \frac{1}{H} \vartheta \left( \xi_{\srf} \right) \notag \\
  = &- \frac{1}{H} \int_y \left[ \parder{}{x} \left( H u_{x \srf}^{\ana} \right) + \parder{H}{t} - \mathring{f} \right] \d{y} - \frac{1}{H} \vartheta \left( \xi_{\srf} \right).
\end{align}
Evaluating expression \cref{analytic_y_velocity} at the lower surface,
\begin{align}
  \label{analytic_y_velocity_b}
  u_y^{\ana}(x,y,z=B,t) 
  = &- \frac{1}{H} \int_y \left[ \cancel{H \parder{u_{x \srf}^{\ana}}{x}} + \parder{H}{t} - \Vert \hat{\rankone{k}} - \nabla S \Vert \smb - \Vert \nabla B - \hat{\rankone{k}} \Vert \bmb + \cancel{u_{x \srf}^{\ana} \parder{S}{x}} - \cancel{u_{x \srf}^{\ana} \parder{B}{x}} \right] \d{y} \notag \\
  &- \frac{1}{H} \int_y
     \left[ H \parder{u_{x \bed}^{\ana}}{x} - \cancel{H \parder{u_{x \srf}^{\ana}}{x}} + \cancel{u_{x \srf}^{\ana} \parder{B}{x}} - u_{x \bed}^{\ana} \parder{B}{x} \right] \d{y} \notag \\
  &+ \frac{1}{H} \int_y
     \left[ \cancel{u_{x \srf}^{\ana} \parder{S}{x}} - u_{x \bed}^{\ana} \parder{S}{x} \right] \d{y} - \frac{1}{H} \vartheta \left( \xi_{\srf} \right) \notag \\
  = &- \frac{1}{H} \int_y \left[ \parder{H}{t} - \Vert \hat{\rankone{k}} - \nabla S \Vert \smb - \Vert \nabla B - \hat{\rankone{k}} \Vert \bmb + H \parder{u_{x \bed}^{\ana}}{x} + u_{x \bed}^{\ana} \parder{H}{x} \right] \d{y} - \frac{1}{H} \vartheta \left( \xi_{\srf} \right) \notag \\
  = &- \frac{1}{H} \int_y \left[ \parder{}{x} \left( H u_{x \bed}^{\ana} \right) + \parder{H}{t} - \Vert \hat{\rankone{k}} - \nabla S \Vert \smb - \Vert \nabla B - \hat{\rankone{k}} \Vert \bmb \right] \d{y} - \frac{1}{H} \vartheta \left( \xi_{\srf} \right) \notag \\
  = &- \frac{1}{H} \int_y \left[ \parder{}{x} \left( H u_{x \bed}^{\ana} \right) + \parder{H}{t} - \mathring{f} \right] \d{y} - \frac{1}{H} \vartheta \left( \xi_{\srf} \right).
\end{align}
Therefore, setting $\vartheta(a_{\srf}) = 0$ and $i = \srf,\bed$, relations \cref{analytic_y_velocity_s,analytic_y_velocity_b} imply relation \cref{analytic_y_velocity_sb}.

\section{Python source code}
\label{sec_sup_source}

All source code used to generate the results of this work are provided in this section.

\pythonexternal[caption={Python source code used to generate Antarctica values in \cref{tab_surface_area,tab_r2_error}.}, label={ant_surface_error_script}]{scripts/antarctica/calc_error_s.py}

\pythonexternal[caption={Python source code used to generate Antarctica \cref{fig_ronne_grad_s,fig_ronne_grad_b}.}, label={ant_surface_image_script}]{scripts/antarctica/paper_calc_melt.py}

\pythonexternal[caption={Python source code used to generate Greenland \cref{fig_jakob_grad_s,fig_jakob_grad_b}.}, label={gre_surface_image_script}]{scripts/greenland/paper_calc_melt.py}

\pythonexternal[caption={Python source code used to generate analytic-$x$ component of velocity \cref{fig_analytic_velocity}.}, label={u_ana_script}]{scripts/plot_u_ana.py}

\pythonexternal[caption={Python source code used to generate analytic solution \cref{fig_r3}.}, label={r3_script}]{scripts/r3_mass_balance/r3_mb.py}

\end{document}